\documentclass[preprint,showpacs,preprintnumbers,amsmath,amssymb]{revtex4}

\usepackage{graphicx}
\usepackage{epsfig}		
\usepackage{dcolumn}
\usepackage{bm}
\usepackage{review}
\usepackage{mathrsfs}
\usepackage[T1]{fontenc} 
\usepackage{hyperref}

\def\bb#1{\mbox{\footnotesize $(#1)$}}

\begin{document}
%

\title{Quantum flavor oscillations extended to the Dirac theory}
\author{A. E. Bernardini}
\email{alexeb@ufscar.br}
\affiliation{Departamento de F\'{\i}sica, Universidade Federal de S\~ao Carlos, PO Box 676, 13565-905, S\~ao Carlos, SP, Brasil}
\author{M. M. Guzzo}
\email{guzzo@ifi.unicamp.br}
\affiliation{Instituto de F\'{\i}sica Gleb Wataghin, Universidade Estadual de Campinas, PO Box 6165, 13083-970,
Campinas, SP, Brasil}
\author{C. C. Nishi}
\email{celso.nishi@ufabc.edu.br}
\affiliation{Universidade Federal do ABC,
Rua Santa Ad\'elia, 166,\linebreak 09.210-170, Santo Andr\'e, SP, Brasil}

\date{\today}

\begin{abstract}
\medskip
\setlength{\baselineskip}{.85\baselineskip}
Flavor oscillations by itself and its coupling with chiral oscillations and/or spin-flipping are the most relevant quantum phenomena of neutrino physics.
This report deals with the quantum theory of flavor oscillations in vacuum, extended to fermionic particles in the several subtle aspects of the first quantization and second quantization theories.
At first, the basic controversies regarding quantum-mechanical derivations of the flavor conversion formulas are reviewed based on the internal wave packet (IWP) framework.
In this scenario, the use of the Dirac equation is required for a satisfactory evolution of {\em fermionic} mass-eigenstates since in the standard treatment of oscillations the mass-eigenstates are implicitly assumed to be {\em scalars} and, consequently, the spinorial form of neutrino wave functions is {\em not} included in the calculations.
Within first quantized theories, besides flavor oscillations, chiral oscillations automatically appear when we set the dynamic equations for a {\em fermionic} Dirac-{\em type} particle.
It is also observed that there is no constraint between chiral oscillations, when it takes place in vacuum, and the process of spin-flipping related to the helicity quantum number, which does not take place in vacuum.
The left-handed chiral nature of created and detected neutrinos can be implemented in the first quantized Dirac theory in presence of mixing; the probability loss due to the changing of initially left-handed neutrinos to the undetected right-handed neutrinos can be obtained in analytic form.
These modifications introduce correction factors proportional to $m_{\nu}^{\2}/E_{\nu}^{\2}$ that are very difficult to be quantified by the
current phenomenological analysis.
All these effects can also be identified when the non-minimal coupling with an external (electro)magnetic field in the neutrino interacting Lagrangian is taken into account.
In the context of a causal relativistic theory of a free particle, one of the two effects should be present in flavor oscillations: (a) rapid oscillations or (b) initial flavor violation.
Concerning second quantized approaches, a simple second quantized treatment exhibits a tiny but inevitable initial flavor violation without the possibility of rapid oscillations.
Such effect is a consequence of an intrinsically indefinite but approximately well defined neutrino flavor.
Within a realistic calculation in pion decay, including the quantum field treatment of the creation process with finite decay width, it is possible to quantify such violation.
The violation effects are shown to be much larger than loop induced lepton flavor violation processes, already present in the standard model in the presence of massive neutrinos with mixing.
For the implicitly assumed {\em fermionic} nature of the Dirac theory, the conclusions of this report lead to lessons concerning flavor mixing, chiral oscillations, interference between positive and negative frequency components of Dirac equation solutions, and the field formulation of quantum oscillations.
\end{abstract}



\maketitle
\setlength{\baselineskip}{.9\baselineskip}
\tableofcontents
\setlength{\baselineskip}{1.11\baselineskip}

\newpage
\section{Introduction}

Particle mixing and flavor oscillations \cite{Gel55,Pic55,Gri69,Bil76,Fri76} continue to stimulate interesting and sometimes fascinating discussions on the many subtleties of quantum mechanics involved in oscillation phenomena \cite{Zra98,akhmedov:paradoxes,glashow:no}.
The flavor mixing models \cite{Nir03}, the quantum field prescriptions \cite{Kob82,Bla95,Beu03,Giu02,GiuBla,Bla03,akhmedov:qft} and, generically, the quantum mechanics of oscillation phenomena \cite{Giunti:91,Ber08B,Ber08A,Vog04,Giu98,Ber05,Akh10}
have been extensively studied in the last years.
In particular, the properties of neutrinos \cite{Zub98,Alb03} obtained in all of these frameworks have become the subject of an increasing number of theoretical constructions.
Notwithstanding the exceptional ferment in this field, the numerous conceptual difficulties in describing accurately the particle mixing and oscillations have renewed the interest in understanding the derivation of the flavor conversion probability formulas and in overcoming the main physical inconsistencies hidden in the standard theoretical approaches.

The flavor oscillation analysis have been supported by compelling experimental evidences which have continuously ratified that neutrinos undergo flavor oscillations in vacuum and in matter.
One can focus, for instance, on the outstanding results of the Super-Kamiokande atmospheric neutrino experiment \cite{Fuk02}, in which a significant up-down asymmetry of the high-energy muon events was observed, the results of the SNO solar neutrino experiment \cite{Ahm02,Ban03}, in which a direct evidence for the transition of the solar electron neutrinos into other flavors was obtained, and also the results of the KamLAND experiment \cite{Egu03} that confirmed that the disappearance of solar electron neutrinos is mainly due to oscillations among active neutrinos and not due to other types of neutrino conversion mechanisms \cite{Guz02,Bar02B}.
The experimental data could be completely interpreted and understood in terms of three neutrino flavors, with the exception of the LSND anomaly\,\cite{Ban03,Ana98,Agu01}.
Such anomaly, although not confirmed by the MiniBoone experiment\,\cite{Agu07,miniboone:prl09}, led to speculations of the existence of (at least) a fourth light neutrino flavor which had to be inert.
The presence of such light sterile neutrinos is largely excluded by the oscillation data\,\cite{minos:10,miniboone:sterile} but depending on their mass scale it may influence certain astrophysical and cosmological phenomena ranging from the thermal evolution of the Universe\,\cite{wmap:10} to supernova explosion, pulsar kicks and even a significant part of dark matter\,\cite{kusenko:09}.
On the other hand, the hypothesis of mixing between known neutrino species (electron, muon and tau) and higher mass neutrinos has much stronger theoretical motivation since it may account for the lightness of the active neutrinos through the seesaw mechanism\,\cite{Gel79,Ma:prl98} and also account for the matter and antimatter asymmetry of the Universe through the mechanism of leptogenesis\,\cite{leptog}. The observation of such mixing at low energies, however, could be extremely difficult.

In parallel, the neutrino spin-flipping attributed to some dynamic external \cite{Oli90} interacting process, which comes from the non-minimal coupling of a magnetic moment with an external electromagnetic field \cite{Vol81}, was formerly supposed to be a relevant effect in the context of the solar-neutrino puzzle.
As a consequence of a non-vanishing magnetic moment interacting with an external electromagnetic field, left-handed neutrinos could change their helicity (to right-handed)\cite{Bar96}.
The effects on flavor oscillations due to external magnetic interactions in a kind of chirality-preserving phenomenon were also studied \cite{Oli96} but they lack a full detailed theoretical analysis.
It was partially provided by some recent theoretical studies where the Dirac/Majorana characteristic of neutrinos becomes relevant \cite{Ber08A,Dvo09,Dvo02,Bla09}.
Only for ultra-relativistic (UR) neutrinos, however, changing helicity approximately means changing chirality.
One of our goals is to demonstrate that, in the context of oscillation phenomena and in the framework of a first quantized theory, the small differences between the concepts of chirality and helicity, which had been interpreted as representing the same physical quantities for massless particles \cite{Oli90,Vol81,Bar96,Oli96,Kim93,DeL98,Dvo02}, can be quantified for massive particles.
It raises the possibility that chirality coupled to flavor oscillations could lead to some small modifications to the standard flavor conversion formula \cite{Ber05}.

On the theoretical background, under the point of view of a first quantized theory, the treatment of the flavor oscillation phenomena in terms of the intermediate wave packet (IWP) approach \cite{Kay81} eliminates the most controversial points arising
with the {\em standard} plane-wave formalism \cite{Kay89,Kay04}.
However, a common argument against the IWPs is that oscillating neutrinos are
neither prepared nor observed \cite{Beu03}.
This point was partially clarified by Giunti\,\cite{Giu02} who suggested a solution
in terms of an improved version of the IWP model where the wave packet (WP) of the
oscillating particle is explicitly computed with field-theoretical methods by means
of the external wave packet (EWP) approach\,\cite{Kob82,Beu03}.

Such approach, in contrast to IWP approaches, became the customary way to avoid the
ambiguities involving the question on how neutrinos are created and detected.
According to Ref. \cite{Beu03}, the IWP treatments are the simpler first quantized
ones treating the propagation of neutrinos as free localized wave packets.
In contrast, EWP approaches consider localized wave packets for the sources and
detection particles while the neutrinos were considered intermediate virtual
particles.

On the other hand, another classification scheme can be used do classify the various
existing treatments considering a more physical criterion irrespective of the use of
wave packets.
It refers to the descriptions of neutrino oscillations that (A) include explicitly the
interactions responsible for the mixing and those (B) that only treat the propagation
of neutrinos, i.\,e., the mixing is an {\it ad hoc} ingredient.
A more subtle aspect in between would be the (explicit or phenomenologically modeled) consideration of the production (and detection) process(es).
In general, the IWP approaches are of type (B).
The EWP approaches are of type (A).
The Blasone and Vitiello approach\,\cite{Bla95}, although in the quantum field theory (QFT), is of type (B) since mixing is introduced without explicitly including the interaction responsible for it.
The type (B) approaches have the virtue that they can be formulated in a way in which
total oscillation probability in time is always conserved and normalized to unity
\cite{DeL04,Bla95,Ber04,Ber04B,ccn:no12}. This feature will be present
in all first quantized approaches
treated here (sections\;\ref{sec:localization} and \;\ref{sec:relativistic}) and in a
second quantized version (section\;\ref{subsec:simpleQFT}).
If different observables are considered, or a modeling of the details of the
production and detection processes is attempted, further normalization is necessary
\cite{Giunti:91,Giu98,BlasoneP:03}.
In such cases, the oscillating observable might differ from the oscillation probability.
On the other hand, type (A) approaches tend to be more realistic and can account for the production and detection processes giving experimentally observable oscillation probabilities \cite{Cardall.00}.
Of course, they are essential to the investigation of how neutrinos are produced and
detected \cite{Kie96,Kie97,Dolgov}.
To rigorously derive a flavor/chiral conversion formula for {\em fermionic} particles
(non-minimally coupled to an external magnetic field), we avoid the field
theoretical methods in the preliminary investigation.
As we are initially interested in the Dirac equation properties, as a first
analysis, the IWP framework is a suitable simplification for the understanding of the
physical aspects concerning the oscillation phenomena.

In section \ref{sec:localization} of this manuscript, the spatial localization is
included in the formulation of flavor oscillations.
Quite generally, the analytical description of the dynamical evolution of a mass-eigenstate do not involve the wave packet limitations.
In particular, the analytical properties of gaussian distributions\,\cite{Giu98,Giu02} enable us to quantify the first and the second order corrections to the oscillating behavior of propagating particles.
We assume sharply peaked momentum distributions and then we approximate the
mass-eigenstate energy in order to analytically obtain the expressions for the
wave packet time evolution and for the flavor oscillation probability.
In the IWP approach a consistent energy expansion is taken up to the second order
term in the wave packet parameter $\sigma_{\ii} \sim (a\,E_{\ii})^{-1}$, that
satisfies $ \sigma_{\ii}\ll 1$ for sharply peaked momentum distributions.
The wave packet spreading as well as the loss of coherence between the propagating wave packets are
quantified in both non-relativistic (NR) and ultra-relativistic (UR) propagation
regimes.
Thus, the preliminary step of our study consists in a self-consistent approximation to the mass-eigenstate energy in order to analytically obtain the expressions for the wave packet time evolution and for the flavor oscillation probability.
We also identify an additional time-dependent phase which changes the {\em
standard} oscillating character of the flavor conversion formula.

The extension to the Dirac theory is introduced in
section \ref{sec:relativistic}, where spin and
relativistic completeness are included into the flavor oscillation formulation.
This section is concerned with the analytical derivation of a flavor conversion
formula where the {\em fermionic} instead of the {\em scalar} character of a
propagating mass-eigenstate is assumed.
To that end we shall use the Dirac equation as the evolution equation for the
mass-eigenstates and show that the Dirac formalism is useful and essential in
keeping clear many of the conceptual aspects of quantum oscillation phenomena that
naturally arise in a relativistic spin one-half particle theory.
More particularly, we show that a superposition of both positive and negative
frequency solutions of the Dirac equation is often a necessary condition to
correctly describe the time evolution of the mass-eigenstate wave packets.
We give, for strictly peaked momentum distributions and UR particles,
an analytic expression for the Dirac flavor conversion probability.
A modified formula for the conversion probability is shown and an
additional \textit{rapid} oscillation term, coming from the interference between
the positive and negative frequency contributions, is found.
To completely disentangle the influence of the initial wave packet to the phenomenon,
an analysis independent of the initial wave packet is conducted through the
calculation of the Dirac time evolution kernel in the presence of flavor mixing. The
properties of completeness and causality are briefly analyzed.
An analogous calculation is performed for relativistic spin zero particles to show
that rapid oscillations are indeed a consequence of the presence of positive and
negative frequency solutions (completeness) for relativistic wave equations and not
of the spin degree of freedom itself.
Within first quantized Dirac theory, we also establish the inextricable relationship
between two phenomena: initial flavor violation and rapid oscillations.

Further consequences of spin structure and relativistic completeness, such as
chiral oscillations, are discussed in
section \ref{sec:consequences}.
Chiral oscillations naturally enter the discussion of neutrino oscillations
because neutrinos are produced and detected through weak interactions that are
chiral in nature, more specifically, left-handed in chirality.
It is shown that the inclusion of chiral oscillation effects, together with the
time-evolution of spinorial wave packets for the mass-eigenstates, can modify the
flavor conversion probability formula.
In particular, the probability loss due to the conversion of left-handed
to right-handed neutrinos is calculated.
The {\em differences} between the dynamics of chirality and
helicity for a neutrino non-minimally coupled to an external
(electro)magnetic field are expressed in terms of the equation of the motion of the
correspondent operators $\gamma^{\5}$ and $h$, respectively.
In particular, the oscillating effects can be explained as an implication of the {\em
zitterbewegung} (ZBW) phenomenon that emerges when the Dirac equation solution is
used for describing the time evolution of a wave packet \cite{Ber04}.
Due to this tenuous relation between ZBW and chiral oscillations, the question to be
analyzed in this section concerns with the {\em immediate} description of chiral
oscillations in terms of the ZBW motion, i.\,e., we shall demonstrate that, in fact,
chiral oscillations are coupled with the ZBW motion so that
they cannot exist independently of each other.
It provides the interpretation of chiral oscillations as very rapid oscillations in position along the direction of motion, i.e., longitudinal to the momentum of the particle.
In a subsequent step, we report about a further class of static properties of neutrinos, namely, the (electro)magnetic moment associated to the Lagrangian with non-minimal coupling.
It allows the comparison between the dynamics of chiral oscillations and the
dynamics of spin-flipping in the presence of an external magnetic field. It is also
verified how the interaction with an external (electro)magnetic field can modify the
neutrino flavor oscillation formula\,\cite{Ber04B}.
To summarize, the basic idea is thus to quantify the modifications that appear in
the flavor-chirality conversion formula, previously obtained for free propagating
particles in vacuum\,\cite{Ber05}, when an external magnetic field can affect
chiral oscillations.

In section \ref{sec:qft} we finally analyze the inclusion of some aspects of
field quantization into the description.
Firstly, a simple second quantized description of the flavor oscillation
phenomenon is devised based on free second quantized Dirac theory.
There is no interference term between positive and negative components, but it
still gives simple normalized oscillation probabilities.
We also review the central issue distinguishing the general IWP and
EWP\,\cite{Beu03,Giu93} approaches: despite its direct unobservability, is the
intermediate neutrino a real (on-shell) particle propagating freely?
The answer is affirmative except for the possibility of contribution
of the antineutrino component in the propagation of virtual
neutrinos which, nevertheless, is very small\,\cite{ccn:no12}.
Thus IWP description is a good approximation of the oscillation
phenomenon\,\cite{ccn:no12,Gri96}.
We also compare the distinct approach of Blasone and Vitiello \cite{Bla95} with the
first quantized description of neutrino oscillations.

We have also considered that a central issue of the phenomenon of flavor oscillations discussed along our manuscript is:
how the coherent superposition of mass-eigenstate neutrinos, i.\,e., the flavor
state, is created and detected\,\cite{Dolgov,Kie97,Ric93}?
To answer such question, it is necessary to explicitly consider the interactions
responsible for creation and/or detection.
Our contribution to such a fruitful discussion is a detailed calculation of the
creation probability of the neutrino produced through pion decay performed using
the full (perturbative) QFT formalism at tree level, with explicit inclusion of pion
localization.
As a result, it is possible to study how the localization properties of neutrinos
follows from the parent particle (pion) that decays.
The calculation then provides the missing ingredients to quantify the new effect of
intrinsic neutrino flavor violation\,\cite{ccn:intrinsic}. Such effect is already
present in first
quantized formulations and it is manifested as an initial flavor violation or flavor
indefinition but its presence was not mandatory and its magnitude could not be
calculated a priori\,\cite{DeL04}.
Moreover, we can show that the coherent creation of neutrino flavor states
follows from the common negligible contribution of neutrino masses to their creation
probabilities.
At the same time, in the strict sense, we can also conclude that
neutrino flavor is only an \textit{approximately} well defined concept.

The manuscript is structured in order to allow the reader to recognize the origin of
each novel ingredient that can be included in the description of the quantum flavor
oscillation phenomenon for neutrinos.
We draw our conclusions in section \ref{sec:conclusion}.

\section{Inclusion of spatial localization}
\label{sec:localization}

This section deals with the quantum-mechanical derivation of the oscillation formula,
as well as the inclusion of spatial localization through the IWP framework, as it has
been extensively discussed in the literature \cite{Kay89,Kay04,Zra98,Beu03}. Our main
contribution concerns the identification, through analytical expressions, of
secondary effects coupled with the wave packet decoherence, which introduce small
modifications to the oscillation pattern.
In particular, a preliminary discussion introducing the possibility of initial flavor
violation and the respective consequences at neutrino creation/propagation/detection,
is contextualized in the IWP framework.

Associating a plane wave with each mass-eigenstate \cite{Kay89,Kay04} is
certainly the simplest and probably the most intuitive way to describe the
interference phenomenon that gives rise to flavor oscillations in terms of an
oscillation length and an oscillation probability.
For oscillations between two different neutrino flavors that we will choose to be $\nue$ and $\numu$ by convenience, the probability for flavor transition is usually expressed in terms of the mixing angle $\theta$ and of the relative phase $\Delta \Phi$ by
\small\eq{
\mathcal{P}(\bs{\nue}\rightarrow\bs{\numu})
 = \sin^{\2} (2 \theta) \, \sin^{\2} \left[\frac{\Delta \Phi}{2}\right]
\label{II:standard}
}\normalsize
where the Lorentz invariant phase difference,
\small\eq{
\label{II00} \Delta \Phi = \Delta \, (E\,T-p\,L),
}\normalsize
sets that initial \textit{pure} flavor states are modified with time and distance.
The mass-eigenstate phase difference $\Delta \Phi$  is then conventionally evaluated
by setting $\Delta T=\Delta L=0$ and considering, for ultra-relativistic (UR)
particles, $T\approx L$ and $p_{\1, \2} \approx E_{\1, \2}$, i.\,e.
\small\eq{
\label{II000}
\Delta \Phi =   T \, \Delta E - L \, \Delta p \approx L
\, \left( \Delta E - \Delta p \right)
\approx  \frac{\Delta m^{\2}}{2 \bar{p}} \, L.
}\normalsize
One thus gets the well-known expression \cite{Kay04}
\small\eq{
\label{II0000}
\mathcal{P}(\bs{\nue}\rightarrow\bs{\numu};L) = \sin^{\2} (2 \theta)
 \sin^{\2} \left[ \frac{\Delta m^{\2}}{4 \bar{p}} \, L \right].
}\normalsize

Considering the plane wave approach, controversial points arises even in the
derivation of formulas containing extra factors in the oscillation length
\cite{Fie03,Giu01,Tak01}.
In particular, the extra factor of two in particle oscillation phases was discussed and refuted in the context of theory and phenomenology of neutral meson-antimeson oscillations \cite{Bil05CC}.
In addition, the simplified view of using wave packets (WPs) allows us to understand the origin of these extra
factors.
It is implicitly assumed that at creation the flavor state is unique
even up to the phase at all points and times of creation. In the wave packet
treatment, at time $T$ and at a fixed position in the overlapping region, one
experiences the interference between space points whose separation at creation is
given by $\Delta v \, T$ and this implies that an additional initial phase is
automatically included in the wave packet formalism \cite{Giu98,DeL04}.
The final result contains the difference of phase given in Eq.~(\ref{II000}).
We do not intend here to re-discuss the many controversies in the plane wave derivations of the oscillation probability formula.
We only remark that an approach strictly considering plane waves leads to conceptual
difficulties and fails to explain fundamental aspects of particle oscillations, such
as localization and coherence length.
Wave packets eliminate some of these problems \cite{Kay81}.
In fact, the use of wave packets for propagating mass-eigenstates (IWP model)
guarantees the existence of a coherence length, avoids the ambiguous approximations
in the plane wave derivation of the phase difference and, under particular conditions
of minimal loss of coherence, recovers the oscillation probability given in
Eq.~(\ref{II0000}).
Moreover, the coherence necessary for neutrino oscillations depends crucially on
localization aspects of the particles involved in the production of neutrinos
\cite{Kay81}.
This point of view can be supported by quantum field theory (QFT) arguments as
well\,\cite{Gri96,Giu02}.

In practice, the loss of coherence is only relevant for
neutrinos traveling cosmological distances\,\cite{farzan:cosmic}.
At the same time,
it is not easy to determine the size of the wave packets at creation
and it is not clear whether it makes sense to consider a unique time of creation
\cite{Ric93,DeL04}.
It configures a common argument against the IWP
formalism, i.\,e., oscillating neutrinos are neither prepared nor  observed.
Consequently, it would be more convenient to write a transition probability between the observable particles involved in the production and detection process.
This point of view characterizes the so-called EWP approach \cite{Giu02,Beu03}.
The oscillating particle, described as an internal line of a Feynman diagram by a relativistic mixed scalar propagator, propagates between the source and target (external) particles represented by wave packets.
The function which represents the overlap of the incoming and outgoing wave packets in the EWP model corresponds to the wave function of the propagating mass-eigenstate in the IWP formalism.
Remarkably, it could be shown that the probability densities for UR stable oscillating particles in both frameworks are mathematically equivalent \cite{Beu03}.
However, the IWP picture brings up a problem, as the overlap function takes into account not only the properties of the source, but also of the detector.
This is unusual for a wave packet interpretation and not satisfying for causality \cite{Beu03}.
This point was clarified by Giunti \cite{Giu02} who evaluates the problem by
proposing an improved version of the IWP model where the wave packet of the
oscillating particle is explicitly computed with field-theoretical methods in terms
of external wave packets. Despite of not being applied in a completely free way, the
(intermediate) wave packet treatment commonly simplifies the discussion of some
physical aspects associated to the oscillation phenomena \cite{DeL04,Tak01}.
Thus, it makes sense, as a preliminary investigation, to consider a wave packet
associated with the propagating particle.

\subsection{The IWP framework}

The main aspects of oscillation phenomena can be understood by studying the two flavor problem.
In this case, by associating the wave packets $\phi_{\1}$ and $\phi_{\2}$ to
mass-eigenstates $\bs{\nu_{\1}}$ and $\bs{\nu_{\2}}$, flavor wave packets can be
described by the $\nue$\,--\,like state vector
\small\eqarr{
\Phi\bb{\bx,t} &=&
\phi_{\1}\bb{\bx,t}\cos{\theta}\,\mbox{\boldmath$\nu_{\1}$} +
\phi_{\2}\bb{\bx,t}\sin{\theta}\,\mbox{\boldmath$\nu_{\2}$}
\nonumber\\
          &=& \left[\phi_{\1}\bb{\bx,t}\cos^{\2}{\theta} +
\phi_{\2}\bb{\bx,t}\sin^{\2}{\theta}\right]\,\bs{\nue}+
		  \left[\phi_{\2}\bb{\bx,t} -
\phi_{\1}\bb{\bx,t}\right]\cos{\theta}\sin{\theta}\,\bs{\numu}
\nonumber\\
          &=& \phi_{\nue}\bb{\bx,t;\theta}\,\bs{\nue} +
\phi_{\numu}\bb{\bx,t;\theta}\,\bs{\numu},
\label{II0}
}\normalsize
where $\bs{\nue}$ and $\bs{\numu}$ are flavor states that are related to
mass-eigenstates {\boldmath$\nu_{\1}$} and
{\boldmath$\nu_{\2}$} by the mixing relation
\small\eq{
\label{II:mixing}
\begin{pmatrix} \bs{\nue} \cr \bs{\numu}\end{pmatrix}
=
\begin{pmatrix}\phantom{-}\cos\theta & \sin\theta \cr
-\sin\theta & \cos\theta \end{pmatrix}
\begin{pmatrix} \bs{\nu_1} \cr \bs{\nu_2}\end{pmatrix}
\,.
}\normalsize
The mixing relation \eqref{II:mixing} can be also written
\small\eq{
\label{II:mixing:2}
\bs{\nu_\alpha}=U_{\alpha i}\bs{\nu_i}\,, ~~\alpha=e,\mu,~~i=1,2\,,
}\normalsize
where the mixing matrix $U$ can be easily extracted.

It is important do emphasize that $\phi_1(\bx,t)$ and $\phi_2(\bx,t)$ are usual
normalizable wave functions, normalized to unity, while $\phi_{\nue}(\bx,t;\theta)$
and $\phi_{\numu}(\bx,t;\theta)$ are not normalizable wave functions because of their
time dependent norms (flavor oscillation). What is normalizable, as we will see, is
the total probability over all flavors,
\small\eq{\nonumber
\int\!d^3\bx\Big[
|\phi_{\nue}\bb{\bx,t;\theta}|^2+|\phi_{\numu}\bb{\bx,t;\theta}|^2
\Big]
=1\,,
}
\normalsize
which is automatic, with $U$ in Eq.\,\eqref{II:mixing:2} being unitary, once
$\phi_{\1}$ and $\phi_{\2}$ are normalized to unity,
\small\eq{\nonumber
\int\!d^3\bx\,|\phi_{\1}\bb{\bx,t}|^2=1\,,\qquad
\int\!d^3\bx\,|\phi_{\2}\bb{\bx,t}|^2=1\,.
}
\normalsize
We should remark that, in general IWP treatments,
automatic normalization of total probability is not guaranteed
if the detection process is modeled by detection wave packets as in
Refs.\,\cite{Giunti:91,Giu98}, i.e., a further normalization procedure is necessary.
Therefore, within IWP treatments, we will not model the detection process and
only consider idealized measurements that are consistent with the above
normalization conditions.

In addition to the restriction to two families, substantial mathematical
simplifications result from the assumption that the space dependence of wave
functions is one-dimensional ($\bb{\bx,t} \rightarrow \bb{z,t}$).
Therefore, we shall use these simplifications to calculate the oscillation probabilities.
The complementary effects of a $3$-dimensional analysis are well explored in
Ref.\,\cite{Beu03}.
The probability of finding a flavor state $\bs{\numu}$ at the instant $t$ is equal to
the integrated squared modulus of the $\bs{\numu}$ coefficient
\small\eq{
\mathcal{P}(\bs{\nue}\rightarrow\bs{\numu};t)=
\int_{_{\infm}}^{^{\infp}}\hspace{-0.5cm}dz \,\left|\phi_{\numu}\right|^{\2}
=
\frac{\sin^{\2}{(2\theta)}}{2}\left\{\, 1 - \mbox{\sc Int}\bb{t} \, \right\},
\label{II1}
}\normalsize
where $\mbox{\sc Int}\bb{t}$ represents the mass-eigenstate interference term given by
\small\eq{
\mbox{\sc Int}\bb{t} = \mathrm{Re}
 \left[\, \int_{_{\infm}}^{^{\infp}}\hspace{-0.5cm}dz
\,\phi^{\dagger}_{\1}\bb{z,t} \, \phi_{\2}\bb{z,t} \, \right].\,
\label{II2}
}\normalsize

For mass-eigenstate wave packets given by
\small\eq{
\phi_{\ii}(z,0) = \left(\frac{2}{\pi a^{\2}}\right)^{ \frac{1}{4}} \exp{\left[- \frac{z^{\2}}{a^{\2}}\right]} \exp{[i p_{\ii} \, z]},
\label{II3}
}\normalsize
at time $t = 0$, where $i = 1,\, 2$, the corresponding time evolution is given by
\small\eqarr{
\phi_{\ii}\bb{z,t} =
\int_{_{\infm}}^{^{\infp}}\hspace{-0.25cm}\frac{dp_{\z}}{2 \pi} \,
\varphi(p_{\z} \mi p_{\ii}) \exp{\left[-i\,E^{(\ii)}_{p_{\z}}\,t +i \, p_{\z}
\,z\right]},
\label{II4}
}\normalsize
where
$E^{(\ii)}_{p_{\z}} = \left(p_{\z}^{\2} + m_{\ii}^{\2}\right)^{ \frac{1}{2}}$
and
$\varphi(p_{\z} \mi p_{\ii}) =  \left(2 \pi a^{\2} \right)^{ \frac{1}{4}}
\exp{\left[- \frac{(p_{\z} \mi p_{\ii})^{\2}\,a^{\2}}{4}\right]}.$
To obtain the oscillation probability, we can calculate the interference term
$\mbox{\sc Int}\bb{t}$ by integrating
\small\eqarr{
\lefteqn{\int_{_{\infm}}^{^{\infp}}\hspace{-0.25cm}\frac{dp_z}{2 \pi} \,  \varphi(p_z \mi p_{ 1}) \varphi(p_z \mi p_{ 2})
\exp{[-i \, \Delta E_{p_z} \, t]} =}\nonumber\\
&& \exp{\left[ \frac{\mi(a \, \Delta{p})^{\2}}{8}\right]}
\int_{_{\infm}}^{^{\infp}}\hspace{-0.25cm}\frac{dp_z}{2 \pi}  \, \varphi^{\2}(p_z
\mi p_{\0})\exp{[-i \, \Delta E_{p_z} \, t]},~~~
\label{II6}
}\normalsize
where we have changed the $z$-integration into a $p_{\z}$-integration and introduced
the quantities $\Delta p\equiv p_{ \1} \mi p_{ \2} ,\,\, p_{\0}\equiv
\frac{1}{2}(p_{\1} + p_{\2})$ and $\Delta E_{p_{\z}}\equiv E^{(\1)}_{p_{\z}} \mi
E^{(\2)}_{p_{\z}}$.
The oscillation term is bounded by the exponential function {\small
$\exp[-(a\,\Delta p)^2/8]$} at any instant of time.
Under this condition we would never observe a {\em pure} flavor state.
Moreover, oscillations are considerably suppressed if $a \, \Delta p > 1$.
Hence, a necessary condition to observe oscillations is that $a \, \Delta p \ll 1$.
This constraint can also be expressed by $\delta p \gg \Delta p$ where $\delta p$ is the momentum uncertainty of the particle.
The overlap between the momentum distributions is indeed relevant only for $\delta p \gg \Delta p$.
Strictly speaking, we are assuming that the oscillation length ($\pi\frac{4 \bar{E}}{\Delta m^{\2}_{\ii\jj}}$) is sufficiently larger than the wave packet width.
It simply says that the wave packet must not extend as wide as the oscillation length, otherwise the oscillations are washed out \cite{Kay81,Gri96,Gri99}.

Turning back to Eq.~(\ref{II6}), without loss of generality, we can assume
\small\eq{
\mbox{\sc Int}\bb{t} = \mathrm{Re}
\left\{\int_{_{\infm}}^{^{\infp}}\hspace{-0.25cm}\frac{dp_{\z}}{2
\pi}
 \, \varphi^{\2}(p_{\z} \mi p_{\0})\exp{[-i \, \Delta E_{p_{\z}} \, t]}\right\}
\label{II9}.
}\normalsize
This equation is often obtained by assuming two mass-eigenstate wave packets described by the same momentum distribution centered around the average momentum $\bar{p} = p_{\0}$.
This hypothesis also guarantees {\em instantaneous} creation of a {\em pure}
flavor state $\bs{\nue}$ at $t = 0$ \cite{DeL04}.
In fact, for $\phi_{\1}(z,0)=\phi_{\2}(z,0)$, we get, from Eq.~(\ref{II0}),
\small\eq{
\phi_{\nue}(z,0,\theta)
=\phi_1(z,0)=\phi_2(z,0)
=\left(\frac{2}{\pi a^{\2}}\right)^{\frac{1}{4}} \exp{\left[-
\frac{z^{\2}}{a^{\2}}\right]}
\exp{[i  p_{\0} \,z]}
\label{II9B}}\normalsize
and $\phi_{\numu}(z,0,\theta) =0$. Therefore, in what follows, we shall use this
simplification.

\subsubsection{Spatial localization versus temporal average}

The flavor conversion probability from flavor $\nue$ to $\numu$ is basically the
squared modulus of the probability amplitude $\phi_{\numu}(\bx,t)$ of the
state in Eq.\,\eqref{II0}, with its initial localization properties determined by
$\phi_{\nue}(\bx,0)$, integrated over all space.
The probability thus depends on time.
But in a real experiment, time $t$ is not a measurable variable; just the distance $L$ between the source and the detector is known.
To rewrite $P(t)$ as $P(L)$, the formula $L = \upsilon t$ describing the trajectory of a free classical particle is, sometimes, inadvertently invoked.
To clarify this point, let us calculate the probability of the beam of particles,
produced at $|\bx| = 0$, to reach a physical detector of volume $V$, at average
distance $|\bx| = L$, by integrating the corresponding current density of probability
$\mathbf{j}(\bx, t)$ over the surface $\partial V$ enclosing the detector and
integrating over the time of observation from $t_{1}$ to $t_{2}$, as
\small\eq{
\mathcal{P}( t_{\1}< t < t_{\2})
=-\int^{^{t_{\2}}}_{_{t_{\1}}} dt\,\int_{\partial
V}\mbox{d}\mathbf{S}\cdot\mathbf{j}(\bx, t),
\label{II10A}}\normalsize
where the minus sign arises because we want to quantify the flux entering $V$ but $\mbox{d}\mathbf{S}$ points outwards the surface $\partial V$.
This procedure, although straightforward and natural, is not generally adopted and more complicated methods are used instead.
The reason for the rejection of Eq.~\eqref{II10A} is very simple: there is a
difficulty in defining correctly the probability current density
$\mathbf{j}_{e,\mu}(\bx, t)$, for flavor defined neutrino states
$\nue$ or $\numu$. Indeed, such neutrino states have undefined masses,
and in general it is not possible to define conserved currents for
them\,\cite{Zra98}. Nevertheless, it is possible to define an approximately conserved
current $\mathbf{j}_{e,\mu}(\bx,t)$ that obeys
\small
\small\eq{
\label{II10:jalpha}
\frac{\partial}{\partial t}|\phi_{e,\mu}(\bx,t)|^2
+\nabla\ponto\mathbf{j}_{e,\mu}(\bx,t)\approx 0\,,
}\normalsize
where we simplified the notation by using $\phi_{e,\mu}(\bx,t)\equiv
\phi_{\nue,\numu}(\bx,t)$.
The terms violating this conservation are proportional to the neutrino mass
differences and can be neglected locally.
The explicit construction for wave functions satisfying the NR Schroedinger equation can be found in Ref.\,\cite{Zra98}.
One can show that the results remain valid for Dirac fermions and spinless particles
if $|\phi_{e,\mu}(\bx,t)|^2$ is replaced by the corresponding probability or
flavor charge density [see Eq.\,\eqref{jf:cons}].

A typical experiment which tries to observe particle oscillations between two
flavors, assumed generically to be the flavors $\nue$ and $\numu$,
measures the flux of $\numu$ particles in the detector localized at some distance $L$
from the source which produces particles of flavor $\nue$.
The time of the measurements is not known. Usually typical experiments last hours, days or even years (like the observation of solar neutrinos).
So the most appropriate way to find the probability (or number of particles) to cross the (theoretically) closed surface $\partial V$ of the detector is to integrate the probability current density over the surface and integrate the result once more over the duration of the measurements.
To that end, we make use of the conservation law for the total current, for two
flavors $\nue$ and $\numu$,
\small\eq{
\frac{\partial}{\partial t} \left(|\phi_{e}(\bx,t)|^{\2} +
|\phi_{\mu}(\bx,t)|^{\2}\right)
+
\mathbf{\nabla}\cdot\left(\mathbf{j}_{e}(\bx, t) +
\mathbf{j}_{\mu}(\bx, t)\right) = 0,
\label{II10B}}\normalsize
Notice Eq.~\eqref{II10B} is exact since
$\mathbf{j}_{e}+\mathbf{j}_{\mu}=\mathbf{j}_1+\mathbf{j}_2$,
despite Eq.~\eqref{II10:jalpha} approximate nature.
Making use of the Gauss theorem and Eq.~\eqref{II10B}, we get for the sum of
probabilities of Eq.~\eqref{II10A}, with flavors $\nue$ and $\numu$,
\small\eqarr{
\mathcal{P}_{\nue+\numu}( t_{\1}< t < t_{\2})
&\equiv& \mathcal{P}_{\nue}( t_{\1}< t < t_{\2})
+\mathcal{P}_{\numu}( t_{\1}< t < t_{\2})
\cr
&=& \int^{^{t_{\2}}}_{_{t_{\1}}}\! dt\, \frac{\partial}{\partial t}
\int_{V}d^{3}\!\bx
\left(|\phi_{e}(\bx,t)|^{\2} + |\phi_{\mu}(\bx,t)|^{\2}\right)
\cr
&=&
\int_{V}d^{3}\!\bx \big(|\phi_{e}(\bx,t)|^{\2} +
|\phi_{\mu}(\bx,t)|^{\2}\big)\big|_{t_2}
-\int_{V}d^{3}\!\bx \big(|\phi_{e}(\bx,t)|^{\2} +
|\phi_{\mu}(\bx,t)|^{\2}\big)\big|_{t_1},~\quad
\label{II10C}}\normalsize
where $\phi_{e,\mu}(\bx,t)$ correspond to the same scalar wave
function considered in Eq.~\eqref{II0}.
However, it is a fact that the above spatial integration in the volume $V$ is bounded by the extension of the detector and, essentially, by the position and localization of the wave packet.
If we consider $t_1\approx 0$ to be the creation time, the last integral gives zero, and the first integral results in a value different from zero only when $t\sim t_{\2}\sim L/ \upsilon$, where $L$ is the source-detector distance and $\upsilon$ is the average velocity.
The above probability expressions could thus be written as
\small\eq{
\mathcal{P}_{\nue+\numu}(t_{\1}< t < t_{\2})=
\mathcal{P}_{\nue+\numu}(t_{\2}\sim T\sim L/\upsilon)=
\int_{V}d^{3}\!\bx
\big(|\phi_{e}(\bx,L/\upsilon)|^{\2} +
|\phi_{\mu}(\bx,L/\upsilon)|^{\2}\big)\,.~\quad
\label{II10D}}\normalsize

In one-dimension analysis, considering appropriate cylindrical surfaces, we have
\small\eq{
\int_{\partial A}\mbox{d}\mathbf{S}\cdot\mathbf{j}(\bx,t)\equiv
J(z,t),
\label{II10E}}\normalsize
and the continuity equation is reduced to
\small\eq{
\left[
\frac{d}{dt}
\int^{^{+\infty}}_{_{-\infty}}\!\! dz\left(|\phi_{e}(z,t)|^{\2} +
|\phi_{\mu}(z,t)|^{\2}\right)\right]
+
\big(J_{e}(z, t) +  J_{\mu}(z,t)\big)\Big|^{^{z=+\infty}}_{_{z=-\infty}}
=0,
\label{II10F}}\normalsize
so that, from approximate conservation of the flavor current \eqref{II10:jalpha}
over distances and time scales much smaller than the oscillation length,
\small\eq{
\mathcal{P}_{\nue,\numu}(z = L, t_{\2}\!\sim\! T\!\sim\! L/\upsilon) =
\mathcal{P}_{\nue,\numu}(t\!\sim\! L/\upsilon) =
\int^{^{+\infty}}_{_{-\infty}}\!\! dz
\left|\phi_{e,\mu}(z,L/\upsilon;\theta)\right|^{\2}
\equiv \int\! dt\,J_{e,\mu}(L,t),
\label{II10G}}\normalsize
where the $z$-integration has been extended from $-\infty$ to $+\infty$ because we
are assuming the detector extension $D$ is much larger than the wave packet width
$a$, i.\,e., $D \gg a$\,\footnote{Otherwise, the problem could be evaluated for
opposite situations where $a >> D$ since the simple existence of a localized
detector, in order to not violate the boundary conditions of localized states,
imposes the necessity of a wave packet approach. It should be mathematically
equivalent to assume wave packets with the shape of a box of dimension $D$.}.

In the literature, the change of variables $\int dz \longrightarrow \int
dt\,\upsilon$ is frequently noticed.
For the spatial integration of the above expression, it is mathematically acceptable when $z\sim \upsilon\,t$ in one-dimension analysis, i.\,e.,
\small\eq{
\mathcal{P}_{\nue,\numu}(t \sim L/\upsilon) =
 \int^{^{+\infty}}_{_{-\infty}}\!\! dz
 \left|\phi_{e,\mu}(z,L/\upsilon;\theta)\right|^{\2} \approx
\upsilon
\int^{^{+\infty}}_{_{-\infty}}\!\! dt
\left|\phi_{e,\mu}(L,t;\theta)\right|^{\2} \equiv
\int dt \,J_{e,\mu}(L, t)\,.
\label{II10H}}\normalsize
However, it is important to remark that automatic normalization is only guaranteed for space integration.

To summarize this point, it seems obvious from the above fundamental quantum
mechanical calculations that the spatial integration is not in confront with time
integration.
In fact, the spatial integration just concerns the wave packet localization.
When one makes some assertion about the measurements, additional integrations over
the measurement time and energy may (or should) be required (in order to obtain
time/energy averaged values for $P_{\nue,\numu}$, and as we have observed,
it can be definitely adequate to the analysis here performed.

Another way of reconciling temporal oscillation with spatial oscillation makes use of the usual textbook connection between position and time for a free particle in Quantum Mechanics: the Ehrenfest's theorem.
A free particle can be found to be located at mean position
{\small $\aver{\bx(t)}=\aver{\mathbf{v}}t$} at time $t$ with error given by
{\small$\Delta x=\sqrt{\aver{(\mathbf{X}-\aver{\bx})^2}}$}\,.
For each mass-eigenstate, when dispersion can be neglected, such quantities are well
defined and obey
\small\eq{
\aver{\bx_i(t)}=\aver{\mathbf{v}_i}t\,,~~ \Delta x_i\approx a_i=\text{constant},
}\normalsize
for initial position {\small $\aver{\bx_i(0)}=0$}, where
{\small$\aver{\mathbf{v}_i}$}  is the group
velocity, the expectation value of {\small $\mathbf{v}_i=\bp/E_i(\bp)$} in momentum
space.
For flavor states, in the regime of total overlap, the mean position should be given
by $\bar{\mathbf{v}}t$, where $\bar{\mathbf{v}}$ is the average of
{\small$\aver{\mathbf{v}_i}$}, with error no larger than $\max(a_i)$.
When neutrinos are detected after traveling a distance $L$, which is the only
experimentally known variable, it is licit to replace $t$ in the formulas by
$L/\bar{v}$ with error given by $\max(a_i)$, as long as such quantity is much smaller
than the detector characteristic size $D$.

\subsubsection{The analytical approach}

Now we turn back to the IWP framework in order to obtain an analytical expression for $\phi_{\ii}\bb{z,t}$.
To evaluate the integral in Eq.~(\ref{II4}), we firstly rewrite the energy $E^{(\ii)}_{p_{\z}}$ as
\small\eq{
E^{(\ii)}_{p_{\z}} = E_{\ii} \left[1 + \frac{ p_{\z}^{\2} \mi p_{\0}^{\2}}{E_{\ii}^{\2}}\right]^{ \frac{1}{2}} = E_{\ii} \left[1 + \sigma_{\ii} \left(\sigma_{\ii} + 2 \mbox{v}_{\ii}\right)\right]^{ \frac{1}{2}},
\label{II11}
}\normalsize
where $E_{\ii}\equiv (m_{\ii}^{\2} + p_{\0}^{\2})^{\frac{1}{2}}$,
$\mbox{v}_{\ii}\equiv\frac{ p_{\0}}{E_{\ii}}$ and
$\sigma_{\ii}\equiv \frac{ p_{\z} \mi p_{\0}}{E_{\ii}}$.
The use of free gaussian wave packets is frequently assumed in NR quantum mechanics
because the calculations can be carried out exactly for these particular functions
and, consequently, the main physical aspects can be easily interpreted from the final
analytical expressions.
The reason lies in the fact that the frequency components of the mass-eigenstate wave
packets, $E^{(\ii)}_{p_{\z}}= p_{\z}^{\2}/2 m_{\ii}$, modify the momentum
distribution into ``generalized'' gaussian functions, easily integrated by well-known
methods.
The term $ p_{\z}^{\2}$ in $E^{(\ii)}_{p_{\z}}$ is then responsible for the variation
in time of the width of the mass-eigenstate wave packets, the so-called spreading
phenomenon.

In relativistic quantum mechanics, however, the frequency components of the
mass-eigenstate wave packets, $E^{(\ii)}_{p_{\z}}=\sqrt{ p_{\z}^{\, \2} +
m_{\ii}^{\2} }$, do not permit  an immediate analytic integration.
This difficulty, however, may be remedied by assuming a sharply peaked momentum distribution, i.\,e., $(a \, E_{\ii})^{-1}\sim\sigma_{\ii} \ll 1$.
Meanwhile, the integral in Eq.~(\ref{II4}) can be {\em analytically} solved only if we consider terms up to order $\sigma_{\ii}^{\2}$ in the series expansion.
In this case, we can conveniently truncate the power series
\small\eqarr{
E^{(\ii)}_{p_{\z}} & = & E_{\ii} \left[1 + \sigma_{\ii} \mbox{v}_{\ii}  + \frac{\sigma_{\ii}^{\2}}{2}\left(1 - \mbox{v}_{\ii}^{\2} \right)\right] + \mathcal{O}(\sigma_{\ii}^{\3})
\nonumber\\  &
\approx &
E_{\ii} +  p_{\0} \sigma_{\ii} + \frac{m_{\ii}^{\2}}{2E_{\ii}} \sigma_{\ii}^{\2}
\label{II12}
}\normalsize
and get an analytic expression for the oscillation probability.
The zeroth-order term in the previous expansion, $E_{\ii}$, gives the standard
plane-wave oscillation phase.
The first-order term, $ p_{\0} \sigma_{\ii}$, is responsible for the {\em slippage}
(loss of overlap) between the mass-eigenstate wave packets due to their
different group velocities.
It represents a linear correction to the standard oscillation phase \cite{DeL04}.
Finally, the second-order term, $\frac{m_{\ii}^{\2}}{2E_{\ii}} \sigma_{\ii}^{\2}$,
which is a (quadratic) secondary correction, will give the well-known spreading
effects in the time propagation of the wave packet and will be also responsible for
an {\em additional} phase to be computed in the final calculation.

For gaussian momentum distributions, all the terms discussed above can be {\em
analytically} quantified.
By substituting (\ref{II12}) in Eq.~(\ref{II4}) and changing the $
p_{\z}$-integration into a $\sigma_{\ii}$-integration, we obtain the explicit form of
the mass-eigenstate wave packet time evolution,
\small\eqarr{
\phi_{\ii}\bb{z,t}
 &=&
 \left[\frac{2}{\pi \,a^{\2}_{\ii}\bb{t}}\right]^{
\frac{1}{4}}\exp{\big[-i\,(E_{\ii}\,t - p_{\0}\,z + \theta_{\ii}\bb{t, z})\big]}
 \exp{\left[-\frac{(z - \mbox{v}_{\ii} \,t)^{\2}}{a_{\ii}^{\2}\bb{t}}\right]}
 ,~~~				
\label{II13}
}\normalsize
where
\small\eq{
\theta_{\ii}\bb{t, z} = \left\{\frac{1}{2}\arctan{\left[\frac{2\,m_{\ii}^{\2}\, t}{a^{\2}\, E_{\ii}^{\3}}\right]} - \frac{2\, m_{\ii}^{\2}\, t} {a^{\2}\, E_{\ii}^{\3}}\,\frac{(z - \mbox{v}_{\ii} \,t)^{\2}}{a_{\ii}^{\2}\bb{t}}\right\}
}\normalsize
and
\small\eq{
a_{\ii}\bb{t} = a \left(1 + \frac{4\, m_{\ii}^{\4}}{a^{\4}\, E_{\ii}^{\6}}\,t^{\2}\right)^{ \frac{1}{2}}.
}\normalsize

The time-dependent quantities $a_{\ii}\bb{t}$ and $\theta_{\ii}\bb{t, z}$ contain all
the physically significant informations which arise from the second-order term in the
power series expansion (\ref{II12}).
The spreading of the propagating wave packet can be immediately quantified by interpreting $a_{\ii}\bb{t}$ as a time-dependent width, i.\,e., the spatial localization of the propagating particle is effectively given by $a_{\ii}\bb{t}$ which increases during the time evolution.
In the NR propagation regime, $a_{\ii}\bb{t}$ reduces to
$a^{\mbox{\tiny $NR$}}{\ii}\bb{t} = a \sqrt{1 + \frac{4}{a^{\4} m_s^{\2}}t^{\2}}$
\cite{Coh77}.
For times $t \gg a^{\2} m_{\ii}$ the effective wave packet width $a^{\mbox{\tiny $NR$}}{\ii}\bb{t}$ becomes much larger than the initial width $a$.
On the other hand, the wave packet spreading in the UR propagation regime
is approximated by $a^{\mbox{\tiny $UR$}}_{\ii}\bb{t} = a \sqrt{1 + \frac{4
m_s^{\4}}{a^{\4} p_o^{\6}}t^{\2}}\approx a$.
The UR spreading is practically negligible if we consider the {\em same} time-scale $T$ for both NR and UR cases, i.\,e., $a^{\mbox{\tiny $UR$}}_{\ii}\bb{T} \ll a^{\mbox{\tiny $NR$}}_{\ii}\bb{T}$.
To illustrate this characteristic, we reproduce from \cite{Ber04} the time-dependence of $a_{\ii}\bb{t}$ in
Fig.\,\ref{an1} where we have assumed a particle with a definite mass value
$m_{\ii}$.
By computing the squared modulus of the mass-eigenstate wave function,
\small\eq{
|\phi_{\ii}\bb{z,t}|^{\2} 	 \approx  \left(\frac{2}{\pi a^{\2}_{\ii}\bb{t}}\right)^{ \frac{1}{2}}
\exp{\left[-\frac{2 (z - \mbox{v}_{\ii} \,t)^{\2}}{a_{\ii}^{\2}\bb{t}}\right]},					
\label{II18}
}\normalsize
we reproduce from \cite{Ber04} the wave packet spreading in both NR and UR propagation regimes in
Fig.\,\ref{an2} which is in correspondence with Fig.\,\ref{an1}.
It confirms that the wave packet spreading is irrelevant for UR particles.

Returning to Eq.~(\ref{II13}), we could interpret another second order effect by observing the time-behavior of the phase $\theta_{\ii}\bb{t, z}$.
By taking into account the wave packet localization, we assume that the amplitude of the wave function is relevant in the interval $|z - \mbox{v}_{\ii}\, t| \leq a_{\ii}\bb{t}$.
Due to the $z$-dependence,  each wave packet space-point $z$ evolves in time in a different way.
If we observe the propagation of the space-point $z = \mbox{v}_{\ii} \, t$, the
increasing function $\theta_{\ii}\bb{t,\, \mbox{v}_{\ii} t}$ assume values
limited by the interval $[0, \frac{\pi}{4}[$.
Otherwise, for any other space-point given by $z = \mbox{v}_{\ii} \,t + K \,a_{\ii}\bb{t}$, $0 < |K| \leq 1$, the phase $\theta_{\ii}\bb{t, z}$ does not have a lower limit.
We shall show in the next subsection that the presence of a time-dependent phase can modify the oscillation character of the flavor conversion formula.
Anyway, the phase $\theta_{\ii}\bb{t, z}$ is not influent on the {\em free} mass-eigenstate wave packet propagation as we can see from Eq.~(\ref{II18}).

\subsubsection{The oscillation probability}

By evaluating the integral (\ref{II9}) with the approximation (\ref{II12}) and
performing some mathematical manipulations, we
obtain a factored expression for the interference term \eqref{II2},
\small\eq{
\mbox{\sc Int}\bb{t} = \mbox{\sc Dmp}\bb{t} \times \mbox{\sc Osc}\bb{t},
\label{II20}
}\normalsize
which contains the damping term
\small\eq{
\mbox{\sc Dmp}\bb{t} = \left[1 + \mbox{\sc Sp}^{\2}\bb{t} \right]^{-\frac{1}{4}}
\exp{\left[-\frac{(\Delta \mbox{v} \, t)^{\2}}{2a^{\2}\left[1 + \mbox{\sc
Sp}^{\2}\bb{t}\right]}\right]}
\label{II21}
}\normalsize
and the oscillation term
\small\eqarr{
\mbox{\sc Osc}\bb{t} &=& \mathrm{Re}\!\left\{\exp{\left[-i\Delta E \, t -i
\Theta\bb{t}\right]}
\right\}\nonumber\\
&=& \cos{\left[\Delta E \, t + \Theta\bb{t}\right]},
\label{II22A}
}\normalsize
where
\small\eq{
\mbox{\sc Sp}\bb{t} = \frac{t}{a^{\2}}\Delta\left(\frac{m^{\2}}{E^{\3}}\right) =
\rho\, \frac{\Delta \mbox{v}\, t}{a^{\2} \,  p_{\0}}
\label{II240}
}\normalsize
and
\small\eq{
\Theta\bb{t} = \left[\ml{\frac{1}{2}}\arctan{\left[\mbox{\sc Sp}\bb{t}\right]} -
\ml{\frac{a^{\2}\,p_{\0}^{\2}}{2 \rho^{\2}}}\frac{\mbox{\sc Sp}^{\3}\bb{t}}{\left[1 +
\mbox{\sc Sp}^{\2}\bb{t}\right]}\right],
\label{II24A}
}\normalsize
with $\rho = 1 - \left[3 + \left(\frac{\Delta E}{\bar{E}}\right)^{\2}\right]
\frac{p_{\0}^{\2}}{\bar{E}^{\2}}$, $\Delta E = E_{\1} - E_{\2}$,  $\bar{E} =
\sqrt{E_{\1} \, E_{\2}}$ and $
\Delta \mbox{v}=\mbox{v}_1-\mbox{v}_2$.
The time-dependent quantities $\mbox{\sc Sp}\bb{t}$ and $\Theta\bb{t}$ carry the
second-order corrections and, consequently, the spreading effect to the oscillation
probability formula.
If $\Delta E \ll \bar{E}$, the parameter $\rho$ is limited by the interval $[1,-2]$ and it assumes the zero value when $\frac{ p_{\0}^{\2}}{\bar{E}^{\2}} \approx \frac{1}{3}$.
Therefore, by considering increasing values of $ p_{\0}$, from NR (NR) to UR (UR)
propagation regimes, and fixing $\frac{\Delta E}{a^{\2} \, \bar{E}^{\2}}$, the time
derivatives of $\mbox{\sc Sp}\bb{t}$ and $\Theta\bb{t}$ have their signals inverted
when $\frac{ p_{\0}^{\2}}{\bar{E}^{\2}}$ reaches the value $\frac{1}{3}$.

The suppression of the oscillating behavior, which is primarily caused by the
separation between the mass-eigenstate wave packets, is quantified by the damping
term $\mbox{\sc Dmp}\bb{t}$.
In order to compare $\mbox{\sc Dmp}\bb{t}$ with the correspondent function without
the second-order corrections (without spreading),
\small\eq{
\mbox{\sc Dmp}_{\mbox{\tiny $WS$}}\bb{t} = \exp{\left[-\frac{(\Delta \mbox{v} \,
t)^{\2}}{2a^{\2}}\right]},
\label{II23A}
}\normalsize
we calculate the ratio between Eq.\,\eqref{II21} and Eq.\,\eqref{II23A}:
\small\eqarr{
\frac{\mbox{\sc Dmp}\bb{t}}{\mbox{\sc Dmp}_{\mbox{\tiny $WS$}}\bb{t}}
&=& \mbox{$\left[1 + \rho^{\2} \left(\frac{\Delta E \, t}{a^{\2} \,
\bar{E}^{\2}}\right)^{\2} \right]^{-\frac{1}{4}}$}
\mbox{$ \exp{\left[\frac{\rho^{\2} \,  p_{\0}^{\2} \,
\left(\Delta E \, t\right)^{\4}}
{2\,a^{\6} \, \bar{E}^{\8}\left[1 + \rho^{\2} \left(\frac{\Delta E \, t}{a^{\2} \,
\bar{E}^{\2}}\right)^{\2}\right]}
\right]}.$}
\label{II23AA}
}\normalsize
The NR limit is obtained by setting $\rho^{\2}=1$ and $p_{\0}=0$ in
Eq.~(\ref{II23A}).
In the same way, the UR limit is obtained by setting $\rho^{\2}=4$ and
$p_{\0}=\bar{E}$.
Between these limits, the minimal deviation from unity of
Eq.\,\eqref{II23AA} occurs when $\frac{p_{\0}^{\2}}{\bar{E}^{\2}} \approx
\frac{1}{3}$ ($\rho \approx 0$).
Returning to the exponential term of Eq.~(\ref{II21}), we observe that the
oscillation amplitude is more relevant when $\Delta \mbox{v} \, t \ll a$.
It characterizes the regime of minimal loss of coherence where the spatial
overlap between the mass-eigenstate wave packets is significant.
In such regime, we always have $\frac{\mbox{\sc Dmp}(t)}{ \mbox{\sc
Dmp}_{\mbox{\tiny $WS$}}(t)} \approx 1$.

We reproduce from \cite{Ber04} the Fig.\,\ref{an3} that shows the ratio of Eq.~(\ref{II23AA}) for different propagation
regimes, where we have arbitrarily set $a\, \bar{E} = 10$.
For asymptotic times, the time-dependent term $\mbox{\sc Sp}(t)$ effectively extends
the interference between the mass-eigenstate wave packets since
\small\eq{
\frac{\mbox{\sc Dmp}(t)}{\mbox{\sc Dmp}_{\mbox{\tiny $WS$}}(t)} \stackrel{t
\rightarrow \infty}{\approx}
\frac{a \, \bar{E}} {(\rho \, \Delta E \, t)^{\frac{1}{2}} }
\exp{\left[\frac{p_o^{\2}\left(\Delta E \, t\right)^{\2}}{2\,a^{\2} \,
\bar{E}^{\4}}\right]} \gg 1,
\label{231}
}\normalsize
but, in this case, the oscillations are almost completely destroyed by $\mbox{\sc
Dmp}(t)$ (see Fig.\,\ref{an5}).

The oscillating function $\mbox{\sc Osc}\bb{t}$ of the interference term $\mbox{\sc Int}\bb{t}$ differs from the {\em standard} oscillating term, $ \cos{[\Delta E \, t]}$, by the presence of the additional phase $\Theta\bb{t}$ which is essentially a second-order correction.
The modifications introduced by the additional phase $\Theta\bb{t}$ are discussed in
Fig.\,\ref{an4} where we have compared the time-behavior of $\mbox{\sc Osc}\bb{t}$ to
$\cos{[\Delta E \, t]}$ for different propagation regimes.
The {\em  effective} bound value assumed by $\Theta \bb{t}$ is determined by the damping behavior of $\mbox{\sc Dmp}\bb{t}$.
To illustrate this flavor oscillation behavior, we plot both the curves representing
$\mbox{\sc Dmp}\bb{t}$ and $\Theta\bb{t}$ in Fig.\,\ref{an5}.
We note the phase changes slowly in the NR regime.
The modulus of the phase $|\Theta\bb{t}|$ rapidly reaches its upper limit when $\frac{ p_{\0}^{\2}}{\bar{E}^{\2}} > \frac{1}{3}$ and, after a certain time, it continues to evolve approximately linearly in time.
Essentially, the oscillation vanishes rapidly.
In case of (ultra-relativistic) neutrino oscillations, the values considered for $a$ in Figs.~\ref{an4}-\ref{an5} are reproduced from \cite{Ber04} and they can be considered unrealistically large.

By superposing the effects of $\mbox{\sc Dmp}\bb{t}$ in Fig.\,\ref{an5} and the
oscillating character $\mbox{\sc Osc}\bb{t}$ expressed in Fig.\,\ref{an4}, we
immediately obtain the flavor oscillation probability
\smallskip
\small\eq{
\mathcal{P}(\bs{\nue}\rightarrow\bs{\numu};t)
 \approx
\frac{\sin^{\2}{(2\theta)}}{2}\left\{1 - \left[1 + \mbox{\sc Sp}^{\2}\bb{t}
\right]^{-\frac{1}{4}}
\exp{\left[-\frac{(\Delta \mbox{v} \, t)^{\2}}{2a^{\2}\left[1 + \mbox{\sc
Sp}^{\2}\bb{t}\right]}\right]}
\cos{\left[\Delta E \, t + \Theta\bb{t}\right]}
  \right\}\,,
\label{II25A}
}\normalsize
which is illustrated in Fig.\,\ref{an8}.

Obviously, the larger is the value of $a\,\bar{E}$, the smaller are the wave packet
effects.
If it was sufficiently larger to not consider the second order corrections expressed
in Eq.~(\ref{II12}), we could  compute the oscillation probability with the leading
corrections due to the decoherence effect,
\small\eqarr{
\mathcal{P}(\bs{\nue}\rightarrow\bs{\numu};t) &\approx&
\frac{\sin^{\2}{(2\theta)}}{2}
\left\{1- \exp{\left[-\frac{(\Delta \mbox{v} \, t)^{\2}}{2\,
a^{\2}}\right]}\cos{[\Delta E \, t ]}\right\}
\label{II26A}
}\normalsize
which corresponds to the same result obtained by \cite{DeL04}.
Under minimal decoherence conditions ($\Delta \mbox{v} \, t \ll a$), the above
expression reproduces the {\em standard} plane wave result,
\smallskip
\small\eqarr{
\mathcal{P}(\bs{\nue}\rightarrow\bs{\numu};t) &\approx&
 \frac{\sin^{\2}{(2\theta)}}{2}
 \left\{1- \cos{[\Delta E \, t ]}\right\} = \sin^{\2} (2 \theta)
\sin^{\2} \left[ \frac{\Delta m^{\2}}{4 \bar{E}} \, L \right]
\,,
\label{II27A}
}\normalsize
since we have assumed $a \, \bar{E} \gg 1$.

\subsection{Initial flavor violation}
\label{subsec:flavorvio:I}

The two family flavor conversion formula, obtained by simply associating scalar wave
functions for neutrino mass-eigenstates, was given by Eq.~\eqref{II1}.
We can rewrite such expression, in the three-dimensional space, as
\small\eq{
\mathcal{P}(\bs{\nue}\rightarrow\bs{\numu};t)=
\frac{\sin^{\2}{(2\theta)}}{4}
\int\hspace{-.5ex}d^3\!\bx\,
\big|\phi_1(\bx,t)-\phi_2(\bx,t)\big|^2\,.
\label{fvio1:1}
}\normalsize
It is evident from Eq.~\eqref{fvio1:1} that $\mathcal{P}(\bs{\nue}\rightarrow\bs{\numu};t)\neq 0$ if
$\phi_1(\bx,t)\neq \phi_2(\bx,t)$ as complex functions of $\bx$.
In particular, for $t=0$, we necessarily have $\mathcal{P}(\bs{\nue}\rightarrow\bs{\numu};0)\neq 0$ if $\phi_1(\bx,0)\neq \phi_2(\bx,0)$.
Therefore, we might have an initial flavor violation, without propagation, if the
initial wave functions associated to the two neutrino mass-eigenstates are different.
This leads to the problem of initial flavor definition in the formulation of neutrino oscillations\,\cite{DeL04}.

Three points of view can be adopted concerning initial flavor definition:
(i) the $\nue\,$-flavor state is indeed well defined by
Eq.~\eqref{II0} and we should have $\phi_1(\bx,0)=\phi_2(\bx,0)$ to guarantee ({\it
instantaneous}) initial flavor definition;
(ii) the definition of neutrino flavor should be generalized to accommodate more general initial conditions for neutrino creation;
or (iii) exact flavor definition can not be achieved and initial flavor violation should be regarded as a genuine physical effect.
Within the simple approach of intermediate scalar wave packets, we can only adopt the positions (i) and (ii), and only estimate the consequences of point of view (iii).
We will see in the following that for a free and first quantized Dirac fermion the strict adoption of (i) leads to the inevitable presence of oscillating terms with short oscillation length
$L^{VHF}_{Osc} = \frac{\pi}{p_{0}} = \frac{\Delta m^2}{4 p^{2}_{0}}\,L^{Std}_{Osc}$, in addition to usual flavor oscillation characterized by the usual oscillation length $L^{Std}_{Osc} = \frac{4\pi p_{0}}{\Delta m^{2}}$.
To avoid such rapid oscillations implies (iii), i.\,e., an initial non-null presence of the wrong flavor.
Indeed, in the simple second quantized theory of Dirac fermion with mixing shown in
section\,\ref{subsec:simpleQFT} the adoption of (iii) is inevitable, although its
consequence could be very small and unobservable in practice.
Indeed, it is clear from the discussion after Eq.~\eqref{II6} that initial flavor violation effects implied by point of view (iii) should be small otherwise the amplitude of flavor oscillations would be highly suppressed.
Moreover, large effects are excluded from neutrino conversion experiments. We can estimate the possible effects of flavor violation from Eq.~\eqref{II6}, where gaussian wave packets were used.
The probability of initial flavor violation is given by $\sin^2\!2\theta(a\Delta p/4)^2$, for $|a\Delta p|\ll 1$. Notice the condition $|a\Delta p|\ll 1$ is a particular version of the condition $\phi_1(\bx,t)\approx\phi_2(\bx,t)$ for gaussian wave packets of equal width.

To properly investigate (iii) and quantify its effects, it is necessary to go beyond the description of neutrino propagation and include the interactions responsible for neutrino creation (and detection). In terms of gaussian wave packets, it amounts to calculate the quantity $|a\Delta p|$ for realistic cases.
Such task will be undertaken in section\,\ref{subsec:intrinsic} for neutrinos created
in pion decay.

\section{Inclusion of spin and relativistic completeness}
\label{sec:relativistic}

This section deals with the inclusion of spin and relativistic completeness in an
relativistic quantum-mechanical derivation of the oscillation formula and some
extensions.
Neutrinos are {\em fermions} and we know that the time-evolution of a massive spin
one-half particle has to be described by the Dirac equation.
Firstly, we show how to include the spatial localization in the description of a
flavor oscillating system when the dynamics of the Dirac equation is considered.
In addition, we identify some common points between our results and some previous results for a quantum field approach for flavor oscillations \cite{Bla95,Giu02,Bla03}.
Finally, we extend the discussion about initial flavor violation introduced in the previous section to the Dirac equation case.
In particular, at the level of a quantum mechanical approach, we identify some relations between initial flavor violation and some predictable rapid oscillations.

It is well known that the Dirac equation, in its first quantized version, can give a significantly good description of a Dirac fermion if its inherent localization is much bigger than its Compton wave length; usually this is associated with weak external fields.
For example, the spectrum for the hydrogen atom can be obtained with the relativistic
corrections included (fine structure)\,\cite{Zub80}.
One of the terms responsible for fine structure, the Darwin term, can be interpreted
as coming from the interference between positive and negative frequency parts ({\it
zitterbewegung}) of the hydrogen eigenfunction in Dirac theory compared to the NR
theory \cite{FV,Sak87}.
On the other hand a situation where the theory fails to give a satisfactory physical
description is exemplified by the Klein paradox\,\cite{Zub80,dombey:klein}:
the transmission coefficient for a electron moving towards a step barrier becomes
negative for certain barrier heights, exactly when the localization of the electron
wave function inside the barrier is comparable with its Compton wave length.

It is also well known that relativistic covariance requires the use of both positive and negative energy solutions for free relativistic particles.
Otherwise, as we will see, functional completeness is lost, i.\,e., we can not expand a general solution of a relativistic wave equation in terms of only positive or negative eigenfunctions.
Both types of solutions are also necessary to ensure relativistic causality, i.\,e., the time evolution of the solutions can not extend in space-time beyond the future lightcone of the initial spatial distribution.

Given the desirable properties of functional completeness and causality, which will be called relativistic completeness for short, we will treat in this section the flavor oscillation problem using the free Dirac theory in presence of two families mixing.
We will see that the use of the Dirac equation will automatically encompass the properties of relativistic completeness.

Additionally, obtaining exact solutions of a generic class of Dirac wave
equations\,\cite{Dir28,Esp99,Alh05,Ein81,Bak95}, specially with the presence of
general interaction terms, is important because in some cases the conceptual
understanding of the underlying physics can only be brought about by such solutions.
These solutions also correspond to valuable means for checking and improving models
and numerical methods for solving complicated physical problems.
In the context in which we intend to explore the Dirac formalism, we can report about the Dirac wave packet treatment which can be useful in keeping clear many of the conceptual aspects of quantum oscillation phenomena that naturally arise in a relativistic spin one-half particle theory.
In parallel, much progress on the theoretical front of the quantum mechanics of neutrino oscillations has been achieved, impelled by a phenomenological pursuit of a more refined flavor conversion formula \cite{Giu98,Zra98,Ber05} and by efforts to give the theory a formal structure within the quantum field formalism \cite{Bla95,Giu02,Bla03}.

To introduce the {\em fermionic} character in the study of quantum oscillation
phenomena\,\cite{Ber05A,Ber04,Ber05,ccn:no12}, we shall introduce the Dirac equation
as the evolution equation for the mass-eigenstates. Hence, Eq.~(\ref{II0}) becomes
\small\eqarr{
\Psi\bb{x} &=& \psi_{\1}\bb{x}\cos{\theta}\,\mbox{\boldmath$\nu_{\1}$} +
\psi_{\2}\bb{x}\sin{\theta}\,\mbox{\boldmath$\nu_{\2}$}\nonumber\\
          &=& \left[\psi_{\1}\bb{x}\cos^{\2}{\theta} +
\psi_{\2}\bb{x}\sin^{\2}{\theta}\right]\,\bs{\nue} +
\left[\psi_{\1}\bb{x} -
\psi_{\2}\bb{x}\right]\cos{\theta}\sin{\theta}\,\bs{\numu}
\nonumber\\
          &=& \psi_{\nue}\bb{x;\theta}\,\bs{\nue} +
\psi_{\numu}\bb{x;\theta}\,\bs{\numu}
\label{III0B}
}\normalsize
where $x = (t, \bx)$ and $\psi_{\ii}\bb{x}$, $i=1,2$, is a spinorial wave function,
normalized to unity, that satisfies the Dirac equation for a mass $m_{\ii}$.
The natural extension of Eq.~(\ref{II9B}) reads
\small\eq{
\psi_{\nue}\bb{\bx,\,t=0;\,\theta}
\equiv \psi_{\nue}(\bx)
=\psi_{1}\bb{\bx,\,t=0}
=\psi_{2}\bb{\bx,\,t=0}\,,
\label{D:pureflavor}
}\normalsize
which guarantees an initial pure flavor $\nue$.
Therefore, the probability that an initial pure $\nue$ flavor be detected as $\numu$ flavor \eqref{II1} is
\small\eq{
\mathcal{P}(\ms{\bs{\nue}\!\rightarrow\!\bs{\numu}};t)=
\Int{3}{x}
\psi_{\numu}^\dag\bb{\bx,t}\psi_{\numu}\bb{\bx,t}
=\sin^22\theta\frac{1}{2}\big[1-\mbox{\sc Dfo}\bb{t}\big]\,,
}
where the interference term generalizing Eq.~\eqref{II2} for Dirac wave packets is
\small\eq{
\label{D:int}
\mbox{\sc Dfo}\bb{t}=
\mathrm{Re}\Int{3}{x}\psi_{1}^\dag\bb{\bx,\,t}\psi_{2}\bb{\bx,\,t}\,.
}
More discussion on initial flavor definition within first quantized Dirac theory can be found in section\,\ref{subsec:IFV:D}.

\subsection{Dirac wave packets and the oscillation formula}
\label{subsec:dirac}

Let us firstly consider the time evolution of a spin one-half free particle given by the Dirac equation
\small\eq{
\left(i \gamma^\mu \partial_{\mu} - m \right)\psi\bb{x} = 0,
\label{III00}
}\normalsize
with the plane wave solutions given by $\psi\bb{x} = \psi_{_{\pl}}\bb{x} + \psi_{_{\mi}}\bb{x}$ where
{\small\eqarr{
&&\psi_{_{\pl}}\bb{x} = \exp{[-i\,p\,x]} \, u\bb{p} ~~~~\mbox{for positive
frequencies and}\nonumber\\
&&\psi_{_{\mi}}\bb{x} = \exp{[+i\,p\,x]} \, v\bb{p} ~~~~\mbox{for negative
frequencies};
\label{III01}
}\normalsize
$p$ is the relativistic {\em quadrimomentum},  $p = (E, \bs{p})$, and the
relativistic energy is represented by $E=\sqrt{m^{\2}+ \bs{p}^{\2}}$.
The free propagating mass-eigenstate spinors are written as \cite{Pes95}
{\small\eqarr{
u_{\s}\bb{p} &=& \frac{\gamma^\mu p_\mu + m}{\left[2\,E\,(m+E)\right]^{\frac{1}{2}}}
\,u^{\s}_0=
\left(\begin{array}{r} \left(\frac{E+m}{2E}\right)^{\foh}\eta_{\s}
\raisebox{-.8em}{}
\\
\frac{\textstyle\bs{\Sigma}\ponto\bs{p}}{\left[2\,E\,(E+m)\right]^{
\foh } }
\eta_{\s} \end{array}\right),\nonumber\\
v_{\s}\bb{p} &=& \frac{-\gamma^\mu p_\mu + m}{\left[2\,E\,(m+E)\right]^{\foh}} \,
\,v^{\s}_0=
\left(\begin{array}{r}
\frac{\textstyle\bs{\Sigma}\ponto\bs{p}}{\left[2\,E\,(E+m)\right]^{\foh}}
\eta_{\s}
\\
\raisebox{1.6em}{}
\left(\frac{E+m}{2E}\right)^{\foh} \eta_{\s} \end{array}\right),~~~
\label{III02}
}\normalsize
where $\eta_{\1,\2} = \left(\begin{array}{c}1 \\ 0\end{array}\right),\,
\left(\begin{array}{c}0 \\ 1\end{array}\right)$,
$u^{\s}_0\equiv u_{\s}\bb{m,\bs{0}}$
$v^{\s}_0\equiv v_{\s}\bb{m,\bs{0}}$ are the spinors for zero momentum (they do not
depend on the mass) and the relations
$u_s^\dag\bb{p}u_r\bb{p}=v_s^\dag\bb{p} v_r\bb{p}=\delta_{rs}$ are satisfied.
The right-hand sides of the second equalities of Eq.\,\eqref{III02} assume the Dirac
representation for the gamma matrices (see Eq.\,\eqref{IV501}).
When more than one mass-eigenstate is in use, the positive and negative frequency
eigenspinors will be respectively denoted by $u^{\s}\bb{\bp,m_i}$ and
$v^{\s}\bb{\bp,m_i}$.

To describe the time evolution of mass-eigenstate Dirac wave packets, we could be inclined to superpose only positive frequency solutions of the Dirac equation.
It seems, at first glance, a reasonable choice.
However, for generic initial states it is usually necessary to superpose both
positive and negative frequency solutions of the Dirac equation.
We can show that both types of solution are necessary if we require the {\em
instantaneous} creation of a {\em pure} $\bs{\nue}$ flavor, i.e., the mass-eigenstate
wave functions satisfy $\psi_{\1}(z,0)=\psi_{\2}(z,0)$ in Eq.\,\eqref{D:pureflavor}.
(See sections \ref{subsec:flavorvio:I} and \ref{subsec:IFV:D}.)
More particularly, we can adopt the initial condition
\small\eq{
\psi_{\nue}\bb{\bx,\,t=0;\,\theta}
= \phi_{\nue}\bb{\bs{x}}w\,,
\label{III22}
}\normalsize
obeying Eq.\,\eqref{D:pureflavor},
where $\phi_{\nue}$ is a scalar wave function and $w$ is a constant spinor which
satisfies the normalization condition $w^{\dagger} w = 1$. That is not the most
general condition because $w$ could depend on spatial coordinates, but it simplifies
the preliminary calculations and allows a direct comparison to the scalar treatment
of section \ref{sec:localization}.

To continue the analysis, we turn back to the one-dimensional space
($\bx \rightarrow  z $) and express the flavor wave
function $\psi_{\nue}(z,t,\theta)$ in terms of
\small\eqarr{
\psi_{\ii}\bb{z,t}
&=& \int_{_{-\infty}}^{^{+\infty}}\!\frac{d p_{\z}}{2\pi} \exp{[ i  p_{\z} z]}
\sum_{s=1,2}\{b^s\bb{p_{\z}, m_{\ii}}\,u^s\bb{p_{\z}, m_{\ii}} \exp{[-iE\bb{p_{\z},
m_{\ii}} t]}
\nonumber\\
&&
\hspace{9.em}
+\ d^{s*}\bb{\mi p_{\z}, m_{\ii}}\,v^s\bb{\mi p_{\z}, m_{\ii}} \exp{[+iE\bb{p_{\z},
m_{\ii}}t]}\}\,.
\label{III23}
}\normalsize
Therefore, since $\psi_{\ii}\bb{z,0}=\psi_{\nue}\bb{z,0,\theta}$ for $t=0$, the
Fourier transform of Eq.\,\eqref{III23} should be
\small\eqarr{
\sum_{s=1,2}\left[b^{s}\bb{p_{\z}, m_{\ii}}\,u^{s}\bb{p_{\z}, m_{\ii}} +
d^{s*}\bb{\mi p_{\z}, m_{\ii}}\,v^{s}\bb{\mi p_{\z}, m_{\ii}}\right]
&=&
\varphi\bb{p_{\z} -  p_{\0}} w\,,\quad i=1,2\,,
\label{III25}
}\normalsize
where we used Eq.\,\eqref{III22} and denoted $\varphi\bb{p_{\z}-p_{\0}}$ as the
Fourier transform of $\phi_{\nue}\bb{z}$.
(In fact, the normalized Fourier transform of $\phi_{\nue}\bb{z}$ is
$\varphi\bb{p_{\z}-p_{\0}}/\sqrt{2\pi}$.)
Using the orthogonality properties of Dirac spinors, we find \cite{Zub80}
\small\eqarr{
b^s\bb{p_{\z},m_{\ii}} &=& \varphi\bb{p_{\z} -  p_{\0}}\,u^{s
\dagger}\bb{p_{\z},m_{\ii}} \, w, \nonumber\\
d^{s*}\bb{\mi  p_{\z},m_{\ii}} &=& \varphi\bb{p_{\z} -  p_{\0}}\,v^{s \dagger}\bb{\mi
 p_{\z},m_{\ii}} \, w.~~
\label{III26}
}\normalsize
These coefficients carry an important physical information.
For {\em any} initial state which has the form given in Eq.~(\ref{III22}), the
negative frequency solution coefficient $d^{s*}\bb{\mi p_{\z}, m_{\ii}}$ necessarily
provides a non-null contribution to the time evolving wave packet.
This obliges us to take the complete set of Dirac equation solutions to construct the
wave packet.
In the general case, one can succeed to choose $\psi_{\nue}\bb{z,0,\theta}$ in such
a way that $\psi_{\1}\bb{z,0}=\psi_{\nue}\bb{z,0,\theta}$ has vanishing coefficients
$d^{s*}\bb{\mi p_{\z}, m_{\1}}=0$ but then
$\psi_{\2}\bb{z,0}=\psi_{\nue}\bb{z,0,\theta}$ will necessarily contain
non-vanishing coefficients $d^{s*}\bb{\mi p_{\z}, m_{\2}}$ as it is shown in
section~\ref{subsec:IFV:D}.

Having introduced the Dirac wave packet prescription, we are now in a position to calculate the flavor conversion formula.
The following calculations do not depend on the gamma matrix representation.
By substituting the coefficients given by Eq.~(\ref{III26}) into Eq.~(\ref{III23}) and using the well-known spinor properties \cite{Zub80},
\small\eqarr{
\sum_{i=1,2}\mbox{$u^s\bb{p_{\z}, m_{\ii}}\overline{u}^s\bb{p_{\z}, m_{\ii}}$} &=&
\frac{\gamma^0 E\bb{p_{\z}, m_{\ii}} - \gamma^{\3}  p_{\z} + m_{\ii}}{2E\bb{p_{\z},
m_{\ii}}}
, \nonumber\\
\sum_{i=1,2}\mbox{$v^s\bb{\mi p_{\z}, m_{\ii}}\overline{v}^s\bb{\mi p_{\z},
m_{\ii}}$} &=&
\frac{\gamma^0 E\bb{p_{\z}, m_{\ii}} + \gamma^{\3}  p_{\z} -m_{\ii}}{2E\bb{p_{\z},
m_{\ii}}},
\label{III28}
}\normalsize
we obtain
\small\eqarr{
\psi_{\ii}\bb{z,t}  = \int_{_{-\infty}}^{^{+\infty}}\!\!\frac{d p_{\z}}{2 \pi} \,
\varphi\bb{p_{\z} - p_{\0}}  \exp{[i p_{\z} z]}
\left\{\cos{[E\bb{p_{\z}, m_{\ii}} t]} -\frac{i\gamma^0\left(\gamma^{\3}  p_{\z}+
m_{\ii}\right)}{E\bb{p_{\z}, m_{\ii}}}\sin{[E\bb{p_{\z}, m_{\ii}} t]}\right\} w
\,.~~
\label{III29}
}\normalsize
By simple mathematical manipulations, the new interference oscillating term in Eq.~\eqref{D:int} will be written as
\small\eqarr{
\mbox{\sc Dfo}\bb{t} & = &\int_{_{-\infty}}^{^{+\infty}}\!\frac{d p_{\z}}{2 \pi} \,
\varphi^{\2}\bb{p_{\z} - p_{\0}}
\left\{\left[1 - f\bb{p_{\z},m_{\1,\2}}\right]
\cos[\epsilon_{\mi}\bb{p_{\z},m_{\1,\2}} \, t ]
\right.
\cr  & & \hspace{9em} \left.
+\ f\bb{p_{\z},m_{\1,\2}}  \cos [\epsilon_{\pl}\bb{p_{\z},m_{\1,\2}} \, t] \right\}
\label{III30}
}\normalsize
where
\[
f\bb{p_{\z},m_{\1,\2}}= f\bb{p_{\z},m_{\1},m_{\2}}=
\frac{E\bb{p_{\z},m_{\1}}E\bb{p_{\z},m_{\2}} -
 p_{\z}^{\2} - m_{\1} m_{\2}}{2\, E\bb{p_{\z},m_{\1}}E\bb{p_{\z},m_{\2}}}\]
and
\[\epsilon_{\pm}\bb{p_{\z},m_{\1,\2}} = \epsilon_{\pm}\bb{p_{\z},m_{\1},m_{\2}} =
E\bb{p_{\z},m_{\1}} \pm
E\bb{p_{\z},m_{\2}}~.
\]
The time-independent term $f \bb{p_{\z},m_{\1},m_{\2}}$ deserves some comments.
It has a minimum  at $p_{\z}=0$ and two maxima at $p_z = \pm\sqrt{m_{\1} m_{\2}}$.
We can readily observe in Fig.\,\ref{fig1} that it goes rapidly to zero when $p_{\z}
\gg m_{1,2}$ (UR limit) as well as when $p_{\0} \ll m_{1,2}$
(NR limit).
It means that when we consider a momentum distribution sharply peaked around $p_{\0}
\gg m_{1,2}$ or $p_{\0} \ll m_{1,2}$ the corrections introduced by $f
\bb{p_{\z},m_{\1},m_{\2}}$ are negligible.
The maximum value of $f \bb{p_{\z},m_{\1},m_{\2}}$ is
\small\eq{
f_{max}
 = \frac{1}{2} - \frac{\sqrt{m_{\1}m_{\2}}}{m_{\1}+m_{\2}}
=\frac{(\sqrt{m_1}-\sqrt{m_2})^2}{2(m_1+m_2)}
\,,
\label{III36}
}\normalsize
which vanishes in the limit $m_{\1} = m_{\2}$.

The effects introduced by $f \bb{p_{\z},m_{\1},m_{\2}}$ are relevant only when $\Delta m \approx m_{\1} \gg m_{\2}$.
Meanwhile, what is interesting about the result in Eq.~(\ref{III30}) is that it was
obtained without any assumption on the initial spinor $w$ or Fourier transform
$\varphi\bb{p_{\z} - p_{\0}}$.
Otherwise, the initial spinor carries some fundamental physical information about the created state.
And this could be relevant in the study of chiral oscillations \cite{Ber06A,Ber06B} where the initial state plays a fundamental role.
In comparison with the standard treatment of neutrino oscillations done by using {\em scalar} wave packets, where the interference term  $\mbox{\sc Int}\bb{t}$ is given by Eq.~(\ref{II9}) with $\Delta E( p_{\z}) \equiv \epsilon_{\mi}\bb{p_{\z},m_{\1},m_{\2}}$, we notice in $\mbox{\sc Dfo}\bb{t}$ two additional terms.
In the first one, the {\em standard} oscillating term $\cos{[\epsilon_{\mi}\bb{p_{\z},m_{\1,\2}}\, t]}$, which arises from the interference between mass-eigenstate components of equal sign frequencies, is multiplied by a {\em new factor} obtained by the products $u^{\dagger}\bb{p_{\z},m_{\1}}\,u\bb{p_{\z},m_{\2}}$, $v^{\dagger}(\mi p_{\z},m_{\1})\,v(\mi  p_{\z},m_{\2})$ and h.c..
The second one is a {\em new oscillating term}, $\cos{[\epsilon_{\pl}\bb{p_{\z},m_{\1,\2}}\, t]}$, which comes from the interference between mass-eigenstate components of positive and negative frequencies.
The factor multiplying such an additional oscillating term is obtained by the products $u^{\dagger}\bb{p_{\z},m_{\1}}\,v\bb{\mi p_{\z}, m_{\2}}$, $v^{\dagger}\bb{\mi p_{\z},m_{\1}}\,u\bb{p_{\z},m_{\2}}$ and h.c..

The new oscillations have very high angular frequencies $\epsilon_+$.
Indeed, we can find the relation $\epsilon_+\epsilon_-=\Delta m^2$, where $\epsilon_-$ is the usual (double of) flavor oscillation angular frequency.
Such a peculiar oscillating behavior is similar to the phenomenon referred to as ZBW.
In atomic physics, the electron exhibits this violent quantum fluctuation in the
position and becomes sensitive to an effective potential which explains the Darwin
term in the hydrogen atom \cite{Sak87}.
We shall see later that, at the instant of creation, such rapid oscillations
introduce a small modification in the oscillation formula.

\subsection{The oscillation formula with first order corrections}
\label{subsec:Dirac+1}

A more satisfactory interpretation of the modifications introduced by the Dirac
formalism is given when we explicitly calculate $\mbox{\sc Dfo}\bb{t}$.
To that end, we resort to the same gaussian wave packet used in section
\ref{sec:localization} and use Eq.~(\ref{II9B}) for the scalar part
$\phi_{\nue}\bb{z}$ of the initial wave packet in Eq.\,\eqref{III22}:
\small\eq{
\psi_{\nue}\bb{z,\,t=0;\,\theta}
=\phi_{\nue}(z)w
=\left(\frac{2}{\pi a^{\2}}\right)^{\frac{1}{4}}
\exp{\left[-\frac{z^{\2}}{a^{\2}}\right]}
\exp{[ i p_{\0}z]}\,w\,.
\label{III34:initial}
}\normalsize
However, in contrast to section \ref{sec:localization}, where we have considered
the energy $E\bb{p_{\z}, m_{\ii}}$ expansion up to the second order terms in
Eq.~(\ref{II12}), so that the spreading effects were included in the analysis, we
focus this preliminary study only on first order corrections.
Thus, we  approximate the frequency components by
\small\eq{
E\bb{p_{\z}, m_{\ii}} \approx \, E_{\ii} \,  + \mbox{v}_{\ii} \, \left(
 p_{\z}- p_{\0}  \, \right) ~.
\label{III34AA}
}\normalsize

Considering the approximation \eqref{III34AA}, we can write
\small\eqarr{
f\bb{p_{\z},m_{\1},m_{\2}}  \approx  \mbox{$\frac{1}{2}$} \, \left\{
1 - \mbox{v}_{\1}\mbox{v}_{\2}\left(1 + \frac{m_{\1}m_{\2}}{p_{\0}^{\2}}\right)+
\mbox{v}_{\1}\mbox{v}_{\2} \, \left[\left(\mbox{v}_{\1}^{\2} +
\mbox{v}_{\2}^{\2}\right)
\left(1 +  \frac{m_{\1}m_{\2}}{p_{\0}^{\2}}\right)
-2 \right] \, \frac{ p_{\z}- p_{\0} }{ p_{\0}} \, \right\}
}\normalsize
and
\small\eqarr{
\epsilon_{\ppm}\bb{p_{\z},m_{\1},m_{\2}} & \approx & E_{\1} \pm
E_{\2} + \left( \, \mbox{v}_{\1} \pm \mbox{v}_{\2} \, \right) \,\left(
 p_{\z}- p_{\0}  \, \right) ~.
}\normalsize
For UR particles ($m_{\ii}\ll  p_{\0}$), we can also use the following expression for the central energy values ($E_{\ii}$) and the group velocities ($\mbox{v}_{\ii}$)  of the mass-eigenstate wave packets,
\[ E_{\ii} \approx \,  p_{\0} +
\frac{m_{\ii}^{\2}}{2\, p_{\0}}~~~~~\mbox{and} ~~~~~ \mbox{v}_{\ii} \approx
\, 1 -  \frac{m_{\ii}^{\2}}{2\, p_{\0}^{\, \2}}~.
\]
This implies
\begin{eqnarray*}
f(p_{\z},m_{\1},m_{\2}) & \approx & \left(\frac{\Delta m}{2 \, p_{\0}}\right)^{\2} \,
\left( \, 1 -  2 \, \,
\frac{ p_{\z}- p_{\0} }{ p_{\0}} \, \right)
\, \, , \\
\epsilon_{\pl}\bb{p_{\z},m_{\1},m_{\2}} & \approx & 2 \,  p_{\0}   \,
\left[ \, 1 \,  + \, \frac{m_{\1}^{\2}+ m_{\2}^{\2}}{4\, p_{\0}^{\,
\2} } \, +
\frac{ p_{\z}- p_{\0} }{ p_{\0}} \, \left( \,  1 -
\frac{m_{\1}^{\2} + m_{\2}^{\2}}{4\, p_{\0}^{\,  \2} }\, \right) \,
\right]\, \, ,\\
\epsilon_{\mi}\bb{p_{\z},m_{\1},m_{\2}} & \approx &  \frac{\Delta
m^{\2}}{2\, p_{\0}}  \, \left[ \, 1  - \, \frac{ p_{\z}- p_{\0}
}{ p_{\0}} \, \right]\, \, .
\end{eqnarray*}
where $(\Delta m)^{\2}= (m_{\1} - m_{\2})^{\2}$ is different from $\Delta m^{\2}=
m_{\1}^{\2} - m_{\2}^{\2}$ which appears in the {\em standard} oscillation phase.

Finally, by simple algebraic manipulations and after gaussian integrations, we find
for Eq.\,\eqref{III30},
\small\eqarr{
\lefteqn{\mbox{\sc Dfo}\bb{t} \approx \mbox{$\exp{\left[-\left(\frac{\Delta m^{\2}\,
t}{2 \sqrt{2}a  p_{\0}^{\2}}\right)^{\2}\right]}\left\{\left[1 - \left(\frac{\Delta
m}{2 p_{\0}}\right)^{\2}\right]\cos{\left[\frac{\Delta m^{\2}}{2  p_{\0}}t\right]} +
\left(\frac{\Delta m}{2 p_{\0}}\right)^{\2} \frac{\Delta m^{\2}}{a^{\2}
p_{\0}^{\3}}t\sin{\left[\frac{\Delta m^{\2}}{2
p_{\0}}t\right]}\right\}$}}\nonumber\\
&&        ~~~~~~~~~~~~~~~~~~~~~~~~~~~~~~~~ + \mbox{$\exp{\left[-\frac{t^{\2}}{2
a^{\2}}\left(2 - \frac{m_{\1}^{\2} + m_{\2}^{\2}}{2
p_{\0}^{\2}}\right)^{\2}\right]}\left(\frac{\Delta m}{2 p_{\0}}\right)^{\2}
\left\{\cos{\left[ p_{\0}t\left(2 + \frac{m_{\1}^{\2} + m_{\2}^{\2}}{2
p_{\0}^{\2}}\right)\right]}%
		 \right.$}\nonumber\\
         & &\mbox{$\left.~~~~~~~~~~~~~~~~~~~~~~~~~~~~~~~~~~~~~~~~~~~~~~~~~~~
		 + \frac{2 p_{\0}t}{(a  p_{\0})^{\2}}\left(2 - \frac{m_{\1}^{\2} +
m_{\2}^{\2}}{2  p_{\0}^{\2}}\right)\sin{\left[ p_{\0}t\left(2 + \frac{m_{\1}^{\2} +
m_{\2}^{\2}}{2  p_{\0}^{\2}}\right)\right]}\right\}$}.~~~~
\label{IIIA14}
}\normalsize
As we have already noticed, the oscillating functions going with the second
exponential function in Eq.~(\ref{IIIA14}) arise from the interference between
positive and negative frequency solutions of the Dirac equation.
It produces very high frequency oscillations which is similar to the quoted
phenomenon of ZBW \cite{Sak87}.
The oscillation length which characterizes the very high frequency oscillations is
given by $L^{VHF}_{Osc} \approx \frac{\pi}{ p_{\0}}$.
Obviously, $L^{VHF}_{Osc}$ is much smaller than the standard oscillation length given
by $L^{Std}_{Osc} = \frac{4 \pi  p_{\0}}{\Delta m^{\2}}$.
It means that the propagating particle exhibits a violent quantum fluctuation of its flavor quantum number around a flavor average value which oscillates with $L^{Std}_{Osc}$.
Meanwhile, except at times $t \sim 0$, it provides a practically null contribution to
the oscillation probability as we can observe in Fig.\,\ref{zwb}.

To explain such a statement, let us suppose that an experimental measurement takes place after a time $t \approx L$ for UR particles.
The observability conditions impose that the propagation distance $L$ must be larger than the wave packet localization $a$.
Since the (second) exponential function vanishes when $L \gg a$, for measurable distances, the effective flavor conversion formula will not contain such very high frequency oscillation terms, and can be written as
\small\eqarr{
P_{\Dirac}(\bs{\nue}\rightarrow\bs{\numu};L)
&\approx&
\frac{\sin^{\2}{(2\theta)}}{2}\!
\left\{ 1 -\exp\!{\left[-\left(\!\frac{\Delta m^{\2}\, L}{2 \sqrt{2}a
p_{\0}^{\2}}\right)^{\!\2}\right]}\right.
\!\left\{\left[1 - \left(\!\frac{\Delta m}{2
p_{\0}}\!\right)^{\!\2}\right]\!\cos\!{\left[\frac{\Delta m^{\2}}{2
p_{\0}}L\right]}+
~~ \right.
\nonumber\\
&& ~~~~~~~~~~~~~~~~~~~~ ~~~~~~~~~~~~~~~~~~~~ ~~~~~~~
\left.\left.
\left(\!\frac{\Delta m}{2 p_{\0}}\!\right)^{\!\2} \!\frac{\Delta m^{\2}}{a^{\2}
p_{\0}^{\3}}L\sin\!{\left[\frac{\Delta m^{\2}}{2  p_{\0}}L\right]}\right\}\right\}
\!.~~~
\label{IIIA14BB}
}\normalsize
For distances which are restricted to the interval $a \ll L \ll a \frac{2 \sqrt{2}
p_{\0}^{\2}}{\Delta m^{\2}}$,
 the interference term is preserved due to minimal {\em slippage} between the wave
packets.
In this case, we could approximate the oscillation probability to
\small\eqarr{
P_{\Dirac}(\bs{\nue}\rightarrow\bs{\numu};L)
&\approx&
\frac{\sin^{\2}{(2\theta)}}{2}\left\{
1 - \left[1-\left(\frac{\Delta m^{\2} L}{2 \sqrt{2}a  p_{\0}^{\2}}\right)^{\2}\right]
\left[1 - \left(\frac{\Delta m}{2 p_{\0}}\right)^{\2}\right]\cos{\left[\frac{\Delta
m^{\2}}{2  p_{\0}}L\right]}\right\}
\,.
\label{IIIA14B}
}\normalsize
However, we reemphasize that it is {\em not} valid for $T \approx L \sim 0$ when the
rapid oscillations are still relevant ($L < a$).
By comparing the result of Eq.~(\ref{IIIA14B}) with the {\em scalar} oscillation
probability of Eq.~(\ref{II20}), we notice a deviation of the order
$\big(\!\frac{\Delta m}{2 p_{\0}}\!\big)^{\!\2}$ that appears as an additional
coefficient of the cosine function.
It is not relevant in the UR limit as we have noticed after studying the function $f(p_{\z},m_{\1},m_{\2})$.

\subsection{Time evolution operator}
\label{subsec:timeev}

\providecommand{\TT}{\mathsf{T}}
\providecommand{\ST}{\mt{ST}}

An initial value problem, described by a differential equation, can be completely
solved if we find a well defined kernel that governs the time evolution. Then from an
initial field (wave function) configuration we can find the subsequent
configurations at all subsequent times. Moreover, it is possible to analyze the
properties of time evolution that are independent of the choice of the initial wave
packet. Analyzing the properties of the kernel, it is also possible to check the
properties of relativistic completeness for the free propagation of Dirac fermions
with flavor mixing.
Therefore, we will explicitly calculate, in this section, the time evolution operator
(kernel) in presence of mixing in order to extract the common features independently
of the initial wave packet. The extension to treat three families of neutrinos is
straightforward. A matricial notation will be used throughout this subsection to
express the mixing.
Since the one-dimensional restriction does not lead to any formal simplification, the
full three dimensional space will be considered here.

In matricial notation the mixing relation that extends Eq.~\eqref{II:mixing},
between flavor wave functions ${\Psi}_f^{\TT}(\bx)\equiv
(\psi_{\nu_e}^{\TT}(\bx),\psi_{\nu_\mu}^{\TT}(\bx))$
and mass wave functions ${\Psi}_m^{\TT}(\bx)\equiv (\psi_1^{\TT}(\bx),
\psi_2^{\TT}(\bx))$,  is
\small\eq{
\label{mixing}
\Psi_{f}(\bx)
\equiv
U\Psi_{m}(\bx)
=
\left(\begin{array}{cc}
\phantom{\!\!\!-}\cos\!\theta & \sin\!\theta \cr
\!\!\!-\sin\!\theta & \cos\!\theta
\end{array}\right)
\Psi_{m}(\bx)
~.
}\normalsize
Each mass wave function is defined as a four-component
spinorial function $\psi_n(\bx,t)$, $n=1,2$ that satisfy the free Dirac
equation\,\eqref{III00}
\small\eq{
\label{nu:n:eq}
i\frac{\partial}{\partial t}\psi_n(\bx,t)=H_n^D\psi_n(\bx,t)~,~~n=1,2~,
}\normalsize
where the free Hamiltonian is the usual
\small\eq{
\label{nu:n:H}
H_n^D\equiv -i\bs{\alpha}\ponto\nabla + \beta m_n~,~~n=1,2~.
}\normalsize

The normalization condition is
\small\eq{
\label{III:norm}
\int d^3\!\bx\,\Psi_m^\dag(\bx)\Psi_m(\bx)=\int
d^3\!\bx\,\Psi_f^\dag(\bx)\Psi_f(\bx)=1\,,
}\normalsize
which implies that the wave functions $\psi_1(\bx)$ and $\psi_2(\bx)$, or, for the
same reason, $\psi_{\nu_e}(\bx)$ and $\psi_{\nu_\mu}(\bx)$, can not be
simultaneously normalized to unity. A different notation and normalization
convention was being used through this review and it deserves a further
clarification. As it can be seen in Eqs.\,\eqref{II0} and \eqref{III0B},
the wave functions for the mass-eigenstates were normalized to unity. To
recover Eq.~\eqref{III0B}, the wave functions $\psi_1$ and $\psi_2$ used in this
subsection, as in Eq.~\eqref{mixing}, should be replaced by
\small\eq{
\psi_1 \rightarrow \, \psi_1/\cos\theta\,,\qquad
\psi_2 \rightarrow \, \psi_2/\sin\theta\,.
}\normalsize
This correspondence should be clear from the first line of Eq.~\eqref{II0}.
Equation \eqref{III:norm} is a restatement of the fact that the
probability to find a neutrino over all space irrespective of mass or irrespective
of flavor is unity. We also see that condition \eqref{III:norm} is time independent.

We will work in the flavor diagonal basis. This choice defines the flavor
basis vectors simply as
\small\eq{
\label{versor}
\bs{\nu_e} \rightarrow
\hat{\nu}_e=(1,0)^{\TT}~, ~~
\bs{\nu_\mu} \rightarrow
\hat{\nu}_{\mu}=(0,1)^{\TT}~,
}
while the flavor projectors are obviously
\small\eq{
\label{proj:f}
\mathsf{P}_{\nu_{\alpha}}\equiv \hat{\nu}_{\alpha}\hat{\nu}_{\alpha}^{\dag}
\,.
}\normalsize
Actually, as an abuse of notation, the equivalence $U \sim U\otimes \id_{D}$ is
implicit, as well as, $\mathsf{P}_{\nu_\alpha}\sim
\mathsf{P}_{\nu_\alpha}\otimes \id_D$; the symbol $\id_D$ refers to the identity
matrix in spinorial space.

The total Hamiltonian governing the dynamics of $\Psi_m$ is
$H^D=\mathrm{diag}(H_1^D,H_2^D)$. From the considerations above, $\Psi_f(\bx,t)$
satisfy the equation \small\eq{
\label{nu:f:eq}
i\frac{\partial}{\partial t}\Psi_f(\bx,t)\equiv
U H^D U^{-1} \Psi_f(\bx,t)~.
}\normalsize
The solution to the equation above can be written in terms of a flavor evolution
operator $K^D$ as
\small\eq{
\label{f:evol:D}
\Psi_f(\bx,t)=K^D(t)\Psi_f(\bx,0)=
\Int{3}{\bx'}K^D(\bx-\bx^{\pr};t)\Psi_f(\bx^{\pr},0)
~,
}\normalsize
where
\small\eq{
\label{KD}
K^D(\bx-\bx^{\pr};t)
=\int\!\frac{d^3\!\bp}{(2\pi)^3}\, K^D(\bp;t)\,e^{i\bp\cdot(\bx-\bx^{\pr})} ~.
}\normalsize
We can calculate $K^D(t)$ in any representation (momentum or position) as
\eqarr{
\label{KD:mtrx}
K^D(t)
&=&
U e^{-iH^Dt}U^{-1}
\cr
&=&
\left(
\begin{array}{cc}
\cos^2\!\theta\,e^{-iH_1^Dt}+\sin^2\!\theta\,e^{-iH_2^Dt} &
-\cos\!\theta\sin\!\theta(e^{-iH_1^Dt}-e^{-iH_2^Dt}) \cr
-\cos\!\theta\sin\!\theta(e^{-iH_1^Dt}-e^{-iH_2^Dt}) &
\sin^2\!\theta\,e^{-iH_1^Dt}+\cos^2\!\theta\,e^{-iH_2^Dt}
\end{array}
\right)~.
}\normalsize
It is important to emphasize that Eq.~\eqref{f:evol:D} leads to a causal
propagation because the kernel in Eq.~\eqref{KD} respects $K^D(\bx-\bx^{\pr};t)=0$
for space-like distances $|\bx-\bx'|>|t|$. This property follows directly from
Eq.~\eqref{KD:mtrx} and the properties of causality of each kernel,
$\bra{\bx}e^{-iH_n^Dt}\ket{\bx'}=0$ for $|\bx-\bx'|>|t|$, associated to the Dirac
Hamiltonian with mass $m_n$, $n=1,2$ [see Eqs.\,\eqref{S} and \eqref{S:D} or
Ref.\,\cite{thaller}].

The conversion probability is then
\eqarr{
\label{prob:em:D}
\mathcal{P}(\ms{\bs{\nu_e}\!\rightarrow\!\bs{\nu_\mu}};t)
&=&
\Int{3}{\bx}
\Psi_f^{\dag}(\bx,0)K^{D\,\dag}(t)
\mathsf{P}_{\nu_\mu}
K^D(t)\Psi_f(\bx,0)
\cr
&=&
\Int{3}{\bp}
\tilde{\psi}_{\nu_e}^{\dag}(\bp)(K_{\mu e}^{D})^\dag
K_{\mu e}^D(\bp,t) \tilde{\psi}_{\nu_e}(\bp)
~,
}\normalsize
satisfying the initial condition $\Psi_f^{\TT}(\bx,0)=(\psi^{\TT}_{\nu_e}(\bx,0)
, 0)$. Such initial condition implies, in terms of mass eigenfunctions,
$\psi_1(\bx,0)=\cos\theta\,\psi_{\nu_e}(\bx)$ and
$\psi_2(\bx,0)=\sin\theta\,\psi_{\nu_e}(\bx)$, as a requirement to obtain an
initial wave function with definite flavor, as seen in
section\,\ref{subsec:flavorvio:I}\,\cite{DeL04}. The function
$\tilde{\psi}_{\nu_e}(\bp)$ denotes the inverse Fourier transform of
$\psi_{\nu_e}(\bx)$ (see Eqs.~\eqref{fourier:xp} and \eqref{fourier:px}).

Before obtaining the conversion probability for Dirac fermions, let us replace
the spinorial functions $\psi_n(\bx)$ by spinless one-component wave functions
$\varphi_n(\bx)$ in the flavor wave function ${\Psi}_f^{\TT}(\bx)\rightarrow
(\varphi_{\nu_e}(\bx),\varphi_{\nu_\mu}(\bx))$ and mass wave function
${\Psi}_m^{\TT}(\bx)\rightarrow (\varphi_1(\bx), \varphi_2(\bx))$. We also
replace  the Dirac Hamiltonian in momentum space $H^D_n(\bp)$ \eqref{nu:n:H} by
the relativistic energy $E_n(\bp)=\sqrt{\bp^2+m^2_n}$.
Inserting these replacements into Eq.~\eqref{prob:em:D} we can recover the
usual oscillation probability \cite{DeL04,Ber05}
\eqarr{
\label{prob:em:S}
\mathcal{P}(\ms{\bs{\nu_e}\!\rightarrow\!\bs{\nu_\mu}};t)&=&
\Int{3}{\bx} |\hat{\nu}_\mu^{\TT}\Psi_{f}(\bx,t)|^2
\cr
&=&
\Int{3}{\bp}|K^S_{\mu e}(\bp,t)\tilde{\varphi}_{\nu_e}(\bp)|^2
\cr
&=&
\Int{3}{\bp}\mathscr{P}(\bp,t)
|\tilde{\varphi}_{\nu_e}(\bp)|^2
~,
}\normalsize
where $\Psi_{f}(\bx,0)^{\TT}=(\varphi_{\nu_e}(\bx)^\TT,0)$,
$K^S_{\mu e}(\bp,t)\equiv(K^S)_{21}=
-\sin\theta\cos\theta(e^{-iE_1(\bp)t}-e^{-iE_2(\bp)t})$ and
\small\eq{
\label{P}
\mathscr{P}(\bp,t)=\sin^2\!2\theta\sin^2\!(\Delta E\mfn{(\bp)}t/2)
~
}\normalsize
is just the standard oscillation formula \eqref{II:standard} with $\Delta
E\bb{\bp}\equiv E_1\bb{\bp}-E_2\bb{\bp}$.
The conversion probability \eqref{prob:em:S} in this case is then the standard
oscillation probability smeared out by the initial momentum distribution. If the
substitution $|\tilde{\varphi}_{\nu_e}(\bp)|^2\rightarrow \delta^3(\bp-\bp_0)$ is
made the standard oscillation formula is recovered: it corresponds to the plane-wave
limit.

After we have checked the standard oscillation formula can be recovered for
spinless particles restricted to positive energies, we can return to the case of
Dirac fermions.
We can obtain explicitly the terms of the mixed evolution
kernel \eqref{KD:mtrx} by using the property of the Dirac Hamiltonian in
momentum space $(H_n^D)^2=E_n^2(\bp)\,\id_D$, which leads to
\eqarr{
\label{KD:em:0}
(K_{\mu e}^{D})^\dag K_{\mu e}^D(\bp,t)
&=&
\label{KD:em}
\mathscr{P}(\bp,t)[1-f(\bp)]\id_D
\cr&&
+\, \sin^2\!2\theta f(\bp)\sin^2(\bar{E}\bb{\bp}\,t)\id_D
~, }
where
\small\eq{
\label{ccn:f}
f(\bp)=\frac{1}{2}[1-\frac{\bp^2+m_1m_2}{E_1E_2}]~,
}\normalsize
$\bar{E}\bb{\bp}\equiv\frac{1}{2}[E_1\bb{\p}+E_1\bb{\p}]$\,\cite{endnote0}
and $\mathscr{P}(\bp,t)$ is the standard conversion probability function
\eqref{P}. The expression \eqref{ccn:f} was already found in Eq.~\eqref{III30} in a
slightly different form.
A unique implication of Eq.~\eqref{KD:em}, which is
proportional to the identity matrix in spinorial space, is that the conversion
probability \eqref{prob:em:D} does not depend on the spinorial structure of
the initial flavor wave function but only on its momentum density as
\begin{align}
\label{prob:em:D2}
\mathcal{P}(\ms{\bs{\nu_e}\!\rightarrow\!\bs{\nu_\mu}};t)=
\Int{3}{\bp}\{\mathscr{P}(\bp,t)[1-f(\bp)] \hs{3em}
\nonumber\\
+~\sin^2\!2\theta f(\bp)\sin^2(\bar{E}t)\}\,
\tilde{\psi}^\dag_{\nu_e}(\bp)\tilde{\psi}_{\nu_e}(\bp)
\,.
\end{align}
(The tilde will denote the inverse Fourier transformed function throughout
this subsection.) Furthermore, the modifications in Eq.~\eqref{prob:em:D2} compared
to the scalar conversion probability \eqref{prob:em:S} are exactly the same
modifications found in Eq.\,\eqref{III30} if we replace
$\varphi^{\2}\bb{p_{\z}-p_{\0}}/2\pi$ by
$\tilde{\psi}^\dag_{\nu_e}(\bp)\tilde{\psi}_{\nu_e}(\bp)$ and use three-dimensional
integration.

The conservation of total probability
\small\eq{
\label{prob:cons}
\mathcal{P}(\ms{\bs{\nue}\!\rightarrow\!\bs{\numu}};t)+
\mathcal{P}(\ms{\bs{\nue}\!\rightarrow\!\bs{\nue}};t)=1
~,
}\normalsize
is automatic in virtue of
\small\eq{
\label{prob:cons:D}
K_{ee}^{D\dag}(\bp,t)K_{ee}^D(\bp,t)+
K_{\mu e}^{D\dag}(\bp,t)K_{\mu e}^D(\bp,t)=\id_D
~,
}\normalsize
and initial normalization in Eq.~\eqref{III:norm}.
The survival and conversion probability for an initial muon neutrino are
identical to the probabilities for an initial electron neutrino because of the
relations
\eqarr{
\label{prob:me:D}
K_{\mu\mu}^{D\dag}(\bp,t)K_{\mu\mu}^D(\bp,t)&=&
K_{ee}^{D\dag}(\bp,t)K_{ee}^D(\bp,t)
~,
\\
K_{\mu e}^{D\dag}(\bp,t)K_{\mu e}^D(\bp,t)&=&
K_{e\mu}^{D\dag}(\bp,t)K_{e\mu}^D(\bp,t)
~.
}\normalsize

\subsection{Spinless neutrinos}
\label{subsec:ST}
\providecommand{\TT}{\mathsf{T}}
\providecommand{\ST}{\mt{ST}}

The derivation of the usual conversion probability \eqref{prob:em:S} takes into
account only the positive frequency contributions.
The mass wave function used to obtain Eq.~\eqref{prob:em:S} corresponds to the
solutions of the wave equation
\small\eq{
\label{we:S}
i\frac{\partial}{\partial t}\varphi(\bx,t)
=\sqrt{-\nabla^2+m^2}\,\varphi(\bx,t)~,
}\normalsize
which is equivalent to the Dirac equation in the Foldy-Wouthuysen
representation \cite{FW}, restricted to positive energies.
The evolution kernel for this equation is not satisfactory from the point of
view of causality \cite{thaller}, i.e, the kernel is not null for spacelike
intervals. Moreover, the eigenfunctions restricted to one sign of energy do not
form a complete set \cite{FV}.

To recover a causal propagation in the spin 0 case, the Klein-Gordon wave
equation must be considered. In the first quantized version, the spectrum of
the solutions have positive and negative energy as in the Dirac case.
However, to take advantage of the Hamiltonian formalism used in
section\,\ref{subsec:timeev}, it is more convenient do work in the Sakata-Taketani (ST)
Hamiltonian formalism \cite{FV} where each mass wave function is formed by two
components
\small\eq{
\label{Phi}
\Phi_n(\bx,t)=
\left(
\begin{array}{c}
\varphi_n(\bx,t) \cr \chi_n(\bx,t)
\end{array}
\right)
~,~~n=1,2\,.
}\normalsize
The components $\varphi$ and $\chi$ are combinations of the usual scalar
Klein-Gordon wave function $\phi(x)$ and its time derivative
$\partial_0\phi(x)$. This is necessary since the Klein-Gordon equation is
a second order differential equation in time and the knowledge of the function
and its time derivative is necessary to completely define the time evolution.
An analysis of the flavor oscillation of scalar wave functions obeying the
Klein-Gordon equation can be found in Ref.\,\cite{dvornikov}.

The time evolution in this formalism is governed by the Hamiltonian \cite{FV}
\small\eq{
\label{H:ST}
H_n^{\mt{ST}}=-(\tau_3+i\tau_2)\frac{\nabla^2}{2m_n}+m_n^2
~,
}\normalsize
which satisfies the condition $(H_n^{\ST})^2=(-\nabla^2+m_n^2)\id_{\ST}$,
like the Dirac Hamiltonian \eqref{nu:n:H}. The $\tau_k$ represents the usual
Pauli matrices and $\id_{\ST}$ is the identity matrix.

A charge density \cite{endnote1} can be defined as
\small\eq{
\label{rho:S}
\bar{\Phi}_n\Phi_n \equiv \Phi_n^\dag\tau_3\Phi_n=|\varphi_n|^2-|\chi_n|^2~,
}
which is equivalent to the one found in Klein-Gordon notation
$i\phi^*\overset{\leftrightarrow}{\partial}_0\phi$. Needless to say, this
density \eqref{rho:S} is only non-null for complex (charged) wave functions.
The charge density $\bar{\Phi}\Phi$ is the equivalent of fermion probability
density $\psi^{\dag}\psi$ in the Dirac case, although the former is not positive
definite as the latter. The adjoint $\bar{\Phi}=\Phi^\dag\tau_3$ were defined to
make explicit the (non positive definite) norm structure of the conserved charge
\small\eq{
\label{Q:ST}
\Int{3}{\bx}\bar{\Phi}_n(\bx,t)\Phi_n(\bx,t) \equiv
(\Phi_n,\Phi_n)=\text{time independent}
~.
}\normalsize
Consequently, the adjoint of any operator $\Omega$ can be defined as
$\bar{\Omega}=\tau_3\Omega^\dag\tau_3$, satisfying
$(\bar{\Omega}\Phi,\Phi)=(\Phi,\Omega\Phi)$. Within this notation, the
Hamiltonians of Eq.~\eqref{H:ST} is self-adjoint, $\bar{H}_n^{\ST}=H_n^{\ST}$,
and the time invariance of Eq.~\eqref{Q:ST} is assured.

We can assemble, as in the previous section, the mass wave functions into
$\Psi_m^\TT \equiv (\Phi_1^\TT,\Phi_2^\TT)$ and the flavor wave functions into
$\Psi_f^\TT \equiv (\Phi_{\nu_e}^\TT,\Phi_{\nu_\mu}^\TT)$, satisfying
the mixing relation $\Psi_f\equiv U\Psi_m$.
Then, the time evolution of $\Psi_f$ can be given
through a time evolution operator $K^{\ST}$ acting in the same form as in
Eq.~\eqref{f:evol:D}.
In complete analogy to the calculations from Eq.~\eqref{KD}
to Eq.~\eqref{prob:em:D}, we can define the conversion probability as
\eqarr{
\label{prob:em:ST}
\mathcal{P}(\ms{\bs{\nu_e}\!\rightarrow\!\bs{\nu_\mu}};t)
&=&\Int{3}{\bx}
\bar{\Psi}_f(\bx,0)\overline{K^{\ST}(t)}\mathsf{P}_{\nu_\mu}K^{\ST}(t)
\Psi_f(\bx,0) \cr &=&
\Int{3}{\bp}\overline{\tilde{\Phi}_e(\bp)}\overline{K_{\mu e}^{\ST}}
K_{\mu e}^{\ST}(\bp,t)\tilde{\Phi}_e(\bp)
~,
}\normalsize
where $\Psi_f(\bx,0)^\TT=(\Phi_e(\bx)^\TT,0)$. The adjoint operation were
also extended to $\bar{\Psi}_f=\Psi_f^\dag(\id_{\theta}\otimes\tau_3)$, where
$\id_\theta$ is the identity in mixing space.

The information of time evolution, hence oscillation, is all encoded in
\eqarr{
\label{KST:em:0}
\overline{K_{\mu e}^{\ST}}K_{\mu e}^{\ST}(\bp,t)
&=&
\label{KST:em}
\mathscr{P}(\bp,t)[1-f(\mu\bp)]\id_{\ST}
\cr &&
+~\sin^2\!2\theta f(\mu\bp)\sin^2(\bar{E}t)\id_{\ST} ~,
}\normalsize
where the function $f(\bp)$ were already defined in Eq.~\eqref{ccn:f} and
\small\eq{
\label{mu}
\mu=\sqrt{\frac{1}{2}(\frac{m_1}{m_2}+\frac{m_2}{m_1})}
~.
}\normalsize
The factor $\mu\ge 1$ determines the difference with the Dirac case in
Eq.~\eqref{KD:em}.
The equality $\mu=1$ holds when $m_1=m_2$, i.\,e., when there is no
oscillation.

Therefore, rapid oscillations are also present in flavor
oscillations of charged (in the sense of Eq.\,\eqref{rho:S}) spin 0 particles, with
contributions slightly different from the Dirac case. Such result reinforces the
fact that rapid oscillations are direct consequences of flavor mixing, (a)
relativistic nature of the wave equations governing time evolution and (b) initial
flavor definition. The presence of a spinorial degree of freedom is not a
requirement to rapid oscillations.

\subsection{Initial flavor violation and rapid oscillations}
\label{subsec:IFV:D}

We saw in Eqs.\,\eqref{III30} and \eqref{prob:em:D2} that the flavor conversion
probability for Dirac fermions presents rapid oscillation terms with frequency
$2\bar{E}=E_1(\bp)+E_2(\bp)$, in addition to the usual oscillation frequency $\Delta
E=E_1(\bp)-E_2(\bp)=\frac{m^2_1-m^2_2}{E_1(\bp)+E_2(\bp)}$. Such frequencies arise
naturally from the propagation of free particles with mixing in a relativistic
classical field theory that contains states with positive and negative frequencies
$\pm E_i(\bp)$, e.g., spin 1/2 fermions and spin 0 bosons.
It was assumed, however, that flavor can be well defined and a pure $\nue$ flavor is
created at $t=0$.
We will elucidate here that such rapid oscillations can be avoided but only at
the expense of not having an initial flavor exactly defined.
More specifically, we will see there is a clash between two choices:
\begin{itemize}
\item \textbf{cond.\,A}: exact initial flavor definition (pure flavor creation) and
\item \textbf{cond.\,B}: standard flavor oscillation, without rapid oscillations.
\end{itemize}

Firstly, following the formalism of subsection \ref{subsec:timeev},
we calculate the probability that a generic superposition of neutrinos
$\nu_1$ and $\nu_2$, denoted simply by $\nu$, described by a generic $\Psi_m$, be
detected as a neutrino $\numu$:
\small\eq{
\label{Fdef:prob}
\mathcal{P}(\bs{\nu}\!\rightarrow\!\bs{\numu};t)=
\int d\bx\,\Psi_m^\dag(\bx)e^{-iH^Dt}\mathsf{P}_{\nu_{\mu}}e^{-iH^Dt}
\Psi_m(\bx)\,,
}\normalsize
where the operator inside, in the mass-basis, can be calculated explicitly
\small\eq{
\label{Fdef:time.dep}
e^{-iH^Dt}\mathsf{P}_{\nu_{\mu}}e^{-iH^Dt}=
\begin{pmatrix}
\sin^2\theta & -\sin\theta\cos\theta e^{iH_1^Dt}e^{-iH_2^Dt} \cr
 -\sin\theta\cos\theta e^{iH_2^Dt}e^{-iH_1^Dt} & \cos^2\theta
\end{pmatrix}
\,.
}\normalsize

Notice in this section we are working in the mass-basis, keeping the same notation
as in section\,\ref{subsec:timeev}, where
$(\mathsf{P}_{\nu_{\mu}})_{ij}=\hat{\nu}_\mu \hat{\nu}_\mu^\dag=U^*_{\mu i} U_{\mu
j}$, since $(\hat{\nu}_\alpha)_i\equiv U^*_{\alpha i}$ and
$(\hat{\nu}_i)=\delta_{ij}$ in this basis.
Only the off-diagonal terms depend on time $t$, and energies $E_1$ and $E_2$, which
means the functional dependence on time and energies, in momentum space, should be
of the form $g(\Delta Et,\bar{E}t)$, where $\Delta E\equiv\,E_1(\bp)-E_2(\bp)$ and
$2\bar{E}\equiv E_1(\bp)+E_2(\bp)$. Such functional form is in accordance to
Eq.~\eqref{prob:em:D2}, for example.
To disentangle the dependencies on $\Delta Et$ and $\bar{E}t$, it is instructive to
rewrite the free Dirac time evolution operator, in momentum space, in the form
\small\eq{
\label{UD}
e^{-iH_n^Dt}=e^{-iE_nt}\Lambda_{n+}^D+e^{iE_nt}\Lambda_{n-}^D
~,
}\normalsize
where
\small\eq{
\label{Lambda+-}
\Lambda_{n\pm}^D=\frac{1}{2}(\id_D\pm \frac{H_n^D}{E_n})
~,
}\normalsize
are the projector operators to positive (+) or negative (-) energy eigenstates
of $H_n^D$.
By using the decomposition above \eqref{UD}, we can separate the different
contributions in the off-diagonal terms of Eq.~\eqref{Fdef:time.dep} by rewriting
\eqarr{
e^{iH_1^Dt}e^{-iH_2^Dt}&=&e^{i\Delta Et}\Lambda_{1+}^D\Lambda_{2+}^D
+e^{-i\Delta Et}\Lambda_{1-}^D\Lambda_{2-}^D
\cr &&
+~e^{i2\bar{E}t}\Lambda_{1+}^D\Lambda_{2-}^D
+e^{-i2\bar{E}t}\Lambda_{1-}^D\Lambda_{2+}^D
\,.~~~
}\normalsize
The same procedure can be applied to its hermitian conjugate. The time evolution
kernel of Eq.~\eqref{KD:mtrx}, in the flavor basis, can be analyzed in the same
fashion.
Since $\Lambda_{1\pm}^D\Lambda_{2\mp}^D\neq 0$, it can be seen that the rapid
oscillating terms come from the interference between, \textit{e.\,g.}, the positive
frequencies of the Hamiltonian $H_1^D$ and negative frequencies of the Hamiltonian
$H_2^D$.

We know that both positive and negative energy states should propagate
properly to preserve causality. With a kernel containing only positive energy
contributions $e^{-iE_nt}$, there is no causal propagation\,\cite{thaller}.
Therefore, to maintain causality,  we can not eliminate, for example, the negative
energy contribution in the kernel of Eq.~\eqref{UD}.
One may, of course, restrict the initial wave functions to contain
only positive energy states. Such restriction eliminates the rapid oscillatory
terms, also known as {\it zitterbewegung}, for a one-particle free Dirac
theory\,\cite{Zub80}.
For two masses, however, the positive energy eigenfunctions
with respect to a basis characterized by a mass $m_1$ necessarily have non-null
components of negative energy with respect to another basis characterized by $m_2$,
as we will see.	

Let us examine exact initial flavor definition (\textbf{cond.\,A}). Firstly, we
introduce wave functions $\psi_1$ and $\psi_2$, normalized to unity, by using
\small\eq{
\label{Fdef:Psi:0}
\Psi_m=
\begin{pmatrix}
\cos\alpha \,\psi_1 \cr
\sin\alpha \,\psi_2
\end{pmatrix}\,,~~0\le\alpha\le \frac{\pi}{2}\,;
}\normalsize
where additional phases can be easily incorporated in $\psi_{1},\psi_2$.
Then Eq.~\eqref{Fdef:prob} yields
\small\eq{
\label{Fdef:prob:0}
\mathcal{P}(\bs{\nu}\!\rightarrow\!\bs{\numu};t)=
\sin^2(\theta+\alpha)\sin^2(\mn{\frac{1}{2}}\Xi_t)+
\sin^2(\theta-\alpha)\cos^2(\mn{\frac{1}{2}}\Xi_t)\,,
}\normalsize
where we used the following parametrization for Eq.\,\eqref{D:int},
\small\eq{
\cos\Xi_t=\mathrm{Re}\int\!d^3\!\bx\,\psi_1^\dag(\bx,t)\psi_2(\bx,t)\,,
~~0\le\Xi_t\le \pi\,,
}\normalsize
which is consistent with $|\cos\Xi_t|\le 1$ for normalized $\psi_1$ and $\psi_2$.
Obviously $\psi_i(\bx,t)=e^{-iH_i^Dt}\psi_i(\bx)$.
Since the expression in Eq.~\eqref{Fdef:prob:0} is a sum of non-negative
terms, to have $\mathcal{P}(\ms{\bs{\nu}\!\rightarrow\!\bs{\numu}};0)=0$ at $t=0$,
we should have
\small\eq{
\label{Fdef:cond}
\theta-\alpha=0\,,~~\text{ and }~~ \Xi_0=0\,.
}\normalsize
The other possibilities, $\theta+\alpha=0$ or $\theta+\alpha=\pi$,
are outside the parameter range of $\alpha$ since $0<\theta<\pi/2$ for non-trivial
mixing.
Moreover, $\alpha=\theta$ corresponds to the only local and global minimum of the
second term of Eq.~\eqref{Fdef:prob:0} as a function of $\alpha$.
The function $\sin^2(\alpha+\theta)$ in the first term of Eq.~\eqref{Fdef:prob:0}
is always non-null and has the global minimum at the borders $\alpha=0$ or
$\alpha=\pi/2$. The second equality in Eq.~\eqref{Fdef:cond} is equivalent to
\small\eq{
\label{Fdef:cond:2}
\psi_1(\bx,0)=\psi_2(\bx,0)=\psi(\bx)\,.
}\normalsize

Thus the necessary and sufficient initial conditions for initial pure $\nue$ flavor
or, equivalently (for two flavors), no $\numu$ flavor, restricts
Eq.~\eqref{Fdef:Psi:0} to
\small\eq{
\Psi_m(\bx)=
\begin{pmatrix}
U^*_{e1}\psi_{1}(\bx) \cr U^*_{e2}\psi_{2}(\bx)
\end{pmatrix}
=
U^*_{ei}\psi_{i}(\bx)\hat{\nu}_i
=
\psi(\bx)\hat{\nu}_e
\,,
}\normalsize
where in the last equality, Eq.~\eqref{Fdef:cond:2} was explicitly assumed.
Although condition \eqref{Fdef:cond:2} can be also inferred
from Eq.~\eqref{prob:em:D}, restricted to Eq.~\eqref{III:norm}, we showed
an alternative derivation without assuming a fixed proportion between
mass-eigenstates, which is a more general initial condition than the one used in
section\,\ref{subsec:flavorvio:I}.

Let us now try to eliminate rapid oscillations (\textbf{cond.\,B}) by
adjusting the initial conditions. This was not attempted so far.
We begin by considering only positive energy states for
\small\eq{
\label{Fdef:cond:+}
\psi_i(\bx)=\psi_i^{(+)}(\bx)
~,~~i=1,2\,,
}\normalsize
in Eq.~\eqref{Fdef:Psi:0}. In momentum space, Eq.~\eqref{Fdef:cond:+} can be
written
\small\eq{
\label{Fdef:cond:p+}
\tilde{\psi}_i^{(+)}(\bp)=\sum_s u^s(\bp,m_i)g_i(\bp,s)\,,
}\normalsize
where the spinors $u^s$ were defined in Eq.\,\eqref{III02}.
By construction, $\Lambda_{i+}\psi_i^{(+)}(\bx)=\psi_i^{(+)}(\bx)$ and
$\Lambda_{i-}\psi_i^{(+)}(\bx)=0$, which means
$\psi_i(\bx,t)=e^{-iH_i^Dt}\psi_i(\bx)$ only contains the positive energy term
$e^{-iE_i(\bp)t}$ in its Fourier transform.
In this case, the time dependent
function $\Xi_t$ in the $\numu$ detection probability\,\eqref{Fdef:prob:0} is
given by
\eqarr{
\label{Fdef:cosXi:+}
\cos\Xi_t&=&
\mathrm{Re}\!\int\!d^3\!\bp\,
\tilde{\psi}_1^\dag(\bp)\tilde{\psi}_2(\bp)e^{i\Delta E(\bp)t}
\\ &=&
\label{Fdef:cosXi:g+}
\mathrm{Re}\sum_s\!\int\!d^3\!\bp\,
g_1^*(\bp,s)g_2(\bp,s)e^{i\Delta E(\bp)t}[1-2f_2(\bp)]
\,,
}\normalsize
where
\small\eq{
f_2(\bp)\equiv
\frac{1}{2}
-\frac{1}{4}\sqrt{1+x_1}\sqrt{1+x_2}
-\frac{1}{4}\sqrt{1-x_1}\sqrt{1-x_2}\,,
}\normalsize
defined with the shorthand $x_i\equiv\frac{m_i}{E_i(\bp)}$.
We used the calculation
\small\eq{
u^{r\dag}(\bp,m_{\1})u^s(\bp,m_{\2})
=
[1-2f_2(\bp)]\,\delta_{rs}\,.
}\normalsize
We achieved our goal: there is no rapid oscillations. The probability
$\mathcal{P}(\bs{\nu}\!\rightarrow\!\bs{\numu};t)$ is only a function of $\Delta
E(\bp)t$ smeared out in momentum.

Let us now show that restriction \eqref{Fdef:cond} for \textbf{cond.\,A} can not be
simultaneously applied with restriction \eqref{Fdef:cond:+} for \textbf{cond.\,B}.

Let us suppose \textbf{cond.\,A} [Eq.~\eqref{Fdef:cond:2}] is valid, i.\,e.,
$\psi_1(\bx)=\psi_2(\bx)=\psi(\bx)$. We can decompose the spinorial function
$\psi(\bx)$ in terms of bases depending on different masses $m_1$ and $m_2$. Equating
\small\eq{
\label{psi:dec}
\psi(\bx)=
\int\! \frac{d^3\!\bp}{\sqrt{2E_i}}\,[u_i^s(\bx;\bp)g_i^{\mt{(+)}}(\bp,s)
+v_i^s(\bx;\bp)g_i^{\mt{(-)}}(\bp,s)]
~,~~i=1,2\,,
}\normalsize
where $\ml{\frac{u_i^s(\bx;\bp)}{\sqrt{2E_i}}}\equiv
u^s\bb{\bp,m_i}\,\ms{e^{i\bp\,\ponto\bx}/(2\pi)^{\mt{3/2}}}$
and
$\ml{\frac{v_i^s(\bx;\bp)}{\sqrt{2E_i}}}\equiv
v^s\bb{\bp,m_i}\,\ms{e^{-i\bp\,\ponto\bx}/(2\pi)^{\mt{3/2}}}$,
the expansion coefficients can be obtained
\eqarr{
g_i^{\mt{(+)}}(\bp,s)&=&
\int\!d^3\!\bx\,\ml{\frac{u_i^{s\dag}\!(\bx;\bp)}{\sqrt{2E_i}}}\psi(\bx) ~,\\
\label{psi:coef}
g_i^{\mt{(-)}}(\bp,s)&=&
\int\!d^3\!\bx\,\ml{\frac{v_i^{s\dag}\!(\bx;\bp)}{\sqrt{2E_i}}}\psi(\bx)
~.
}\normalsize
From Eq.~\eqref{psi:coef}, and $\tilde{\psi}(\bp)$ being the inverse Fourier
transform of $\psi(\bx)$, we see that imposing the conditions
\eqarr{
g_1^{\mt{(-)}}(\bp,s)&=&
v^{s\dag}\bb{\bp,m_{\1}}\tilde{\psi}(-\bp)=0
~,
\\
g_2^{\mt{(-)}}(\bp,s)&=&
v^{s\dag}\bb{\bp,m_{\2}}\tilde{\psi}(-\bp)=0
~,
}\normalsize
for all $\bp$, leads to the conditions
\eqarr{
\ms{[(m_1+E_2)-(m_2+E_2)]}\,v_0^{s\dag}\tilde{\psi}(-\bp)&=&0
~,~s=1,2\,,
\label{cond1}
\\
\label{cond2}
\ml{\Big[\frac{1}{m_1+E_2}-\frac{1}{m_2+E_2}\Big]}\,
v_0^{s\dag}\bs{\gamma}.\bp\,\tilde{\psi}(-\bp) & = & 0 ~,~s=1,2\,,
}
where the property of Eq.~\eqref{v:p} and $\gamma_0v_0^s=-v_0^s$ were
used.
In case $m_1\neq m_2$, we can use the decomposition
$\tilde{\psi}(\bp)=\tilde{\psi}_+(\bp)+\tilde{\psi}_-(\bp)$, where
$\tilde{\psi}_\pm(\bp)=(\id\pm\gamma_0)\tilde{\psi}(\bp)$/2,
and obtain from Eqs.~\eqref{cond1} and \eqref{cond2} the conditions
\eqarr{
v_0^{s\dag}\tilde{\psi}_-(-\bp)&=&0
~,~s=1,2\,,
\label{cond1:2}
\\
\label{cond2:2}
u_0^{s\dag}\bs{\Sigma}\ponto\bp\,\tilde{\psi}_+(-\bp) & = & 0 ~,~s=1,2\,,
}\normalsize
where the properties $\bs{\gamma}=\gamma_0\gamma_5\bs{\Sigma}$ and
$u_0^s=\gamma_5v_0^s$ were used in Eq.~\eqref{cond2:2}.
Since $\bs{\Sigma}\ponto\bp$ has only non-null eigenvalues and it commutes with
$\gamma_0$, the equations \eqref{cond1:2} and \eqref{cond2:2} are only satisfied if
$\tilde{\psi}_+(\bp)=\tilde{\psi}_-(\bp)=0$, i.\,e., $\psi(\bx)=0$. It is easier to
reach this conclusion in the helicity basis $\{u_0^{\mt{(\pm)}},v_0^{\mt{(\pm)}}\}$
characterized by $\bs{\Sigma}\ponto\bp\,u_0^{\mt{(\pm)}}=\pm |\bp| u_0^{\mt{(\pm)}}$,
but the result is basis independent.

We can then conclude that rapid oscillations are an inevitable consequence of a
relativistic Dirac fermion description of flavor oscillations if initial flavor
definition is exactly required. Quantitatively, however, it was seen in
Eqs.\,\eqref{III30} and \eqref{KD:em:0} that the contribution of
rapid oscillations to the probability
$\mathcal{P}(\ms{\bs{\nue}\!\rightarrow\!\bs{\numu}};t)$ is
negligible for momentum distributions around UR values because rapid
oscillation is quantified by the function $f$ in Eq.~\,\eqref{ccn:f}, which behaves
as $f\sim (\Delta m)^2/(4\bar{E}^2)$ for UR momenta.
On the other hand, it was already shown that avoiding rapid oscillations is
possible to the detriment of having initial flavor violation.

Let us quantify the amount of initial flavor violation for neutrinos states of the form
\eqref{Fdef:cond:+} satisfying \textbf{cond. B}. Let us relax the conditions of
Eq.~\eqref{Fdef:cond}, and consider only the first one, $\alpha=\theta$, which is a
minimum in $\alpha$.
Thus the initial probability to detect the wrong flavor $\numu$ can be rewritten by
using Eq.~\eqref{Fdef:cosXi:g+} as
\eqarr{
\label{Fdef:fvio:+}
\frac{\mathcal{P}(\ms{\bs{\nue}\!\rightarrow\!\bs{\numu}};0)}{\sin^22\theta}&=&
\frac{1}{4}\sum_{s}\int\!d^3\!\bp\,\Big\{
|g_1(\bp,s)-g_2(\bp,s)|^2
+4f_2(\bp)\mathrm{Re}\big[g_1^*(\bp,s)g_2(\bp,s)\big]
\Big\},~~~~
}\normalsize
where we used the expansion \eqref{Fdef:cond:p+}.

We immediately see that taking equal momentum distributions
for the two mass-eigenstates (for two spin states), i.\,e.,
\small\eq{
\label{g1=g2}
g_1(\bp,s)=g_2(\bp,s)=g(\bp,s)\,,
}\normalsize
makes the first term of Eq.~\eqref{Fdef:fvio:+} vanish. Initial flavor
violation is then proportional to
\eqarr{
\label{Fdef:fvio:+2}
\frac{\mathcal{P}(\ms{\bs{\nue}\!\rightarrow\!\bs{\numu}};0)}{\sin^22\theta}&=&
\int\!d^3\!\bp\,|g(\bp)|^2f_2(\bp)\,,
}\normalsize
where $|g(\bp)|^2=\sum_s|g(\bp,s)|^2$ is the spin-independent momentum distribution.
Notice the normalization
\small\eq{
\int\!d^3\!\bp\,|g(\bp)|^2=1\,.
}\normalsize
As it should be, $f_2(\bp)=0$ if $m_1=m_2$ and it behaves as $f_2\sim (\Delta
m)^2/(4\bar{E})^2$ for UR momenta. Moreover, $f_2(\bp)$ has the
unique maximum value at $|\bp|=\sqrt{m_1m_2}$, the same point as $f(\bp)$.
In fact, the function $f_2$ and
the function $f(\bp)$ in Eq.~\eqref{ccn:f} are closely related by the
relation
\small\eq{
[1-2f_2(\bp)]^2=1-f(\bp)\,.
}\normalsize
Therefore initial flavor violation in Eq.~\eqref{Fdef:fvio:+2} is also negligible for momentum distributions around UR momenta, provided that Eq.~\eqref{g1=g2} is valid.

We can conclude that the description of flavor oscillations for a free Dirac fermion subjected to mixing, within first quantization, is satisfactory from the quantitative point of view but in the strict sense one should consider either one of the following phenomena as inevitably present: (i) initial flavor violation or (ii) rapid oscillations.
Both effects might be as negligible as Eqs.\,\eqref{Fdef:fvio:+2} or \eqref{KD:em:0}
(or Eq.~\eqref{III30}), quantified by the functions $f_2(\bp)$ or $f(\bp)$,
respectively. We will see (i) is supported by second quantization.
In that case, it is possible that such effect could be still small but larger than
Eq.~\eqref{Fdef:fvio:+2}. All depends on how different are $\psi_1(\bx)$ and
$\psi_2(\bx)$, associated to the two mass-eigenstates, when a neutrino is created as
a superposition of mass-eigenstates.
Furthermore, the presence of (i) suggests that neutrino flavor is not an exact concept but only an approximately well defined one.

\section{Consequences of spin structure and relativistic completeness}
\label{sec:consequences}

This section deals with some intrinsic consequences of spin structure and relativistic completeness, namely  chiral oscillations and the consequences for a non-minimal coupling with external magnetic fields.
Chiral oscillations naturally enter the discussion of neutrino oscillations since neutrinos are produced and detected through weak interactions that are chiral in nature.
We succeed in showing that the inclusion of chiral oscillation effects, together with the
time-evolution of spinorial wave packets for the mass-eigenstates, can modify the flavor conversion probability formula.
In particular, it provides an interpretation of chiral oscillations as very
rapid oscillations in position along the direction of motion, i.e., longitudinal to
the momentum of the particle.
All the ingredients for decoupling chiral oscillations from the spin-flipping in the presence of an external magnetic field are discriminated.
It allows for depicting all the relevant effects due to the inclusion of spin structure and relativistic completeness that we have introduced.

According to the SM at tree level, neutrinos interact with itself and other leptons
only through weak charged current and neutral current interactions which are
chiral in nature and only contain the contribution of neutrinos with negative
chirality\,\cite{DeL98}.
Consequently, positive chirality neutrinos become sterile with respect
to V-A weak charged currents, independently of any external electromagnetic field.
Despite the experimental circumstances not being favorable to such an
interpretation, the quantum transitions that produces a final flavor state
corresponding to an active-sterile quantum mixing is perfectly acceptable from the
theoretical point of view.
Even focusing specifically on the chiral oscillation mechanism, we have also observed a generic interest in identifying the physical meaning of variables which coexist with the interference between positive and negative frequency solutions of Dirac equation \cite{Bra99,Rup00,Wan01,Bol04} in a Dirac wave packet treatment.
If we follow an almost identical line of reasoning as that applied for rapid oscillations of the position, it is also possible to verify that the average value of the Dirac chiral operator $\gamma^{\5}$ also presents an oscillating behavior.

We intend to clarify some conceptual relations between helicity and chirality by constructing the dynamics of the processes of chiral oscillation and spin-flipping.
The {\em differences} between the dynamics of chirality and the
helicity for a neutrino non-minimally coupled to an external
(electro)magnetic field are expressed in terms of the equation of the motion of the
correspondent operators $\gamma^{\5}$ and $h$, respectively.
In order to determine the relevant physical observables and confront the dynamics of chiral oscillations with the dynamics of spin-flipping in the presence of an external magnetic field, we report about a further class of static characteristics of neutrinos, namely, the (electro)magnetic moment which appears in a Lagrangian with non-minimal coupling.
We shall demonstrate that the oscillating effects can be explained as an implication of the ZBW phenomenon that emerges when the Dirac equation solution is used for describing the time evolution of a wave packet \cite{Ber04}.
Due to this tenuous relation between ZBW and chiral oscillations, one question to be
answered in this manuscript concerns with the {\em immediate} description of chiral
oscillations in terms of the ZBW motion, which shows that, in fact, chiral
oscillations are coupled with the ZBW motion so that they cannot exist independently
of each other.
One shall conclude that chiral oscillations can be interpreted as very rapid
oscillations in position along the direction of motion.

In order to make clear the confront of ideas involving chirality and helicity, we
first review such concepts and their common points. At the same time we point
out their differences by assuming a Lagrangian with minimal or non-minimal coupling.
The latter interaction, more precisely, the non-minimal coupling of a magnetic moment
with an external electromagnetic field \cite{Vol81}, which induces neutrino
spin-flipping\,\cite{Oli90}, was formerly supposed to be relevant in the context of
the solar-neutrino puzzle.
As a consequence of such interaction, left-handed neutrinos could change their
helicity (to right-handed)\,\cite{Bar96}.
The effects on flavor oscillations due to external magnetic interactions in a kind of
chirality-preserving phenomenon were also studied \cite{Oli96} but they lacked a
full detailed theoretical analysis.
We observe, however, that only for UR particles such as neutrinos, changing helicity
approximately means changing chirality.
In the context of oscillation phenomena and in the framework of a first quantized theory, the small differences between the concepts of chirality and helicity, which had been interpreted as representing the same physical quantities for massless particles \cite{Oli90,Vol81,Bar96,Oli96,Kim93}, can be quantified for massive particles.

\subsection{Chirality and helicity}

By considering the gamma matrices (in the Dirac representation),
\small\eq{
\gamma^{\0} = \left(\begin{array}{rr} \mathbf{1} & 0 \\ 0  &
-\mathbf{1}\end{array}\right),
~~~~~~~~ \gamma^i = \left(\begin{array}{cc} 0 & \sigma^i \\ - \sigma^i  &  0
\end{array}\right)
~~~~~~~~ \mbox{and} ~~~~~~~~ \gamma^{\5} = \left(\begin{array}{rr}  0  & \mathbf{1}
\\ \mathbf{1} & 0 \end{array}\right),
\label{IV501}
}\normalsize
where $\sigma^{i}$ are the Pauli matrices, we can observe the following $\gamma^{\5}$
matrix properties,
\small\eqarr{
\left(\gamma^{\5}\right)^{\dagger} = \gamma^{\5},~~~~~~
\left(\gamma^{\5}\right)^{\2} = 1~~~~~~\mbox{and}~~~~~~
\left\{\gamma^{\5}, \gamma^{\mu}\right\}=0,
\label{IV502}
}\normalsize
which lead to the commuting relation $\left[\gamma^{\5}, S^{\mu\nu}\right] = 0$,
where $S^{\mu\nu}$ are the generators of the {\em continuous Lorentz
transformations}.
Among the Dirac field bilinears constructed with the gamma matrices, it is
convenient to define the {\em vector} and
the {\em axial vector} (or {\em pseudo-vector}) currents respectively by
\small\eq{
j^{\mu}\bb{x} = \overline{\psi}\bb{x} \gamma^{\mu} \psi\bb{x}
~~~~~~~~\mbox{and}~~~~~~~~ j^{\mu \5}\bb{x} = \overline{\psi}\bb{x}
\gamma^{\mu}\gamma^{\5} \psi\bb{x}.
\label{IV503}
}\normalsize

By assuming that $\psi\bb{x}$ satisfies the free Dirac equation, we verify that the
vector current density $j^{\mu}\bb{x}$ is conserved, i.e.,
\small\eqarr{
\partial_{\mu} j^{\mu}\bb{x} &=&
(i\,m\,\overline{\psi}\bb{x})\psi + \overline{\psi}(-i\, m\, \psi\bb{x})
= 0\,.
\label{IV504}
}\normalsize
When we couple the Dirac field to the electromagnetic field, $j^{\mu}\bb{x}$ will
become the electric current density.
In a similar way, we can compute
\small\eqarr{
\partial_{\mu} j^{\mu \5}\bb{x} &=& 2 \, i \, m \,\overline{\psi}\bb{x} \gamma^{\5}
\psi\bb{x}
\label{IV505}.
}\normalsize
If $m = 0$, the {\em axial vector current} $j^{\mu \5}\bb{x}$ is also conserved.

The two currents $j^{\mu}\bb{x}$ and $j^{\mu \5}\bb{x}$ are the Noether currents
corresponding respectively to the two gauge transformations
\small\eq{
\psi\bb{x} \rightarrow \exp{[ i \,\alpha]} \psi\bb{x}
\label{IV508}
}\normalsize
and
\small\eq{
\psi\bb{x} \rightarrow \exp{[ i \,\alpha\, \gamma^{\5}]} \psi\bb{x}
\label{IV509}.
}\normalsize
The former is a symmetry of the Dirac Lagrangian.
The latter, called {\em chiral transformation}, is a symmetry of the kinetic part
of the Dirac Lagrangian \cite{Pes95}.
The Noether theorem confirms that an {\em axial vector current} related to a {\em
chiral transformation} is conserved, at the classical level, only if $m = 0$.

In general, it is useful to define the chiral {\em left}-$L$ (negative chiral) and
{\em right}-$R$ (positive chiral) currents through the combinations of the operator
$\gamma^{\5}$ by
\small\eq{
j^{\mu}_L \bb{x} = \overline{\psi}\bb{x} \gamma^{\mu} \frac{1 - \gamma^{\5}}{2}
\psi\bb{x} ~~~~~~~~\mbox{and}~~~~~~~~ j^{\mu}_R\bb{x} = \overline{\psi}\bb{x}
\gamma^{\mu} \frac{1 + \gamma^{\5}}{2} \psi\bb{x}.
\label{IV506}
}\normalsize
Just when $m = 0$, these are respectively the current densities of {\em
left} and {\em right-handed} particles and they are separately conserved.

In parallel, we can define the operator $h = \frac{1}{2}\mbox{\boldmath$\Sigma \cdot
\hat{p}$}$, so that, by conveniently choosing the two component spinors $\eta_{\s}$,
the spinors $u_{\s}\bb{p}$ and $v_{\s}\bb{p}$ become eigenstates of $h$ with
eigenvalues $h = + \frac{1}{2}$ for $s=1$ and $h = - \frac{1}{2}$ for $s=2$ ($\hbar =
c =1$).
Although the free-particle plane wave solutions can always be chosen to be
eigenfunctions of $h$, it is not possible to find solutions which are eigenfunctions
of {\boldmath$\Sigma \cdot \hat{n}$} with an arbitrary unitary vector
{\boldmath$\hat{n}$}.
It occurs because the operator {\boldmath$\Sigma \cdot \hat{n}$} does not commute
with the free particle Hamiltonian unless $\mbox{\boldmath$\hat{n}$} = \pm
\mbox{\boldmath$\hat{p}$}$ or $\bs{p} = 0$.
The operator $h$ that can be diagonalized simultaneously with the free particle
Hamiltonian is called the {\em helicity} operator \cite{Sak87}.
It is exactly the particle spin projection onto the momentum direction.
The particle with $h = + \frac{1}{2}$ is called {\em right-handed} and that one with
$h = - \frac{1}{2}$ is called {\em left-handed}.
The helicity of a massive particle depends on the reference frame, since one can
always {\em boost} to a frame in which its momentum is in the opposite direction (but
its spin is unchanged).
For a massless particle, which travels at speed of light, it is not possible to
perform such a {\em boost}.
In a certain way, helicity and chirality quantum numbers carry a kind of
complementary information. In vacuum, the chirality is invariant under a {\em
continuous Lorentz transformation} but it is not constant in time if the particle is
massive. The helicity is constant in time but it is not Lorentz invariant. A chiral
eigenstate can always be written as a linear combination of two helicity eigenstates.

To focus our attention on the central point of this section, where we intend to
compare the dynamic processes of chiral oscillation and spin-flipping, let us analyze
the possibilities of observing them via interaction terms in the Lagrangian language.
Initially, we can construct a gauge invariant Lagrangian in a theory with initially
massless fermions.
We can couple a Dirac fermion to a gauge field by assigning the chiral-$L$ fields
$\psi_{i_{L}}\bb{x}$ to one nontrivial representation of a gauge group $G$ and
assigning the chiral-$R$ fields $\psi_{i_{R}}\bb{x}$ to a singlet representation.
For such a model, we can write
\small\eq{
\mathcal{L} = \overline{\psi}\bb{x}
\gamma^\mu \left[\partial_\mu - igA^a_\mu\bb{x}\,T^a
\left(\frac{1 - \gamma^{\5}}{2}\right)\right] \psi\bb{x},
\label{IV511}
}\normalsize
where it is evident that only negative chiral fields couple with gauge boson fields
$A^a_\mu\bb{x}$.
It is straightforward to verify that the Lagrangian (\ref{IV511}) is invariant by the
following infinitesimal local gauge transformation \cite{Pes95}
\small\eqarr{
\psi\bb{x}  & \rightarrow & \left[1 +  i \alpha^a_\mu\bb{x}\, T^a \left(\frac{1 -
\gamma^{\5}}{2}\right)\right] \psi\bb{x}, \nonumber\\
A^a_\mu\bb{x}\, & \rightarrow & A^a_\mu\bb{x} + g^{-1} \partial_\mu \alpha^a\bb{x} +
f^{abc}A^b_\mu\bb{x}\, \alpha^c\bb{x},
\label{IV512}
}\normalsize
where the non-abelian character of $G$ is summarized by the commutation relations
\small\eq{
\left[T^a, T^b\right]= i f^{abc} T^c.
\label{IV513}
}\normalsize
Since the chiral-$R$ fields are free fields, we can even eliminate these fields and
write a gauge invariant Lagrangian for purely chiral-$L$ fermions.
The idea of gauge fields that couple only to chiral-$L$ fermions plays a central role
in the construction of  the theory of weak interactions.

To work out the general properties of chirally coupled fermions, it is useful to
analyze one further transformation.
Let us take the charge conjugate operator $C = i \gamma^{\2} \gamma^{\0}$ in order to
write a charge conjugate field $\psi^c$ as
\small\eq{
\psi^c = C \bar{\psi}^{T} = i \gamma^{\2} \psi^{*}
\label{IV514}
}\normalsize
so that the conjugate chiral-$R$ component of a field $\psi$, which would be given by
$\psi^c_L$, transforms as a chiral-$L$ quantity under a continuous Lorentz
transformation, i.\,e.,
\small\eq{
\psi^c_L = (\psi^c)_L = i \left(\frac{1 - \gamma^{\5}}{2}\right) \gamma^{\2} \psi^{*}
= i \gamma^{\2} \left(\frac{1 + \gamma^{\5}}{2}\right)  \psi^{*} = i \gamma^{\2}
\psi_R^{*}\,,
\label{IV515}
}\normalsize
where we have omitted the $x$ dependence for simplicity.
In the same way, the conjugate chiral-$L$ component of a field $\psi$, which
would be given by $\psi^c_R$, transforms as a chiral-$R$ quantity, i.\,e.,
\small\eq{
\psi^c_R  = i \gamma^{\2} \psi_L^{*}\,.
\label{IV516}
}\normalsize
This property is important for observing what occurs in a gauge model with {\em L-R}
symmetry \cite{Moh75,Ple93}.
In this class of models, chiral-$R$ neutrinos interact with additional vector gauge
bosons.

As a general feature, due to the vectorial character of the interactions which are
mediated by vector gauge bosons, the possibility of a left to right (or vice-versa)
chiral conversion via gauge coupling is impossible.
As we can observe in the following, the probability amplitude for a vector coupling
between fields with opposite chiral quantum numbers is null,
\small\eqarr{
4(\overline{\psi}_L \,A_{\mu}\gamma^{\mu}\,\psi_R)^\dag \equiv
4\overline{\psi_R}\,A_{\mu}\gamma^{\mu}\,\psi_L  &=&
\left((1 +\gamma^{\5})\psi\right)^{\dagger}\,\gamma^{\0}\,A_{\mu}\gamma^{\mu} (1
-\gamma^{\5})\psi\nonumber\\
&=& \psi^{\dagger} \,\gamma^{\0}\,A_{\mu}\gamma^{\mu} \,(1 + \gamma^{\5})(1
-\gamma^{\5})\psi = 0,
\label{IV516B}
}\normalsize
which forbids the possibility of chiral changes (oscillations) in this way.
Alternatively, we cannot abandon the idea of spin-inversion which does {\em not}
correspond to chiral conversion but is a very well established mechanism which
emerges via an external vector field interaction.

At last, we can analyze what happens to the vector currents in the presence
of flavor mixing.
Let us define the flavor currents
\eq{
j_{\nu_\alpha}^\mu=\bar{\psi}_{\nu_\alpha}\gamma^\mu\psi_{\nu_\alpha}\,,
}
where the mixed fields are
\eq{
\psi_{\nu_\alpha}\equiv U_{\alpha i}\psi_i, \quad \alpha=e,\mu.
}
Then, the flavor currents are almost conserved because
\eq{
\partial_\mu j_{\nu_\alpha}^\mu=
i\sum_{i\neq j}U^*_{\alpha j}U_{\alpha i}(m_j-m_i)\bar{\psi}_{j}\psi_i\,.
}
For two flavors, we have
\eq{
\label{jf:cons}
\partial_\mu j_{\nue,\numu}^\mu=
\pm (m_2-m_1)
\sin 2\theta\,\im(\bar{\psi}_1\psi_2)
\,,
}
where $U$ is given by Eq.\,\eqref{II:mixing}. Therefore, approximate flavor
conservation is valid locally. The non-conservation is proportional to the mass
difference and to the amount of mixing.

\subsection{Chiral oscillations coupled with the zitterbewegung motion}
\label{subsec:chiral}

If a massless particle represented by a Dirac spinor is produced in a chiral
($\gamma^{\5}$) eigenstate, its chirality is a constant of the motion in addition to
being Lorentz invariant.
On the other hand, independently of the existence of external interacting fields,
these properties no longer hold for Dirac particles with mass.
In vacuum, chiral oscillations can be formally related to the {\em
zitterbewegung} (ZBW) motion.

We begin by calculating the average value of the spatial part of the vector current
in Eq.~(\ref{IV503}). Such average value appears quite reasonable \cite{Sak87} when
we calculate it using the positive frequency solutions ($\psi_{_{\pl}}\bb{x}$),
\small\eqarr{
\langle\bs{\alpha}\bb{t}\rangle _{\pl} &=&
\int\hspace{-0.1 cm}
d^{\3}\hspace{-0.1cm}\bx\,\psi_{\pl}^{\dagger}\bb{x}
\,\mbox{\boldmath$\alpha$} \, \psi_{\pl}\bb{x}
=\int\hspace{-0.1 cm}
\frac{d^{\3}\hspace{-0.1cm}\bp}{(2\pi)^{\3}}\frac{\bs{p}}{E}\sum_{\s \ig
\1,\2} |b_{\s}\bb{p}|^{\2}\,,
\label{IV120}
}\normalsize
which represents exactly the average value of the relativistic velocity
$\frac{\bs{p}}{E}$.
However, for general wave functions $\psi(x)$ containing both positive and negative
frequency components, an extremely rapid oscillation term appears in
Eq.\,\eqref{IV120}.
To understand such behavior, note that the velocity operator
$\mbox{\boldmath$\alpha$}\bb{t}$, in the Heisenberg picture, does not represent a
constant of the motion since\,\footnote{
The velocity operator $\bs{\alpha}$ has also two more peculiarities:
its components $\alpha_{\kk}$ have eigenvalues $\pm 1$, corresponding to
classically forbidden velocities $\pm c$, and different components are incompatible
observables because they do not commute with each other.
}
\small\eq{
\frac{d~}{dt}\bs{\alpha}\bb{t}\,=\, i\left[
H_{\0} ,\bs{\alpha}\bb{t}\right]
\,=\,2\,i\left(\bs{p}-\bs{\alpha}\bb{t}H_{\0}\right)\,.
\label{IV130}
}\normalsize
Keeping in mind that $\bs{p}$ and $\mathit{H}_{\0}$ are constants of the
motion, we can easily solve Eq.~(\ref{IV130}) as a differential equation for
$\mbox{\boldmath$\alpha$}\bb{t}$ and obtain \cite{Sak87}
\small\eq{
\bs{\alpha}\bb{t} =
\bs{p}H_{\0}^{\mi\1}+\left(\bs{\alpha}\bb{0} -
\bs{p}H_{\0}^{\mi\1} \right)
\exp{[-2\,i\, \mathit{H}_{\0} \, t]}
\,.
\label{IV14}
}\normalsize
The time dependence of $\ms{\bs{\alpha}\bb{t}}$ in Eq.\,\eqref{IV14}
translates into the trembling motion (ZBW motion) in the position operator $\bx(t)$,
since $\frac{d\bx(t)}{dt}=\bs{\alpha}(t)$. Such motion is characterized by high
frequencies $|\omega|\ge 2m$, $\omega$ being the eigenvalues of $2H_0$.

We have already seen in the previous section that the particle {\em helicity}
$h = \frac{1}{2}\bs{\Sigma}\cdot\hat{\bs{p}}$ is a constant of
the motion while the chiral operator $\gamma^{\5}$ is not.
At time $t = 0$, the coefficients $b_{\s}\bb{p}$ and $d^*_{\s}\bb{p}$ used in the
construction of the Dirac wave packet $\psi\bb{x}$ can be chosen to provide a
negative (positive) chirality eigenstate \cite{Ber05,Ber05A} or, in the same way, to
provide a helicity eigenstate [when $h u_{\s}\bb{p}(v_{\s}\bb{p}) \equiv \pm
\frac{1}{2}u_{\s}\bb{p}(v_{\s}\bb{p})$].
We can consider the initial chiral eigenstate $\psi\bb{0,\bx}$ is not only a
superposition of momentum eigenstates weighted by a momentum distribution centered
around $\bpp_{\0}$ (like in Eq.~(\ref{III29})), but also a helicity eigenstate
obtained through the production of a spin-polarized particle.
Such condition formally occurs when we assume that the constant spinor $w$ in the
wave packet expression (\ref{III29}) is a simultaneous eigenspinor of $\gamma^{\5}$
and $h$\,\footnote{When we establish that $\psi\bb{0,\bx}$ is a $h$ and/or a $\gamma^{\5}$ eigenstate,
we are referring to the choice of the fixed spinor $w$. The {\em breaking} of the
Lorentz symmetry is not specifically related to the choice of $w$, but more
generically to the choice of the momentum distributions ${b_{\s}\bb{p}}$ and
$d^*_{\s}\bb{\tilde{p}}$ written in terms of $w$ and
$\varphi\bb{\bs{p}-\bs{p}_{\0}}$.
Once one has established an analytical shape for the momentum distribution (as in
\cite{DeL98} and \cite{Zub80}), it is valid only for one specific
reference frame.}.
We then obtain
\small\eqarr{
\frac{d~}{dt}\langle \gamma_{\5}\bb{t}\rangle &=&
\frac{d~}{dt}\langle\mbox{\boldmath$\alpha$}\bb{t} \cdot
\mbox{\boldmath$\hat{p}$}\left(\bs{\Sigma} \cdot
\mbox{\boldmath$\hat{p}$}\right)\rangle
 \nonumber\\
&=&\left(\frac{d~}{dt}\langle\mbox{\boldmath$\alpha$}\bb{t} \cdot
\mbox{\boldmath$\hat{p}$}\rangle\right)\langle\bs{\Sigma} \cdot
\mbox{\boldmath$\hat{p}$}\rangle
+ \langle \mbox{\boldmath$\alpha$}\bb{t} \cdot
\mbox{\boldmath$\hat{p}$}\rangle\,\left(\frac{d~}{dt}\langle
\bs{\Sigma} \cdot \mbox{\boldmath$\hat{p}$}\rangle\right)
\,.~~~~
\label{IV18}
}\normalsize
By observing that the helicity $h$ and the momentum
$\mbox{\boldmath$\hat{p}$}$ are constants of the motion, we can establish a
subtle relation between the chirality operator $\gamma_{\5}\bb{t}$ and the velocity
operator $\mbox{\boldmath$\alpha$}\bb{t}$ expressed in terms of
\small\eq{
\frac{d~}{dt}\langle \gamma_{\5}\bb{t}\rangle
\,=\,
(2 h)\,
\left(\frac{d~}{dt}\langle\mbox{\boldmath$\alpha$}\bb{t}\cdot
\mbox{\boldmath$\hat{p}$}\rangle\right)
=
(2h)\,
\big\langle\frac{d~}{dt}\mbox{\boldmath$\alpha$}\bb{t}
\cdot\mbox{\boldmath$\hat{p}$}
\big\rangle
\,.
\label{IV19}
}\normalsize

The time evolution of $\gamma_{\5}\bb{t}$ presents an oscillating character which can
be interpreted as a direct and natural consequence of the oscillating character of
$\mbox{\boldmath$\alpha$}\bb{t}$ in Eq.\,\eqref{IV14}.
Considering Eqs.~(\ref{IV130}) and (\ref{IV19}) leads to the following explicit
dependence
\small\eq{
\frac{d~}{dt}\langle\gamma_{\5}\bb{t}\rangle =
\aver{
4i\,h\left(\bs{p} -
\mbox{\boldmath$\alpha$}\bb{t}\mathit{H}_{\0}\right){\cdot\,}
\mbox{\boldmath$\hat{p}$}
}
\,.
\label{IV20}
}\normalsize
Since $h$, $\bs{p}$ and $\mathit{H}_{\0}$ are constants of
the motion, we conclude that there will not be chiral oscillations without the
``quivering motion'' of the position.

The relation between the chirality operator $\gamma_{\5}\bb{t}$ and the velocity
operator $\mbox{\boldmath$\alpha$}\bb{t}$ becomes more interesting when we take into
account the complete expression for the vector current density $\overline{\psi}\bb{x}
\gamma_{\mu}\, \psi\bb{x}$ and apply the {\em Gordon decomposition}
\cite{Pes95},
\small\eqarr{
\overline{\psi}\bb{x} \gamma_{\mu} \psi\bb{x} &=&
-\frac{i}{2m}\left[\left(\partial^{\mu}\overline{\psi}\bb{x}\right)
\psi\bb{x} - \overline{\psi}\bb{x} \left(\partial^{\mu}\psi\bb{x}\right)\right]
 +\frac{1}{2m}\partial_{\nu}\left(\overline{\psi}\bb{x} \sigma^	{\mu\nu}
\psi\bb{x}\right),
\label{IV22}
}\normalsize
where  $\sigma_{\mu\nu} = \frac{i}{2}[\gamma_{\mu},\gamma_{\nu}]$.
The spatial integration of the vector components of Eq.~(\ref{IV22}) gives us
\small\eqarr{
\int\hspace{-0.1 cm} d^{\3}\hspace{-0.1cm}\bx\,\psi^{\dagger}
\mbox{\boldmath$\alpha$}  \psi &=&
\frac{1}{2m}\int\hspace{-0.1 cm} d^{\3}\hspace{-0.1cm}\bx \left\{-i
\left[\overline{\psi} \left(\mbox{\boldmath$\nabla$} \psi\right) -
\left(\mbox{\boldmath$\nabla$} \overline{\psi}\right)  \psi\right]
+ \,\left[\mbox{\boldmath$\nabla$}{\times\!}\left(\overline{\psi}
\bs{\Sigma} \psi\right) - i \partial_{t}\left(\overline{\psi}
\mbox{\boldmath$\alpha$} \psi\right)\right]\right\}
\label{IV23},
}\normalsize
where we have suppressed the $x$ dependence.
We can write the decomposed components of
$\langle\mbox{\boldmath$\alpha$}\rangle$ as
\small\eqarr{
\int\hspace{-0.1 cm}
d^{\3}\hspace{-0.1cm}\bx\,\mbox{\boldmath$\nabla$}{\times\!}
\left(\overline{\psi} \bs{\Sigma} \psi\right) &=& 0,
\label{IV24}\\
\frac{i}{2m}\int\hspace{-0.1 cm} d^{\3}\hspace{-0.1cm}\bx\,
\left[\overline{\psi} \left(\mbox{\boldmath$\nabla$} \psi\right) -
\left(\mbox{\boldmath$\nabla$} \overline{\psi}\right)  \psi\right] &=&
\int\hspace{-0.1 cm}
\frac{d^{\3}\hspace{-0.1cm}\bp}{(2\pi)^{\3}}\bigg\{\frac{\bs{p}}{E}
\sum_{\s \ig \1,\2} \left[|b_{\s}\bb{p}|^{\2}+|d_{\s}\bb{p}|^{\2}\right]
 \nonumber\\
&&\left.
~~~~ +\sum_{\s \ig
\1,\2}\left(\frac{m}{E}-\frac{E}{m}\right)a_{\s}\mbox{\boldmath$\hat{p}$}
\left[b^*_{\s}\bb{p}\,d^*_{\s}\bb{\tilde{p}}\, \exp{[+2 \,i\,E\,t]}
\right.\right. \nonumber\\
&&\left.
~~~~~~~~~~~~~~~~~~~~~~~~~~~~ - d_{\s}\bb{p}\,b_{\s}\bb{\tilde{p}}\, \exp{[-2
\,i\,E\,t]}\right]\bigg\}
\label{IV25},
}\normalsize
where $\tilde{p} = (E,-{\bs{p}})$ and we have assumed that $a_{\s}
= \eta_{\sp}^\dag\, \bs{\Sigma}{\cdot}\,\mbox{\boldmath$\hat{p}$}\,\eta_{\s} =
(-1)^{\s \pl \1} \delta_{s s^{\prime}}$.
Finally, the last term is written as
\small\eqarr{
-\frac{i}{2m}\int\hspace{-0.1 cm}
d^{\3}\hspace{-0.1cm}\bx\,\partial_{t}\left(\overline{\psi}
\mbox{\boldmath$\alpha$} \psi\right) &=& \int\hspace{-0.1 cm}
\frac{d^{\3}\hspace{-0.1cm}\bp}{(2\pi)^{\3}}\left\{\sum_{\s \ig
\1,\2}\left(\frac{E}{m}\right)a_{\s}\mbox{\boldmath$\hat{p}$}
\left[b^*_{\s}\bb{p}\,d^*_{\s}\bb{\tilde{p}}\, \exp{[+2 \,i\,E\,t]} \,-\,
d_{\s}\bb{p}\,b_{\s}\bb{\tilde{p}}\, \exp{[-2 \,i\,E\,t]}\right]\right.\nonumber\\
&& ~~~~+ \left.\sum_{\s \neq \sp} \mbox{\boldmath$\hat{n}$}_{\s}
\left[b^*_{\s}\bb{p}\,d^*_{\sp}\bb{\tilde{p}}\, \exp{[+2 \,i\,E\,t]} \,-\,
d_{\s}\bb{p}\,b_{\sp}\bb{\tilde{p}}\, \exp{[-2 \,i\,E\,t]}\right]\right\},~~~
\label{IV260}
}\normalsize
where $\mbox{\boldmath$\hat{n}$}_{\1(\2)} = \hat{1}\pm i \hat{2}$ when
$\mbox{\boldmath$\hat{p}$} = \hat{3}$, with $\hat{1}$, $\hat{2}$ and $\hat{3}$
corresponding to arbitrary mutually orthonormal vectors.

Eq.~(\ref{IV24}) allows us to reach the conclusion that the ZBW
does not get a contribution from the intrinsic spin dependent
($\bs{\Sigma}$) magnetic moment component which couples to the external
magnetic field $\mbox{\boldmath$B$}\bb{x}$.
In fact, the ZBW originates from the current strictly related to the
internal electric moment.
By taking into account the (momentum direction components of the)
Eqs.~(\ref{IV25}-\ref{IV260}), we can turn back to Eq.~(\ref{IV20}), which carries
the main idea of this subsection, and observe that chiral oscillations can be
essentially constructed in terms of the longitudinal components of
$\langle\mbox{\boldmath$\alpha$}\rangle$. By calculating the mean value of
$\langle\mbox{\boldmath$\alpha$}\cdot \mbox{\boldmath$\hat{p}$}\mathit{H}_0\rangle$, we
obtain
\small\eqarr{
\langle\mbox{\boldmath$\alpha$}\bb{t}\cdot \mbox{\boldmath$\hat{p}$}\,\mathit{H}_{\0}\rangle &=&
\int\hspace{-0.1 cm}
\frac{d^{\3}\hspace{-0.1cm}\bp}{(2\pi)^{\3}}\left\{\frac{\bs{p}}{E}\cdot\mbox{\boldmath$\hat{p}$}
\sum_{\s \ig \1,\2} \left[(E)|b_{\s}\bb{p}|^{\2}+(E)|d_{\s}\bb{p}|^{\2}\right]
\right. \nonumber\\
&&\left.
+\ \sum_{\s \ig \1,\2} \frac{m}{E}\,a_{\s}
\left[(E)b^*_{\s}\bb{p}\,d^*_{\s}\bb{\tilde{p}}\, \exp{[+ 2 \,i\,E\,t]} \,-\,(E)
d_{\s}\bb{p}\,b_{\s}\bb{\tilde{p}}\, \exp{[-2 \,i\,E\,t]}\right]\right\}\nonumber\\
&=& \aver{|\bs{p}|} -
\int\hspace{-0.1 cm} \frac{d^{\3}\hspace{-0.1cm}\bp}{(2\pi)^{\3}}\, m \sum_{\s \ig
\1,\2}
a_{\s} \left[b^*_{\s}\bb{p}\,d^*_{\s}\bb{\tilde{p}}\, \exp{[+2 \,i\,E\,t]}
\,-\,d_{\s}\bb{p}\,b_{\s}\bb{\tilde{p}}\, \exp{[-2 \,i\,E\,t]}\right]
\label{IV27},~~~~
}\normalsize
which can be substituted in (\ref{IV20}) in order to give
\small\eqarr{
\frac{d~}{dt}\langle\gamma_{\5}\bb{t}\rangle  =
\int\hspace{-0.1 cm} \frac{d^{\3}\hspace{-0.1cm}\bp}{(2\pi)^{\3}}\,
\frac{m}{E} \sum_{\s \ig \1,\2} (2 \,h \,a_{\s})  (2 \, i \, E)
\left[b^*_{\s}\bb{p}\,d^*_{\s}\bb{\tilde{p}}\,
\exp{[+2 \,i\,E\,t]}
-\, d_{\s}\bb{p}\,b_{\s}\bb{\tilde{p}}\, \exp{[-2 \,i\,E\,t]}\right],~~
\label{IV28}
}\normalsize
so that the time evolution of the chirality operator can be written as
\small\eqarr{
\langle\gamma_{\5}\bb{t}\rangle  =
\langle\gamma_{\5}\bb{0}\rangle +
\int\hspace{-0.1 cm} \frac{d^{\3}\hspace{-0.1cm}\bp}{(2\pi)^{\3}}\,
\frac{m}{E}\,\sum_{\s \ig \1,\2}( 2\, h \, a_{\s})\,
\left[d_{\s}\bb{p}\,b_{\s}\bb{\tilde{p}}\, \left(\exp{[2 \,i\,E\,t]} - 1\right) +
h.c.
\right]
\nonumber\\
=
\langle\gamma_{\5}\bb{0}\rangle +
\int\hspace{-0.1 cm} \frac{d^{\3}\hspace{-0.1cm}\bp}{(2\pi)^{\3}}\,
\frac{m}{E}\,\sum_{\s \ig \1,\2}
\left[d_{\s}\bb{p}\,b_{\s}\bb{\tilde{p}}\, \left(\exp{[2 \,i\,E\,t]} - 1\right) +
h.c.
\right]
\label{IV29}
}\normalsize
since $2 h a_{\s}$ is equal to one for well defined spin up or spin down (helicity)
states.

More subtle physical interpretations of the above results are considered in
\cite{Ber06A,Ber06B,DeL98}.
For a gaussian initial wave function with null average chirality, the oscillations of
the {\em left-right} (L-R) chiralities cancel and there is
again no overall oscillation.
It could be the origin of an apparent paradox.
We observe that for any mass-eigenstate represented by plane wave solutions of the
massive Dirac equation, the rest-frame wave function is always an equal mixture of
both chiralities.
This is easily verified when we write
\small\eq{
\psi = \frac{1 - \gamma^{\5}}{2} \psi + \frac{1 + \gamma^{\5}}{2} \psi = \psi_L +
\psi_R,
\label{IV533}
}\normalsize
where $\psi_{L,R}$ correspond respectively to chirality quantum numbers $\mp 1$.
It could be shown that, in the rest frame of a particle, we have
\small\eq{
|\psi_L|^{\2} = |\psi_R|^{\2}.
\label{IV534}
}\normalsize
Note that this result is not Lorentz invariant since a Lorentz {\em boost} is not a
unitary transformation.
Thus, while the cross section is Lorentz invariant the chiral probabilities are not.
This seems to suggest that probability measurements are chiral independent.
We seem to have an argument against the physical significance of chiral
oscillations.
The reply to this objection based upon the Lorentz invariance is simply that in any
given
Lorentz frame chiral oscillations are manifestly important because of the chiral
projection form (V-A) of the charged weak currents.
The chiral probability variations produced by Lorentz transformations (even if
$\gamma^5$ commutes with the Lorentz generators) are automatically compensated by the
wave function normalization conditions\footnote{In fact, the analytical form of
localization is not frame independent.} and the Lorentz transformations of the
intermediate vector bosons and other participating particles.

In the context of electroweak interactions, the determination of the explicit form of
the helicity $h$, although possible, is not as relevant as the
determination of $\gamma_{\5}$, since, as already pointed out,
the helicity is not a Lorentz invariant quantity.
On the other hand, electroweak interactions single out the negative chirality
state as the interacting state.
The same is not true for the helicity quantum number (spin projection quantum
number).
From the phenomenological point of view, chiral oscillations as well as
spin-flipping can independently take place in combination with a third physical
process, for instance, the flavor conversion of a propagating neutrino.
In fact, chiral oscillations coupled with flavor oscillations can induce some small
modifications to the standard flavor conversion formula \cite{Ber05}. More
specifically, the spinorial form and the interference between positive and
negative frequency components of the mass-eigenstate wave packets can lead to the
coupling of chiral oscillations and flavor oscillations which effectively introduce
some small modifications to the {\em standard} flavor conversion formula \cite{Ber04}
when applied to the study of NR neutrinos.
Therefore, in some cases, the proper use of the Dirac equation as evolution equation
for the mass-eigenstates might be necessary for a satisfactory description of
fermionic (spin one-half) particles,
even if one discusses Weyl or Majorana spinors\,\cite{elko}.

\subsection{Fermionic particles non-minimally coupled to an external magnetic
field}
\label{subsec:externalB}

The spin-flipping attributed to some dynamic external interacting
processes\,\cite{Oli90}
was formerly supposed to be a relevant effect in the phenomenological
context of neutrino physics, where the external interaction had its origin in
the non-minimal coupling of a magnetic moment to an external electromagnetic field
\cite{Vol81}.
On the other hand, independently of any external electromagnetic field, since
neutrinos are detected essentially via V-A charged weak currents, the chiral
oscillation mechanism by itself may even suggest an explanation for the LSND anomaly
\cite{Ana98,Agu01,Ban03}.
Such an erroneous correspondence between chiral oscillation and spin-flipping causes
confusion in the literature where the concept of helicity has
been currently used in the place of chirality.
Consequently, neutrinos with positive chirality are sterile with respect to weak
interactions.
We are now interested in pointing out the modifications of the chirality dynamics
which are observed when the neutrino electrodynamics is accurately discussed.
In particular, we are interested in the (electro)magnetic moment that appears
in a Lagrangian with non-minimal coupling.
We determine the equation of the motion for the relevant physical observables
obtained from such a Lagrangian so that we can confront the dynamics of chiral
oscillations with the dynamics of spin-flipping.

Despite their electric charge neutrality, neutrinos can interact with a photon
through loop (radiative) diagrams\,\cite{mohapatra:book,Giunti:electromagnetic}.
The Lagrangian for the interaction between a fermionic field $\psi\bb{x}$ and an
electromagnetic field written in terms of the field-strength tensor
$F^{\mu\nu}\bb{x}= \partial^{\mu}A^{\nu}\bb{x} - \partial^{\nu}A^{\mu}\bb{x}$ is
given by
\small\eq{
\mathcal{L} = \frac{1}{2}\,\overline{\psi}\bb{x} \,\sigma_{\mu\nu}\left[\mu\,
F^{\mu\nu}\bb{x} - d \,\mathcal{F}^{\mu\nu}\bb{x}\right] \psi\bb{x} + h.c.
}\normalsize
where $x = \bb{t, \bx}$, $\sigma_{\mu\nu} =
\frac{i}{2}[\gamma_{\mu},\gamma_{\nu}]$, the {\em dual} field-strength tensor
is given by $\mathcal{F}^{\mu\nu}\bb{x}=
\frac{1}{2}\epsilon^{\mu\nu\lambda\delta}
F_{\lambda\delta}\bb{x}$\,\footnote{$\epsilon^{\mu\nu\lambda\delta}$ is the totally
antisymmetric tensor.} and the coefficients $\mu$ and $d$ represent,
respectively, the magnetic and the electric dipole moment which establish the
neutrino electromagnetic coupling.
One can notice that we have not distinguished the flavor/mass mixing elements in the
above interacting Lagrangian because we are indeed interested in the physically
observable dynamics governed by the Hamiltonian
\small\eqarr{
\mathit{H} &=& \mbox{\boldmath$\alpha$}\cdot \bs{p} + \beta m
		- \beta\left[\frac{\sigma_{\mu\nu}}{2}\left(\mu \,F^{\mu\nu}\bb{x}- d
\mathcal{F}^{\mu\nu}\bb{x}\right) + h.c.\right]\nonumber\\
 		   &=& \mbox{\boldmath$\alpha$}\cdot \bs{p} + \beta
\left[m - \mu\, \bs{\Sigma}\cdot \mbox{\boldmath$B$}\bb{x} + d \,
\bs{\Sigma}\cdot \mbox{\boldmath$E$}\bb{x}\right]
\label{IVa01},~~
}\normalsize
where, in terms of the Dirac matrices, $\mbox{\boldmath$\alpha$} = \sum_{\kk \ig
\1}^{\3} \alpha_{\kk}\hat{\ms{k}} = \sum_{\kk \ig \1}^{\3}
\gamma_{\0}\gamma_{\kk}\hat{\ms{k}}$, $\beta = \gamma_{\0}$, and
$\mbox{\boldmath$B$}\bb{x}$ and $\mbox{\boldmath$E$}\bb{x}$ are respectively the
magnetic and electric fields.
In fact, Eq.~(\ref{IVa01}) could be extended to an equivalent matrix representation
with flavor and mass mixing elements where the diagonal (off-diagonal) elements
described by $\mu_{\ii,\jj}(m_{\ii,\jj})$ and $d_{\ii,\jj}(m_{\ii,\jj})$, where $i, \, j$
are mass indices, would be called diagonal (transition) moments.
In this context, for both Dirac and Majorana neutrinos, we could have transition
amplitudes with non-vanishing magnetic and electric dipole moments
\cite{Sch1,Mar1,Akh1}.
Otherwise, CP invariance ensures vanishing diagonal electric dipole
moments\,\cite{Kim93}.
Specifically for Majorana neutrinos, it can be demonstrated that the diagonal
magnetic and electric dipole moments vanish if $CPT$ invariance is assumed
\cite{Sch1}.

Turning back to the simplifying hypothesis of diagonal moments and assuming CP and
CPT invariance, we can restrict our analysis to the coupling with only an external
magnetic field $\mbox{\boldmath$B$}\bb{x}$ by setting $d = 0$.
In a minimal extension of the standard model in which neutrinos are described as
massive Dirac particles and without new interactions accessible below the weak scale,
the neutrino magnetic moment arises at the one-loop level, as does the weak
contribution to the anomalous magnetic moment of a charged lepton \cite{Zub80}.
The value of $\mu$ can be read off from general formulas for the electromagnetic
vertex, to one-loop order, of an arbitrary fermion ($\ell$).
To leading order in $m_{\ell}^{\2}/m_{W}^{\2}$, the results are independent of
$m_{\ell}$ and of the mixing matrix so that it turns out to be proportional to the
neutrino mass (matrix),
\small\eq{
\mu = \frac{3\, e \,G}{8 \sqrt{2}\pi^{\2}} m = \frac{3\, m_e \,G}{4
\sqrt{2}\pi^{\2}}\, \mu_{\B} \, m_{\nu}
= 2.7 \times 10^{\mi \1\0}\,\mu_{\B}\,\frac{m_{\nu}}{m_N}
}\normalsize
where $G$ is the Fermi constant and $m_{N}$ is the nucleon mass\,\footnote{We are
using some results of the standard $SU(2)_{L} \otimes U(1)_{Y}$ electroweak theory
\cite{Gla61}.}.
In particular, for $m_{\nu}\approx 1 \, eV$, the magnetic moment introduced by the
above formula is exceedingly small to be detected or to affect physical or
astrophysical processes.

Since we are interested in constructing the dynamics governed by the Hamiltonian of
Eq.~(\ref{IVa01}), we firstly observe that the free propagating momentum is not a
conserved quantity,
\small\eq{
\frac{d~}{dt}\langle\bs{p}\rangle \,=\, i\langle\left[ \mathit{H} ,
\bs{p}\right]\rangle\,=\, \mathrm{Re}\bb{\mu} \langle \beta\,
\mbox{\boldmath$\nabla$} \left(\bs{\Sigma}\cdot
\mbox{\boldmath$B$}\bb{x}\right)\rangle\,.
\label{IVa02}
}\normalsize
In the same way, the particle velocity given by
\small\eq{
\frac{d~}{dt}\langle\bx\rangle \,=\, i\langle\left[ \mathit{H} ,
\bx\right]\rangle\,=\,\langle\mbox{\boldmath$\alpha$}\rangle
\label{IVa03}
}\normalsize
comes out as a non-null value.
After solving the differential equation for $\bx\bb{t}$~(or for
$\mbox{\boldmath$\alpha$}\bb{t}$), it is possible to observe that, in addition to a
uniform motion, the fermionic particle executes very rapid oscillations known as {\em
zitterbewegung} \cite{Sak87}.

By following an analogous procedure for the Dirac chiral operator $\gamma_{\5}$,
recurring again to the equation of the motion, it is possible to have the chirality
and the helicity dynamics respectively given by
\small\eqarr{
\frac{d~}{dt}\langle \gamma^5 \rangle &=& 2\,i \langle\gamma_{\0}\,\gamma_{\5}
\left[m - \mu\, \bs{\Sigma}\cdot \mbox{\boldmath$B$}\bb{x}\right]\rangle
\label{IVa04}
}\normalsize
and
\small\eqarr{
\frac{d~}{dt}\langle h \rangle &=& \frac{1}{2}\,\mu\,\langle \gamma_{\0}
\left[(\bs{\Sigma}\cdot\mbox{\boldmath$\nabla$})(\mbox{\boldmath$\Sigma$
}\cdot\mbox{\boldmath$B$}\bb{x})
+ 2(\bs{\Sigma}\times \mbox{\boldmath$B$}\bb{x})\cdot
\bs{p}\right]\rangle
\label{IVa05}
}\normalsize
where we have alternatively defined the particle helicity as the projection of the
spin angular momentum onto the vector momentum, $h =
\frac{1}{2}\bs{\Sigma}\cdot{\bs{p}}$ (with
${\bs{p}}$ in place of $\hat{\bs{p}}$).
From Eqs.~(\ref{IVa04}-\ref{IVa05}) we can state that if a neutrino has an intrinsic
magnetic moment and passes through a region filled by an magnetic field, the neutrino
helicity can flip in a completely different way from how chiral oscillations evolve
in time.

In the non-interacting case, it is possible to verify that the time-dependent
expectation value of the Dirac chiral operator $\gamma_{\5}$ has an oscillating
behavior \cite{DeL98} very similar to the rapid oscillations of the position.
The Eqs.~(\ref{IVa04}-\ref{IVa05}) can be reduced to the non-interacting case by
setting $\mbox{\boldmath$B$}\bb{x} = 0$ so that
\small\eq{
\frac{d~}{dt}\langle h \rangle \,=\, i\langle\left[ \mathit{H} ,
h\right]\rangle\,=\,- \langle(\mbox{\boldmath$\alpha$} \times
\bs{p})\cdot\hat{\bs{p}}\rangle \,=\ 0
\label{IVa06}
}\normalsize
and
\small\eq{
\frac{d~}{dt}\langle\gamma^{\5}\rangle \,=\, i\langle\left[ \mathit{H} ,
\gamma^{\5}\right]\rangle\,=\,2 \,i\,m \langle\gamma^{\0}\gamma^{\5}\rangle
\label{IVa07}.
}\normalsize
from which we confirm that the chiral operator $\gamma^{\5}$ is {\em not} a constant
of the motion \cite{DeL98}.
The effective value of Eq.~(\ref{IVa07}) appears only when both positive and negative
frequencies are taken into account to compose a Dirac wave packet, i.\,e., the
non-null
expectation value of $\langle\gamma_{\0}\gamma_{\5}\rangle$ is revealed by the
interference between Dirac equation solutions with opposite sign frequencies.
The effective contribution due to this interference effect lead us to resort to the
Dirac wave packet formalism in order to quantify neutrino chiral oscillations in the
presence of an external magnetic field.

Effectively, the above discussion represents the first step for accurately deriving
the expression for the neutrino spin-flipping in magnetic fields which can be related
to chiral oscillations in the limit of a massless particle (UR limit).
By correctly distinguishing the concepts of helicity and chirality, we can determine
the origin and the influence of chiral oscillations and spin-flipping in the complete
flavor conversion formula.

In the previous section \ref{sec:relativistic},
we have confirmed that the {\em fermionic} character of the particles, more
specifically, the presence of positive and negative frequency components and the
requirement of pure flavor creation, modify the standard oscillation probability
obtained by assuming a {\em scalar} nature of the mass-eigenstates.
The new oscillation probability contained an additional oscillating term
with very high frequency, proportional to the sum of energies of the
mass-eigenstates.
Such modifications, with neutrinos being described by free Dirac wave packets,
introduced correction factors that, under the current experimental point of view,
were not effective in the UR limit of neutrino propagation in vacuum.
Nevertheless, the case of neutrinos moving in a background
matter, where spin/chiral effects become more relevant\,\cite{Kim93,MSW,Guz03},
deserves, at least, a careful investigation.
The physical consequences in environments such as supernova can be theoretically
studied \cite{Oli99}.
For instance, it was observed that neutrinos propagating in matter achieve an
effective electromagnetic vertex which affects the flavor conversion process in a
situation where chirality can be preserved\,\cite{Oli96}.

\subsubsection{Chiral oscillations in the presence of a magnetic field}

Assuming the simplifying hypothesis of a uniform magnetic field
$\mbox{\boldmath$B$}$, the physical consequences of the non-minimal coupling to an
external magnetic field can be studied by means of the eigenvalue problem
expressed by the Hamiltonian equation
\small\eq{
H\bb{\bs{p}}\,\varphi_{\n}
= \left\{\mbox{\boldmath$\alpha$}\cdot \bs{p} +
\beta \left[m - \mu\, \bs{\Sigma}\cdot
\mbox{\boldmath$B$}\right]\right\}\varphi_{\n}
= \,E_{\n}\bb{\bs{p}}\,\varphi_{\n}
\label{IVa10},~~
}\normalsize
for which the explicit $4 \times 4$ matrix representation is given by
\small\eq{
H\bb{\bs{p}} \varphi_{\n} = \left[\begin{array}{cccc}
m - \mu B_{\z} & -\mu (B_{\x} - i B_{\y})& p_{\z} & p_{\x} - i p_{\y}\\
-\mu (B_{\x} + i B_{\y})& m + \mu B_{\z} & p_{\x} + i p_{\y} & -p_{\z}\\
p_{\z} & p_{\x} - i p_{\y}& - (m - \mu B_{\z}) & \mu (B_{\x} - i B_{\y})\\
 p_{\x} + i p_{\y} & -p_{\z} & \mu (B_{\x} + i B_{\y})& -(m + \mu B_{\z})
\end{array}\right] \varphi_{\n}
\label{IVa10a}.~~
}\normalsize
The most general eigenvalue ($E_{\n}\bb{\bs{p}}$) solution of the above
problem is given by
\small\eqarr{
E_{\n}\bb{\bs{p}} = \,  \pm E_{\s}\bb{\bs{p}}
&=& \pm \sqrt{m^{\2} + \bs{p}^{\2} + \mbox{\boldmath$a$}^{\2} +\bb{\mi
1}^{\s}2
\sqrt{m^{\2}\mbox{\boldmath$a$}^{\2} + \bb{\bs{p} \times
\mbox{\boldmath$a$}}^{\2}}}, ~~~~s\,=\, 1,\,2
\label{IVa11},
}\normalsize
where we have denoted $E_{\n \ig \1,\2,\3,\4} = \pm E_{\s \ig \1,\2}$ and we have set
$\mbox{\boldmath$a$} = \mu\, \mbox{\boldmath$B$}$.
The complete set of orthonormal eigenstates $\varphi_{\n}$ thus can be written in
terms of the eigenfunctions $\mathcal{U}\bb{p_{\s}}$ with positive energy eigenvalues
($+ E_{\s}\bb{\bs{p}}$) and the eigenfunctions $\mathcal{V}\bb{p_{\s}}$
with negative energy eigenvalues ($- E_{\s}\bb{\bs{p}}$),
\small\eqarr{
\mathcal{U}\bb{p_{\s}} &=& -N\bb{p_{\s}}\,
\left(\sqrt{\frac{A^{\mi}_{\s}}{A^{\pl}_{\s}}},\,\sqrt{\frac{\alpha^{\pl}_{\s}}{
\alpha^{\mi}_{\s}}},\,\sqrt{\frac{A^{\mi}_{\s}\alpha^{\pl}_{\s}}{A^{\pl}_{\s}\alpha^{
\mi}_{\s}}},\,-1\right)^{\dagger}\,,\nonumber\\
\mathcal{V}\bb{p_{\s}} &=& -N\bb{p_{\s}}\,
\left(\sqrt{\frac{A^{\mi}_{\s}}{A^{\pl}_{\s}}},\,-\sqrt{\frac{\alpha^{\mi}_{\s}}{
\alpha^{\pl}_{\s}}},\,-\sqrt{\frac{A^{\mi}_{\s}\alpha^{\mi}_{\s}}{A^{\pl}_{\s}\alpha^
{\pl}_{\s}}},\,-1\right)^{\dagger}\,.
\label{IVa12}
}\normalsize
We denoted $p_{\s}$ as the relativistic {\em quadrimomentum}, $p_{\s} =
(E_{\s}\bb{\bs{p}}, \bs{p})$, $N\bb{p_{\s}}$ as the normalization constant and
\small\eq{
A^{\ppm}_{\s} = \Delta_{\s}^{\2}\bb{\bs{p}} \pm 2 m
|\mbox{\boldmath$a$}|  - \mbox{\boldmath$a$}^{\2},~~~
\alpha^{\ppm}_{\s}= 2 E_{\s}\bb{\bs{p}} |\mbox{\boldmath$a$}| \pm
(\Delta_{\s}^{\2}\bb{\bs{p}} + \mbox{\boldmath$a$}^{\2})
\nonumber
}\normalsize
with
\small\eq{
\Delta_{\s}^{\2}\bb{\bs{p}} = E_{\s}^{\2}\bb{\bs{p}} -
(m^{\2} + \bs{p}^{\2}) + \mbox{\boldmath$a$}^{\2}.
}\normalsize
We can observe that the above spinorial solutions are free of any additional
constraint, namely, at a given time $t$, they are independent functions of
$\bs{p}$ and they do not represent chirality/helicity eigenstates.

In order to describe the above Hamiltonian dynamics for a generic observable
$\mathcal{O}\bb{t}$ we can firstly seek a generic plane-wave decomposition as
\small\eqarr{
&& \exp{[- i(E_{\s}\bb{\bs{p}}\,t -\bs{p} \cdot
\bx)]}\,\mathcal{U}\bb{p_{\s}},
~~~~\mbox{for positive frequencies and}\nonumber\\
&& \exp{[~ i(E_{\s}\bb{\bs{p}}\,t -\bs{p} \cdot
\bx)]}\,\mathcal{V}\bb{p_{\s}},
 ~~~~\mbox{for negative frequencies},
\label{IVa13}
}\normalsize
so that the time-evolution of a plane-wave packet $\psi\bb{t, \bx}$
can be written as
\small\eqarr{
\psi\bb{t, \bx}
&=& \int\hspace{-0.1 cm} \frac{d^{\3}\hspace{-0.1cm}\bs{p}}{(2\pi)^{\3}}
\sum_{\s \ig \1,\2}\{b\bb{p_{\s}}\mathcal{U}\bb{p_{\s}}\, \exp{[-
i\,E_{\s}\bb{\bs{p}}\,t]}
\nonumber\\
&&~~~~~~~~~~~~~~~~
+ d^*\bb{\tilde{p}_{\s}}\mathcal{V}\bb{\tilde{p}_{\s}}\,
\exp{[+i\,E_{\s}\bb{\bs{p}}\,t]}\}
\exp{[i \, \bs{p} \cdot \bx]},
\label{IVa14}
}\normalsize
with $\tilde{p}_{\s} = (E_{\s},-\bs{p})$.
We see Eq.~(\ref{IVa14}) requires some extensive mathematical manipulations to
explicitly construct the dynamics of an operator $\mathcal{O}\bb{t}$ in the form
\small\eq{
\mathcal{O}\bb{t} = \int{d^{\3}\bx\,\psi^{\dagger}\bb{t,
\bx}\,\mathcal{O}\,
\psi\bb{t, \bx}}
\label{IVa15}.
}\normalsize

However, if the quoted observables like the chirality $\gamma^{\5}$, the helicity
$h$ or even the spin projection onto $\mbox{\boldmath$B$}$ commuted with the
Hamiltonian $H$, we could reconfigure the above solutions to simpler ones.
To illustrate this point, let us limit our analysis to very restrictive spatial
configurations of $\mbox{\boldmath$B$}$ so that, as a first attempt, we can calculate
the observable expectation values which appear in Eq.~(\ref{IVa04}).
Let us then assume that the magnetic field $\mbox{\boldmath$B$}$ is either orthogonal
or parallel to the momentum $\bs{p}$.
For both of these cases the spinor eigenstates can be decomposed into orthonormal
bi-spinors as
\small\eq{
\mathcal{U}\bb{p_{\s}} = N^{\pl}\bb{p_{\s}}\left[\begin{array}{r}
\varphi^{\pl}\bb{p_{\s}}\\ \chi^{\pl}\bb{p_{\s}}\end{array}\right]
}\normalsize
and
\small\eq{
\mathcal{V}\bb{p_{\s}} = N^{\mi}\bb{p_{\s}}\left[\begin{array}{r}
\varphi^{\mi}\bb{p_{\s}}\\ \chi^{\mi}\bb{p_{\s}}\end{array}\right]
\label{IVa16}.
}\normalsize
Eventually, in order to simplify some subsequent calculations involving chiral
oscillations, we could set $\varphi^{\ppm}_{\1,\2}$ and $\chi^{\ppm}_{\1,\2}$ as
eigenstates of the spin projection operator
$\bs{\Sigma}\cdot\mbox{\boldmath$B$}$, i.\,e., besides being energy
eigenstates, the general solutions $\mathcal{U}\bb{p_{\s}}$ and
$\mathcal{V}\bb{p_{\s}}$ would become eigenstates of the operator
$\bs{\Sigma}\cdot\mbox{\boldmath$B$}$ and, equivalently, of
$\bs{\Sigma}\cdot\mbox{\boldmath$a$}$.

Now Eq.~(\ref{IVa10}) can be decomposed into the pair of coupled equations
\small\eqarr{
\left(\pm E_{\s} - m + \bs{\Sigma}\cdot\mbox{\boldmath$a$}
\right)\varphi^{\ppm}_{\s} &=& \pm
\bs{\Sigma}\cdot\bs{p}\,\chi^{\ppm}_{\s},\nonumber\\
\left(\pm E_{\s} + m - \bs{\Sigma}\cdot\mbox{\boldmath$a$}
\right)\chi^{\ppm}_{\s} &=& \pm
\bs{\Sigma}\cdot\bs{p}\,\varphi^{\ppm}_{\s},
\label{IVa17}
}\normalsize
where we have suppressed the $p_{\s}$ dependence.
By introducing the commuting relation
$[\bs{\Sigma}\ponto\bs{p},\,\bs{\Sigma}\ponto\mbox{\boldmath$B$}] = 0$ which is
valid when $\bs{p}\times\mbox{\boldmath$B$} = 0$, the eigenspinor representation
can be reduced to
\small\eq{
\mathcal{U}\bb{p_{\s}} = \sqrt{\frac{E_{\s} + m_{\s}}{2E_{\s}}}
\left[\begin{array}{r} \varphi^{\pl}_{\s}\\
\frac{\bs{\Sigma}\cdot\bs{p}}{E_{\s}+
m_{\s}}\,\varphi^{\pl}_{\s}\end{array}\right]}\normalsize and
\small\eq{
\mathcal{V}\bb{p_{\s}} = \sqrt{\frac{E_{\s} + m_{\s}}{2E_{\s}}}
\left[\begin{array}{r} \frac{\bs{\Sigma}\cdot\bs{p}}{E{\s}+
m_{\s}}\,\chi^{\mi}_{\s}\\ \chi^{\mi}_{\s}\end{array}\right]
\label{IVa25},
}\normalsize
with $m_{\s} = m - \bb{\ms{-}1}^{\s}|\mbox{\boldmath$a$}|$ and the energy eigenvalues
\small\eq{
\pm E_{\s} = \pm \sqrt{\bs{p}^{\2} + m_{\s}^{\2}}
\label{IVa26}.
}\normalsize
In this case, the closure relations can be constructed in terms of
\small\eqarr{
\sum_{\s \ig
\1,\2}{\mathcal{U}\bb{p_{\s}}\otimes\mathcal{U}^{\dagger}\bb{p_{\s}}\gamma_{\0}}&=&
\sum_{\s \ig \1,\2}{\left\{\frac{\gamma_{\mu}p_{\s}^{\mu} + m_{\s}}{2 E_{\s}}
\left[\frac{1-\bb{\mi
1}^{\s}\bs{\Sigma}\ponto\hat{\bs{a}}}{2}\right]\right\}}
\nonumber\\
-\sum_{\s \ig
\1,\2}{\mathcal{V}\bb{p_{\s}}\otimes\mathcal{V}^{\dagger}\bb{p_{\s}}\gamma_{\0}}&=&
 \sum_{\s \ig \1,\2}{\left\{\frac{-\gamma_{\mu}p_{\s}^{\mu} + m_{\s}}{2 E_{\s}}
\left[\frac{1-\bb{\mi
1}^{\s}\bs{\Sigma}\ponto\hat{\bs{a}}}{2}\right]\right\}}.
\label{IVa27}
}\normalsize

Analogously, by introducing the anti-commutation relation
$\{\bs{\Sigma}\ponto\bs{p},\,\bs{\Sigma}\ponto\mbox{\boldmath$B$}\}$ when
$\bs{p}\cdot\mbox{\boldmath$B$} = 0$,
the eigenspinor representation can be reduced to
\small\eq{
\mathcal{U}\bb{p_{\s}} = \sqrt{\frac{\varepsilon_{\0} + m}{2\varepsilon_{\0}}}
\left[\begin{array}{r} \varphi^{\pl}_{\s}\\
\frac{\bs{\Sigma}\cdot\bs{p}}{\varepsilon_{\0}+
m}\,\varphi^{\pl}_{\s}\end{array}\right]}\normalsize and
\small\eq{
\mathcal{V}\bb{p_{\s}} = \sqrt{\frac{\varepsilon_{\0} + m}{2\varepsilon_{\0}}}
\left[\begin{array}{r}
\frac{\bs{\Sigma}\cdot\bs{p}}{\varepsilon_{\0}+
m}\,\chi^{\mi}_{\s}\\ \chi^{\mi}_{\s}\end{array}\right]
\label{IVa21},
}\normalsize
with $\varepsilon_{\0} = \sqrt{\bs{p}^{\2} + m^{\2}}$ and the energy
eigenvalues
\small\eq{
\pm E_{\s} = \pm\left[\varepsilon_{\0} +
\bb{\ms{-}1}^{\s}|\mbox{\boldmath$a$}|\right]
\label{IVa22}.
}\normalsize
In this case, the closure relations can be constructed in terms of
\small\eqarr{
\sum_{\s \ig
\1,\2}{\mathcal{U}\bb{p_{\s}}\otimes\mathcal{U}^{\dagger}\bb{p_{\s}}\gamma_{\0}}&=&
\frac{\gamma_{\mu}p_{\0}^{\mu} + m}{2 \varepsilon_{\0}} \sum_{\s \ig \1,\2}
{\left[\frac{1-\bb{\mi
1}^{\s}\gamma_{\0}\bs{\Sigma}\ponto\hat{\bs{a}}}{2}\right]}
\nonumber\\
-\sum_{\s \ig
\1,\2}{\mathcal{V}\bb{p_{\s}}\otimes\mathcal{V}^{\dagger}\bb{p_{\s}}\gamma_{\0}}&=&
\frac{-\gamma_{\mu}p_{\0}^{\mu} + m}{2 \varepsilon_{\0}} \sum_{\s \ig \1,\2}
{\left[\frac{1-\bb{\mi
1}^{\s}\gamma_{\0}\bs{\Sigma}\ponto\hat{\bs{a}}}{2}\right]},
\label{IVa23}
}\normalsize
where $p_{\0} = (\varepsilon_{\0}, \bs{p})$.

Since we can set $\varphi^{\pl}_{\1,\2} \equiv \chi^{\mi}_{\1,\2}$ as the components
of an orthonormal basis, we can immediately deduce the orthogonality relations
\small\eqarr{
&&\mathcal{U}^{\dagger}\bb{p_{\s}} \, \mathcal{U}\bb{p_{\rr}} =
\mathcal{V}^{\dagger}\bb{p_{\s}} \, \mathcal{V}\bb{p_{\rr}} = \delta_{\s\rr},
\nonumber\\
&&\mathcal{U}^{\dagger}\bb{p_{\s}} \,\gamma_{\0}\, \mathcal{V}\bb{p_{\rr}} =
\mathcal{V}^{\dagger}\bb{p_{\s}} \,\gamma_{\0}\, \mathcal{U}\bb{p_{\rr}} = 0
\label{IVa20}
}\normalsize
which are valid for both of the above cases.

Finally, the calculation of the expectation value of $\gamma_{\5}\bb{t}$ is
substantially simplified when we substitute the above closure relations into the wave
packet expression of Eq.~(\ref{IVa14}).
To clarify this point, we suppose the initial condition over
$\psi\bb{t,\bx}$ can be set in terms of the Fourier transform of the
weight function
\small\eq{
\varphi\bb{\bs{p}-\bs{p}_{\ii}}\,w \,=\,
\sum_{\s \ig \1,\2}{\{b\bb{p_{\s}}\mathcal{U}\bb{p_{\s}} +
d^*\bb{\tilde{p}{\s}}\mathcal{V}\bb{\tilde{p}_{\s}}\}}
\label{IVa28}
}\normalsize
so that
\small\eq{
\psi\bb{0, \bx}
= \int\hspace{-0.1 cm} \frac{d^{\3}\hspace{-0.1cm}\bs{p}}{(2\pi)^{\3}}
\varphi\bb{\bs{p}-\bs{p}_{\ii}}\exp{[i \,
\bs{p} \cdot \bx]}
\,w
\label{IVa29}
}\normalsize
where $w$ is some fixed normalized spinor.
By using the orthonormality properties of Eq.~(\ref{IVa20}), we find \cite{Zub80}
\small\eqarr{
b\bb{p_{\s}} &=& \varphi\bb{\bs{p}- \bs{p}_{\ii}} \,
\mathcal{U}^{\dagger}\bb{p_{\s}} \, w, \nonumber\\
d^*\bb{\tilde{p}_{\s}} &=& \varphi\bb{\bs{p}-
\bs{p}_{\ii}}\,\mathcal{V}^{\dagger}\bb{\tilde{p}_{\s}}\, w.
\label{IVa30}
}\normalsize
For {\em any} initial state $\psi\bb{0, \bx}$ given by
Eq.~(\ref{IVa29}), the negative frequency solution coefficient
$d^*\bb{\tilde{p}_{\s}}$ necessarily provides a non-null contribution to the
time-evolving wave packet.
It obliges us to take the complete set of Dirac equation solutions to construct a
complete and correct wave packet solution.
Only if we consider the initial spinor $w$ being a positive energy
($E_{\s}\bb{\bs{p}}$) and momentum ($\bs{p}$) eigenstate, the
contribution due to the negative frequency solutions $d^*\bb{\tilde{p}_{\s}}$ will
become null and we will have a simple expression for the time-evolution of any
physical observable.
By substituting the closure relations of Eqs.~(\ref{IVa27}) and (\ref{IVa23}) into
the time-evolution expression for the above wave packet, Eq.~(\ref{IVa14}) can thus
be rewritten as
\small\eqarr{
\hspace{-0.1 cm}\psi\bb{t, \bx}
&\hspace{-0.1 cm}=&
\hspace{-0.1 cm}\int\hspace{-0.1 cm}
\frac{d^{\3}\hspace{-0.1cm}\bs{p}}{(2\pi)^{\3}}
\varphi\bb{\bs{p}-\bs{p}_{\ii}}\exp{[i \,
\bs{p} \cdot \bx]}
\sum_{\s \ig \1,\2}\mbox{$\left\{\left[\cos{[E_{\s}\,t]}
-i\frac{H_{\s}}{E_{\s}}\sin{[E_{\s}\,t]}\right]\left(\frac{1-(\mi
1)^{\s}\bs{\Sigma}\ponto\hat{\bs{a}}}{2}\right)\right\}$}
w~~
\label{IVa14A}
}\normalsize
for the first case where $E_{\s}$ is given by Eq.~(\ref{IVa26}) and $H_{\s} =
\mbox{\boldmath$\alpha$}\cdot \bs{p} + \gamma_{\0} m_{\s}$, or as
\small\eqarr{
\hspace{-0.1 cm}\psi\bb{t, \bx}
&\hspace{-0.1 cm}=&
\hspace{-0.1 cm}\int\hspace{-0.1 cm}
\frac{d^{\3}\hspace{-0.1cm}\bs{p}}{(2\pi)^{\3}}
\varphi\bb{\bs{p}-\bs{p}_{\ii}}\exp{[i \,
\bs{p} \cdot \bx]}
\sum_{\s \ig \1,\2}\mbox{$\left\{\left[\cos{[E_{\s}\,t]}
-i\frac{H_{\0}}{\varepsilon_{\0}}\sin{[E_{\s}\,t]}\right]\left(\frac{1-(\mi
1)^{\s}\gamma_{\0}\bs{\Sigma}\ponto\hat{\bs{a}}}{2}
\right)\right\}$}
w~~~~
\label{IVa14B}
}\normalsize
for the second case where $\varepsilon_{\0}$ is given by Eq.~(\ref{IVa22}) and
$H_{\0} = \mbox{\boldmath$\alpha$}\cdot \bs{p} + \gamma_{\0} m$.

Once we have assumed the neutrino electroweak interactions at the source and detector
are ({\em left}) chiral $\left(\overline{\psi} \gamma^{\mu}(1 - \gamma^{\5})\psi
W_{\mu}\right)$, only the component with negative chirality contributes to the
propagation.
It was already demonstrated that, in vacuum, chiral oscillations can introduce very
small modifications to the neutrino conversion formula \cite{DeL98,Ber05}.
The probability of a neutrino produced as a negative chiral eigenstate to be detected
after a time $t$ can be summarized by the expression
\small\eqarr{
\mathcal{P}(\mbox{\boldmath$\nu_{\Ll}$}\rightarrow\mbox{\boldmath$\nu_{\Ll}$};t)
& = &
\int{d^{\3}\bx\,\psi^{\dagger}\bb{t, \bx}\,\frac{1 -
\gamma_{\5}}{2}\,
\psi\bb{t, \bx}} = \frac{1}{2}\left(1 -
\langle\gamma_{\5}\rangle\bb{t}\right)
\label{IVa15A}.
}\normalsize
From this integral, it is readily seen that an initial $\ms{-1}$ chiral
mass-eigenstate will evolve in time changing its chirality.

By assuming the fermionic particle is created  at time $t=0$ as a $\ms{-1}$ chiral
eigenstate ($\gamma_{\5} w = \mi w$), for the case where
$[\bs{\Sigma}\ponto\bs{p},\,\bs{\Sigma}\ponto\mbox{\boldmath$B$}] = 0$
({\boldmath$B$} parallel to {\boldmath$p$}), we can write
\small\eqarr{
\langle\gamma_{\5}\bb{t}\rangle &=&
\int\hspace{-0.1 cm}
\frac{d^{\3}\hspace{-0.1cm}\bs{p}}{(2\pi)^{\3}}\varphi^{\2}\bb{\mbox{
\boldmath$p$}-\bs{p}_{\ii}}\times\nonumber\\
&&~~
w^{\dagger}\sum_{\s \ig \1,\2}\mbox{$\left\{\left[\gamma_{\5}\cos^{\2}{[E_{\s}\,t]}
+i \frac{[H_{\s},\,\gamma_{\5}]}{2E_{\s}}\sin{[2\,E_{\s}\,t]} +
\frac{H_{\s}\,\gamma_{\5}\,H_{\s}}{E_{\s}^{\2}}\sin^{\2}{[E_{\s}\,t]}\right]
\left(\frac{1-(\mi
1)^{\s}\bs{\Sigma}\ponto\hat{\bs{a}}}{2}\right)\right\}$}\,
w\nonumber\\
&=&(\ms{-}1) \int\hspace{-0.1 cm}
\frac{d^{\3}\hspace{-0.1cm}\bs{p}}{(2\pi)^{\3}}\varphi^{\2}\bb{\mbox{
\boldmath$p$}-\bs{p}_{\ii}}
\sum_{\s \ig \1,\2}\mbox{$\left\{\left[\cos^{\2}{[E_{\s}\,t]} +
\frac{\bs{p}^{\2} -
m_{\s}^{\2}}{E_{\s}^{\2}}\sin^{\2}{[E_{\s}\,t]}\right]\,w^{\dagger}\left(\frac{1-(\mi
1)^{\s}\bs{\Sigma}\ponto\hat{\bs{a}}}{2}\right)w\right\}$}
\nonumber\\
&=&(\ms{-}1) \int\hspace{-0.1 cm}
\frac{d^{\3}\hspace{-0.1cm}\bs{p}}{(2\pi)^{\3}}\varphi^{\2}\bb{\mbox{
\boldmath$p$}-\bs{p}_{\ii}}
\sum_{\s \ig \1,\2}\mbox{$\left\{\left[\frac{\bs{p}^{\2}}{E_{\s}^{\2}} +
\frac{m_{\s}^{\2}}{E_{\s}^{\2}}\cos{[2\,E_{\s}\,t]}\right]\,w^{\dagger}\left(\frac{
1-(\mi
1)^{\s}\bs{\Sigma}\ponto\hat{\bs{a}}}{2}\right)w\right\}$}
\label{IVa15B},
}\normalsize
where we have used the wave packet expression of Eq.~(\ref{IVa14A}). In the
second equality, we made use of
\small\eq{
w^{\dagger} \gamma_{\5} w = \mfn{-}1,~~~~  w^{\dagger} [H_{\s},\,\gamma_{\5}]w = 0
~~~~\mbox{and}~~~~ H_{\s}\,\gamma_{\5}\,H_{\s} = \bs{p}^{\2} -
m_{\s}^{\2}.
\label{IVa15B1}
}\normalsize
The expression \eqref{IVa15B} can be reduced to a simpler one in the non-interacting
case, Eq.\,\eqref{IV29}.

Due to a residual interaction with the external magnetic field {\boldmath$B$} we
could also observe chiral oscillations in the UR limit.
However, from the phenomenological point of view, the coefficient of the oscillating
term goes with $\frac{m_{\s}^{\2}}{E_{\s}^{\2}}$ which makes chiral oscillations
negligible for UR neutrinos \cite{Ber05,Ber05A}.
As a {\em toy model} illustration, by assuming a highly peaked momentum distribution
centered around a NR momentum $p_{\ii}\ll m_{\s}$, where the wave
packet effects are practically ignored, the chiral conversion formula can be written
as
\small\eq{
\mathcal{P}(\mbox{\boldmath$\nu_{\Ll}$}\rightarrow\mbox{\boldmath$\nu_{\R}$};t)
\approx  \frac{1}{2}\left(1 - \cos{[2\,m\,t]}\cos{[2\,|\mbox{\boldmath$a$}|\,t]}
-\sin
{[2\,m\,t]}\sin{[2\,|\mbox{\boldmath$a$}|\,t]}w^{\dagger}\bs{\Sigma}
\ponto\hat{\bs{a}}\,w\right)
\label{IVa15A222}
}\normalsize
where all the oscillating terms come from the interference between positive and
negative frequency solutions which compose the wave packets.

Turning back to the case where
$\{\bs{\Sigma}\ponto\bs{p},\,\bs{\Sigma}\ponto\mbox{\boldmath$B$}\} = 0$
({\boldmath$B$} orthogonal to {\boldmath$p$}), we
could have a phenomenologically more interesting result.
By following a similar procedure and mathematical manipulations, we could write
\small\eqarr{
\langle\gamma_{\5}\bb{t}\rangle &=&
\int\hspace{-0.1 cm}
\frac{d^{\3}\hspace{-0.1cm}\bs{p}}{(2\pi)^{\3}}\varphi^{\2}\bb{\mbox{
\boldmath$p$}-\bs{p}_{\ii}}
w^{\dagger}\mbox{$\left\{\gamma_{\5}\cos{[E_{\1}\,t]}\cos{[E_{\2}\,t]} +
\frac{H_{\0}\,\gamma_{\5}\,H_{\0}}{\epsilon_{\0}^{\2}}
\sin{[E_{\1}\,t]}\sin{[E_{\2}\,t]} +\right.$}\nonumber\\
&&~~~~~~~~~~~~~~~~~~~~~~~~~\left.\mbox{$\frac{i}{2}\left[[H_{\0},\,\gamma_{\5}]\sin{[
E_{\1}+E_{\2}]} -
\{H_{\0},\,\gamma_{\5}\}\gamma_{\0}\bs{\Sigma}\cdot\hat{\mbox{
\boldmath$a$}}\sin{[E_{\1}-E_{\2}]} \right]$}\right\}w\nonumber\\
&=&(\ms{-}1) \int\hspace{-0.1 cm}
\frac{d^{\3}\hspace{-0.1cm}\bs{p}}{(2\pi)^{\3}}\varphi^{\2}\bb{\mbox{
\boldmath$p$}-\bs{p}_{\ii}}
\mbox{$\left\{\cos{[E_{\1}\,t]}\cos{[E_{\2}\,t]} + \frac{\bs{p}^{\2} -
m^{\2}}{\epsilon_{\0}^{\2}} \sin{[E_{\1}\,t]}\sin{[E_{\2}\,t]}\right\}$}\nonumber\\
&=&(\ms{-}1) \int\hspace{-0.1 cm}
\frac{d^{\3}\hspace{-0.1cm}\bs{p}}{(2\pi)^{\3}}\varphi^{\2}\bb{\mbox{
\boldmath$p$}-\bs{p}_{\ii}}
\mbox{$\left\{\frac{\bs{p}^{\2}}{\epsilon_{\0}^{\2}}\cos{[2\,|\mbox{
\boldmath$a$}|\,t]}+ \frac{ m^{\2}}{\epsilon_{\0}^{\2}}
\cos{[2\,\epsilon_{\0}\,t]}\right\}$}
\label{IVa15C},
}\normalsize
where we have used the wave packet expression of Eq.~(\ref{IVa14B}) and, in addition
to $w^{\dagger} \gamma_{\5} w = \ms{-}1$, we have also observed that
$\{H_{\0},\,\gamma_{\5}\} = 2
\gamma_{\5}\bs{\Sigma}\cdot\hat{\bs{p}}$ and, subsequently,
$w^{\dagger}\bs{\Sigma}\cdot\hat{\bs{p}}\gamma_{\0}\mbox{
\boldmath$\Sigma$}\cdot\hat{\mbox{\boldmath$a$}} w = 0$.
Now, in addition to the non-interacting oscillating term $\frac{
m^{\2}}{\epsilon_{\0}^{\2}} \cos{[2\,\epsilon_{\0}\,t]}$, which comes from the
interference between positive and negative frequency solutions of the Dirac equation,
we have an extra term which comes from the interference between equal sign
frequencies. For very large time scales, such term can substantially change the
oscillating results.
In this case, it is interesting to observe that the UR limit of
Eq.~(\ref{IVa15C}) leads to the following expressions for the chiral conversion
formulas,
\small\eq{
\mathcal{P}(\mbox{\boldmath$\nu_{\Ll}$}\rightarrow\mbox{\boldmath$\nu_{\R}$};t)
\approx  \frac{1}{2}\left(1 - \cos{[2\,|\mbox{\boldmath$a$}|\,t]}\right)
~~~~\mbox{and}~~~~
\mathcal{P}(\mbox{\boldmath$\nu_{\Ll}$}\rightarrow\mbox{\boldmath$\nu_{\Ll}$};t)
\approx  \frac{1}{2}\left(1 + \cos{[2\,|\mbox{\boldmath$a$}|\,t]}\right)
\label{IVa15A2}
}\normalsize
which, differently from chiral oscillations in vacuum, can be phenomenologically
relevant.
Obviously, we are reproducing the consolidated results already attributed to neutrino
spin-flipping \cite{Kim93} where, by taking the UR limit, the
chirality quantum number can be approximated by the helicity quantum number, but now
it was accurately derived from the complete formalism with Dirac spinors.

\subsection{Restrictive possibility of analysis}

Assuming that neutrinos are produced with completely random spin orientation before
interacting with the magnetic fields, we can take the neutrino initial state as
a collection of particles where $50\%$ are characterized by spin-up states
and the remaining $50\%$ by spin-down states.
We can understand the result
provided by Eq.~(\ref{IVa15B}) as a much more interesting phenomenological
instrument\,\cite{Ber08B} than the result expressed by Eq.~(\ref{IVa15C}).
By averaging both the above results for $\langle\gamma_{\5}\rangle$ on time,
we can easily observe that the corresponding average value for the case where
{\boldmath$p$} is perpendicular to {\boldmath$B$} [Eq.\,\eqref{IVa15C}] is zero,
while the same quantity for the case where {\boldmath$p$} is parallel to
{\boldmath$B$} [Eq.\,\eqref{IVa15B}] depends on the polarization of the initial state
as follows:
\small\eqarr{
\langle\gamma_{\5}\rangle_{time}&=&
(\ms{-}1) \int\hspace{-0.1 cm}
\frac{d^{\3}\hspace{-0.1cm}\bs{p}}{(2\pi)^{\3}}\varphi^{\2}\bb{\mbox{
\boldmath$p$}-\mbox{\boldmath$p_{\ii}$}}
\sum_{\s \ig \1,\2}\mbox{$\left[\frac{\bs{p}^{\2}}{E_{\s}^{\2}}
\,w^{\dagger}\left(\frac{1-(\mss{-}1)^{\s}\bs{\Sigma}\ponto\hat{\bs{a}}}{2}\right)
w\right]$}
\label{15BB},
}\normalsize

By supposing the feasibility of the construction of a detection apparatus which
identify the neutrino polarization state\,\footnote{We are referring to the helicity
(spin-up and spin-down) states and not to the chirality states which have to be
negative-chiral in the detection process.}, the last result becomes
phenomenologically relevant since it distinguishes the chirality conversion rates for
spin-up ($s=1$) and spin-down ($s=2$) states.
Once we have assumed the neutrino electroweak interactions at the source and detector
are {\em left} (or negative) chiral, only the component with negative chirality
contributes to the effective detection result.
The {\em residual} contribution due to the chiral oscillation mechanism appears in
the final time averaged result where it vanishes.

Let us compare the initial polarization distribution with the distribution
after traversing the magnetic field region with $\bs{p}\parallel \bs{B}$.
Initially, i.e., before interacting with the magnetic fields, the spin-up and
spin-down states would be equally probable for any detection process and the ratio
between the probabilities to detect the different spin eigenstates would be unity.
After traversing the region of parallel magnetic fields,
the chiral oscillation mechanism can modify the expectation values for the opposite
polarization states in different ways, which reflects in a
ratio between the detection probabilities of spin-up and spin-down states
different from unity. These ratios are illustrated in Figs.~\ref{Chi1} and
\ref{Chi2}.

As one can notice, since the coefficient of the oscillating term
goes with $\frac{m_{\s}^{\2}}{E_{\s}^{\2}}$, the chiral oscillation effects become
negligible for UR neutrinos \cite{Ber05,Ber05A}.

To have a significant impact on the quantum process,
there is also a natural scale that is required for the magnetic field strength,
i.\,e., a critical value of $4.41 \times 10^{\1\3}\, G$.
There are reasons to expect that magnetic fields of such or even larger magnitudes
can arise in cataclysmic astrophysical events such as supernova explosions or
coalescing neutron stars, i.e., situations where an enormous neutrino outflow should
be expected.
Two classes of stars, the soft gamma-ray repeaters (SGR) \cite{SGR} and the anomalous
x-ray pulsars (AXP) \cite{AXP} are supposed to be remnants of such events which form
magnetars \cite{MAG}, neutron stars with magnetic fields of $10^{\1\4} - 10^{\1\5} \,
G$.
From the theoretical point of view, the possibility of larger magnetic fields
$10^{\1\6} - 10^{\1\7} \, G$ have not been discarded yet.
The early universe between the QCD phase transition ($\sim 10^{-\5} \, s$) and the
nucleosynthesis epoch ($10^{-\2} - 10^{\2} \, s$) is believed to be yet another
natural environment where strong magnetic fields and large neutrino densities could
coexist.

Concerning the polarization measurements, the observation of the
dependence of the neutrino chirality conversion rate on the neutrino polarization
state (measurements) can be converted into a clear signal of the presence of
right-handed (positive chiral) neutrinos in the neutrino-electron scattering.
In fact, in a more extended scenario, the scattering of neutrinos on a polarized
electron target \cite{Mis05} was proposed as a test for new physics beyond the
Standard Model (SM).
To search for exotic right-handed weak interactions \cite{Mis05,Cie05} the strong
polarized neutrino beam and the polarized neutrino target is required
\cite{Mis05,Cie05}.
It has been shown how the presence of the right-handed neutrinos changes the spectrum
of recoil electrons in relation to the expected standard model prediction, using the
current limits on the non-standard couplings.
In this framework, the interference terms between the standard and exotic couplings
in the differential cross section depend on the angle between the transverse incoming
neutrino polarization and the transverse electron polarization of the target, and do
not vanish in the limit of massless neutrino.

Finally, we would have been dishonest if we had ignored the complete analysis of the
general case comprised by Eqs.~(\ref{IVa10}-\ref{IVa12}) where we had not yet assumed
an arbitrary (simplified) spatial configuration for the magnetic field.
We curiously notice the fact that those complete (general) expressions for
propagating wave packets do not satisfy the standard dispersion relations like
$E^{\2} = m^{\2}+{\bs{p}}^{\2}$ excepting the two particular cases where
$E_{\s}\bb{\bs{p}}^{\2} = m_{\s}^{\2}+ {\bs{p}}^{\2}$ for
$\bs{p}\times\mbox{\boldmath$B$} = 0$ or $\epsilon_{\0}^{\2} = m^{\2}+
{\bs{p}}^{\2}$ for $\bs{p}\cdot\mbox{\boldmath$B$} = 0$.
In particular, the process of neutrino propagation through an active medium
consisting of magnetic field and plasma and the consequent modifications to the
neutrino dispersion relations have been studied in the literature \cite{Kuz06}.
In addition, such a general case leads to the formal connection between quantum
oscillation phenomena and a very different field.
In principle, it could represent an inconvenient obstacle forbidding the extension of
these restrictive cases to a general one. However, we believe that it can also
represent a starting point in discussing dispersion relations which may be
incorporated into frameworks encoding the breakdown (or the validity) of Lorentz
invariance.

\subsection{Absolute neutrino mass from helicity measurements}

After the confirmation that neutrinos have non-null masses and non trivial
mixing among the various types, the knowledge of the absolute scale of neutrino
masses is one of the most urgent questions in neutrino physics.
In recent times, the greatest advances in the understanding of neutrino
properties were boosted by neutrino oscillation experiments which are only
capable of accessing the two mass-squared differences and, in principle,
three mixing angles and one Dirac CP violating phase of the
Maki-Nakagawa-Sakata (MNS) leptonic mixing matrix \cite{MNS}.

Too large masses for the light active neutrinos may alter significantly the recent
cosmological history of the Universe.
One stringent bound for the value of the sum of neutrino masses come from cosmology
as
\small\eq{
\label{mnu<}
\sum_\nu m_\nu < 0.17 \mathrm{eV}
}\normalsize
at 95\% confidence level\,\cite{seljak.06:nu}.
However, it is regarded as too optimistic by most cosmologists because
it has been derived from data sets that are inconsistent among themselves. In
particular, the influence of Lyman-$\alpha$ forest data may suffer from large
systematical errors.
Nevertheless, a value of around $1\rm eV$ would be a conservative upper bound for
the sum of neutrino masses coming from cosmology.

Another issue of great interest refers to the existence of heavy right-handed
neutrinos which could explain at the same time the tiny active neutrinos masses
through the seesaw mechanism as well as the matter-antimatter asymmetry of
the Universe through the implementation of the mechanism of
bariogenesis through leptogenesis\,\cite{leptog}.

Despite of the stringent bound \eqref{mnu<} coming from cosmological analysis,
terrestrial direct search experiments establish much more looser
bounds on the effective neutrino masses\,\cite{pdg}.
These bounds are based on ingeniously planned experiments\,\cite{giunti}, but
their intrinsic difficulties rely on the fact that they should probe,
essentially, the kinematical effects of neutrino masses which are negligible
compared to other typical quantities involved in processes with neutrino
emission.

Although the terrestrial bounds are not as stringent as the cosmological bound in Eq.~\eqref{mnu<}, it is always
desirable to have a direct measurement of neutrino masses. Two more reasons can
be listed in favor of direct terrestrial searches: (a) cosmological bounds may
be quite model dependent and (b) we may have access to mixing parameters through
the effective neutrino masses.
For the electron neutrino, there are ongoing experiments planning to reduce the
respective bound to 0.2 eV\,\cite{katrin}.

The main goal of this section is to review the possibility of accessing the
absolute neutrino mass scale through one of the most natural consequences of
massive fermions, i.\,e., the dissociation of \textit{chirality} and
\textit{helicity}.
Such possibility was originally suggested in Refs.\,\cite{shrock,shrock2} and it was
recently considered in Ref.\,\cite{ccn:helicity} in order to reanalyze the
possibility taking into account the present experimental bounds.

Let us consider the pion decay $\pi^-\rightarrow
\mu^-\bar{\nu}_\mu$. Since the pion is a spin zero particle, in its rest frame,
the decaying states should have the following form from angular momentum
conservation,
\small\eq{
\label{RR+LL}
\ket{\pi}\rightarrow\quad
\ket{\mu\!:\LL}\ket{\bar{\nu}\!:\RR}
+\delta \ket{\mu\!:\RL}\ket{\bar{\nu}\!:\LR}
\,,
}\normalsize
where the arrows represent the momentum direction (longer arrow) and the spin
direction (shorter arrow); the combination $\RR$, for example, means a
rightgoing fermion with a right helicity. On later calculations
the right (left) helicity will be denoted simply as $h=+1$ ($h=-1$).
The normalization of the state is arbitrary and the coefficient $\delta$ is of
the order of $m_\nu/E_\nu$, which will be calculated in the following.
For massless neutrinos there is no second term in Eq.~\eqref{RR+LL} since the
antineutrino is strictly right-handed in helicity and chirality. Thus by
measuring the wrong helicity contribution of the charged lepton it is possible
to have access to the neutrino mass.
The precision in the polarization measurement necessary to extract the wrong
helicity will be also calculated.

The pion decay $\pi^-\rightarrow l_i^-\bar{\nu}_j$ can be effectively described by
the four-point Fermi interaction Lagrangian
\small\eq{
\label{L:CC}
\lag_{\rm CC}=
2\sqrt{2}G_F \bar{l}_i\gamma^\mu LU_{ij}\nu_j J_\mu + h.c.,
}\normalsize
where $L=\frac{1}{2}(1-\gamma_5)$, $\{U_{ij}\}$ denotes the MNS matrix while
$J_\mu$ is the hadronic current
that in the case of pion decay reads
\small\eq{
J_\mu=V_{ud} \bar{u}_L\gamma_\mu d_L
\,.
}\normalsize

From Eq.~\eqref{L:CC} it is straightforward to calculate the amplitude for
$\pi(\bp)\rightarrow l_i(\bq)\nu_j(\bk)$, $i,j=1,2$,
\small\eq{
\label{pi:amp:1}
-i\mathcal{M}(\pi\rightarrow l_i\bar{\nu}_j)=
2G_FF_\pi V_{ud}
\bar{u}_{l_i}(\bq)\sla{p}LU_{ij}v_{\nu_j}(\bk)
}\normalsize
by using\,\cite{DGH}
\small\eq{
\bra{0}\bar{u}\gamma_5\gamma_\mu d\ket{\pi^-}=ip_\mu\sqrt{2} F_\pi
\,,
}\normalsize
where $F_\pi\approx 92 \rm MeV$ is the pion decay constant.
Let us denote the spinor dependent amplitude as
\small\eq{
\tM_{ij}=
\bar{u}_{l_i}(\bq)\sla{p}Lv_{\nu_j}(\bk)
\,.
}\normalsize
\eqarr{
\label{M2:spins}
\sum_{\rm spins}|\tM_{ij}|^2
&=&
4(\ip{p}{q_i})(\ip{p}k_j)-2p^2(\ip{q_i}k_j)\\
&=&
\label{p=q+k}
M^2_i(M^2_\pi-M^2_i)
\cr&& +\
m^2_j(M^2_\pi+2M^2_i-m^2_j)
\,.
}\normalsize
The last expression \eqref{p=q+k} is exact and follows when $p=q_i+k_j$
(four-vector), $p^2=M^2_\pi$, $q_i^2=M^2_i$ and $k_j^2=m^2_j$.
The first expression \eqref{M2:spins} does not assume energy-momentum
conservation.

We can calculate the squared amplitude, summed over the neutrino spin, but
dependent on the polarization $\hat{\bn}$ of the charged lepton in its rest
frame:
\eqarr{
\label{A^2:n}
P_{ij}(n_i)&\equiv&
\sum_{\nu_j \text{ spin}}|\tM_{ij}|^2
\\&=&
M^2_i[\ip{q_i}k_j+M_i(\ip{k_j}n_i)]
+ 2M^2_im^2_j
\cr&&
+\ m^2_j[\ip{q_i}k_j-M_i(\ip{k_j}n_i)]
\,,~~~
}\normalsize
where
\small\eq{
n_i^\mu=\Big(\frac{\ip{\bq}\hn}{M_i},
\hn+(\frac{E_{l_i}}{M_i}-1)(\ip{\hn}\hat{\bq})\hat{\bq}\Big)
}\normalsize
For the particular directions $\hn= h_i\hat{\bq}$ we single out the
positive ($h_i=1$) and negative ($h_i=-1$) helicity for the charged
lepton\,\cite{Zub80}.

In the helicity basis
\small\eq{
\ip{q_i}k_j\pm M_i(\ip{k_j}n_i)=
[E_{l_i}(\bk)\pm h_i|\bq|][E_{\nu_j}\mp h_i\ip{\hat{\bq}}\bk]
}\normalsize
Therefore, for the pion at rest we obtain
\eqarr{
P_{ij}(h_i=1)&=&
M^2_i(M^2_\pi-M^2_i) + O(m^2_j)
\\
P_{ij}(h_i=-1)&=&
m^2_j\frac{M^4_\pi}{M^2_\pi-M^2_i}+ O(m^4_j)
}\normalsize
Without approximations we obtain
\small\eq{
P_{ij}(h_i)-P_{ij}(-h_i)=
h_iM_\pi(M^2_i-m^2_j)|\bq|
\,,
}\normalsize
while the sum is given by Eq.~\eqref{p=q+k}. One can see this results
are in accordance with Eq.~(2.38) of Ref.\,\cite{shrock2} where the
polarization $[P_{ij}(+)-P_{ij}(-)]/[P_{ij}(+)+P_{ij}(-)]$ is calculated.

The ratio between the squared amplitudes for left-handed and right-handed
helicities is
\small\eq{
\label{ratio}
R_{ij}=\frac{P_{ij}(h_i=-1)}{P_{ij}(h_i=1)}=
\frac{m^2_j}{M^2_i}\frac{M^4_\pi}{(M^2_\pi-M^2_i)^2}
\,.
}\normalsize
Considering numerical values we obtain for $M_i=M_\mu$,
\small\eq{
R_{\mu j}=\frac{m^2_j}{(100\mathrm{keV})^2}\times 4.92\times 10^{-6}
\,,
}\normalsize
while for $M_i=M_e$,
\small\eq{
R_{ej}=\frac{m^2_j}{(1\mathrm{eV})^2}\times 3.83\times 10^{-6}
\,.
}\normalsize
Considering the actual direct bounds for the neutrino masses \cite{pdg}, we
need a precision of $10^{-6}$ in the helicity measurement to reach those bounds
either in the case of muons or electrons. Although the branching ratio to produce
muons from pion decay is much larger than to produce electrons, the required dominant
versus wrong helicity probability ratios are similar.

Therefore, the coefficient $\delta$ in Eq.~\eqref{RR+LL} has exactly the
modulus
\small\eq{
\label{delta2}
|\delta_{\mu j}|^2=R_{\mu j} =
\frac{m^2_j}{M^2_\mu}\frac{M^4_\pi}{(M^2_\pi-M^2_\mu)^2}
\,.
}\normalsize
If we rewrite
\small\eq{
|\delta_{\mu j}|=\frac{m_j}
{2E_{\nu_j}}\frac{M_\pi}{M_\mu}
\,,
}\normalsize
where $E_{\nu_j}=\frac{M^2_\pi-M^2_\mu}{2M_\pi}+O(m^2_j)$, we see that
$|\delta_{\mu j}|$ is modified by the factor $\frac{M_\pi}{M_\mu}$ when
compared to the naive estimate $m_\nu/2E_\nu$\,\cite{langacker}. We can also
conclude that for the channel $\pi\rightarrow e\bar{\nu}$ the real factor is
enhanced considerably ($\sim 274\times$).

In general, experiments can not achieve a perfect accuracy in helicity
measurements because they usually involve polarization
distributions\,\cite{nuehelicity,roesch,fetscher}.
Even though the measurement of the helicity quantum number as a spin projection can
only result in two discrete values, the projection direction is only defined within
a finite accuracy.
Thus we have to consider the accuracy necessary to be able to measure the
wrong, non-dominant helicity amplitude.
Parameterizing Eq.~\eqref{A^2:n} using $\ip{\hn}\hat{\bq}=-\cos\theta$ yields
\eqarr{
P_{ij}(\theta)&=&
M^2_i(M^2_\pi-M^2_i)\mn{\frac{1}{2}}(1-\cos\theta)
\\&&
+\ \frac{m^2_j}{2}
\Big\{
M^2_\pi+2M^2_i
\cr&& +\
\cos\theta\big[M^2_\pi+\frac{2M^2_i}{M^2_\pi-M^2_i}\big]
\Big\}
\,.
}\normalsize
For $\cos\theta=1$ ($\cos\theta=-1$) we recover the negative (positive)
helicity for the charged lepton.
Expanding around $\theta=0$, we obtain
\small\eq{
P_{ij}(\delta\theta)
\approx
M^2_i(M^2_\pi-M^2_i)\frac{\delta\theta^2}{4}
+m^2_j\frac{M^4_\pi}{M^2_\pi-M^2_i}
\,.
}\normalsize
Therefore we need an angular resolution of
\small\eq{
\delta\theta=2 \sqrt{R}\sim 10^{-3}
}\normalsize
to be able to probe the ratio $R=R_{ij}$ in Eq.~\eqref{ratio}.

Considering the leptonic mixing the measurement of the wrong helicity for
muons probes
\small\eq{
|\mathcal{M}(\pi\rightarrow \mu\bar{\nu}:h_\mu=-1)|^2=
|C|^2\frac{M^4_\pi}{M^2_\pi-M^2_\mu} (m^2_{\nu_\mu})_{\rm eff}
\,,~~
}\normalsize
where $C\equiv 2G_FF_\pi V_{ud}$ and
\small\eq{
\label{mnumu}
(m^2_{\nu_\mu})_{\rm eff}\equiv
\sum_j|U_{\mu j}|^2m^2_j\,,
}\normalsize
is an effective mass for the muon neutrino, analogous to
$m^2_\beta$\,\cite{giunti} inferred from the tritium beta decay experiments for
the electron neutrino. The effective electron neutrino mass $m^2_\beta$ is
defined as the expression in Eq.~\eqref{mnumu} using $|U_{ej}|^2$ instead of
$|U_{\mu j}|^2$.
In fact, $m^2_\beta$ can be extracted from $\pi^-\rightarrow e^-\bar{\nu}$ by
measuring the electron with negative helicity.

To obtain the decay rate, we must multiply the amplitude squared by the factor
\small\eq{
\frac{1}{4\pi}\frac{1}{2M_\pi}
\Big[\frac{v_{\nu_j}v_{l_i}}{v_{\nu_j}+v_{l_i}}\Big]
\approx
\frac{1}{4\pi}\frac{1}{4M_\pi}
\Big(1-\frac{M^2_i}{M^2_\pi}\Big)
+ O(m^2_\nu)
\,,
}\normalsize
arising from the phase space.
We then obtain
\eqarr{
\label{Gamma+}
\Gamma_\alpha\ms{(h_\alpha\!=\!1)}
\!\!&=&\!\!
\frac{G^2_F}{4\pi}F^2_\pi|V_{ud}|^2M_\pi
M^2_{\alpha}\Big(1-\!\frac{M^2_\alpha}{M^2_\pi}\Big)^2
\!\!+O(m^2_\nu)
,\quad~~
\\
\label{Gamma-}
\Gamma_\alpha\ms{(h_\alpha\!=\!-1)}
\!\!&=&\!\!
\frac{G^2_F}{4\pi}F^2_\pi|V_{ud}|^2M_\pi(m^2_{\nu_\alpha})_{\rm eff}
+O(m^4_\nu)
\,,
}\normalsize
where $\alpha=e,\mu$ and $(m^2_{\nu_e})_{\rm eff}=m^2_\beta$.
The expression in Eq.~\eqref{Gamma+} is the ordinary tree level decay rate for the pion decaying into $l_\alpha\nu$\,\cite{DGH}.

Obviously, if new interactions involving the right-handed (chirality) components of
neutrino were present at low energy, the production of the wrong helicity states
would be different. Therefore, by measuring the wrong helicity contribution it is
also possible to constrain new interactions. Such exercise is carried out in
Ref.\,\cite{ccn:helicity} in the context of the two-Higgs-doublet model (2HDM).
In addition to model dependence, the constraints are not very restrictive in the
context of pion decay.

\subsection{Flavor coupled to chiral oscillations}
\label{subsec:flavor+chiral}

The existence of chiral oscillations, studied in section \ref{subsec:chiral}, remind
us that, when treating the time-evolution of the spinorial mass-eigenstate wave
packets, we have overlooked an important feature. We have {\em completely}
disregarded the chiral nature of charged weak currents and the non-conservation
in time of chirality. The flavor oscillation formula obtained in sections
\ref{subsec:dirac} and \ref{subsec:timeev} were obtained  by considering the
oscillation of general flavor ``particle number'' for general Dirac neutrinos.
However, due to the left handed nature of weak interactions only left-handed
components are produced and detected and then the flavor oscillation probability
could be modified by the parallel effect of chiral oscillations.

It is well known that from the Heisenberg equation, we can immediately determine
whether or not a given observable is a constant of the motion.
If neutrinos have mass, the operator $\gamma^{\5}$ does not commute with the
mass-eigenstate Hamiltonian.
This means that for massive neutrinos chirality is {\em not} a constant of the
motion.
We have already seen that localized states contain, in general,  plane-wave
components of negative and positive frequencies.
As an immediate consequence, the interference between positive and negative
frequencies, responsible for the additional oscillatory term in $\mbox{\sc
Dfo}\bb{t}$\,\eqref{III30}, will also imply an oscillation in the average chirality.
Thus, the use of Dirac equation as evolution equation for neutrino mass-eigenstate
wave packets will lead to an oscillation formula containing both
``flavor-appearance'' (neutrinos of a flavor not present in the original source) and
``chiral-disappearance'' (neutrinos of wrong chirality) probabilities.

As a simple treatment, we can use the evolution kernels of section
\ref{subsec:timeev}.
To incorporate the fact that neutrinos are created and detected as left-handed
fermions into, for example, the conversion probability in
Eq.~\eqref{prob:em:D}, it is sufficient to use initial left-handed wave packets and
replace the kernel of Eq.~\eqref{KD:em:0} by the projected counterpart
\begin{align}
\label{KD:em:L}
LK_{\mu e}^{D\dag}(\bp,t)LK_{\mu e}^D(\bp,t)L
=
\mathscr{P}^D(\bp,t)L
\hs{6em}
\nonumber \\
-\frac{1}{4}\sin^22\theta\bigg(
\frac{m_1}{E_1}\sin E_1t-\frac{m_2}{E_2}\sin E_2t
\bigg)^2L
~,~~
\end{align}
where $\mathscr{P}^D(\bp,t)=K_{\mu e}^{D\dag}(\bp,t)K_{\mu e}^D(\bp,t)$ is the
conversion kernel of Eq.~\eqref{KD:em} and $L=(1-\gamma_5)/2$ is the projector to
left chirality.

For an initial chirally left-handed $\nu_e$, i.\,e.,
$\psi_1(\bx,0)=\psi_2(\bx,0)=\psi_{\nue}(\bx)=L\psi_{\nue}(\bx)$, the
flavor conversion probability can be explicitly written
\small\eq{
\label{quiral:prob:emuLL}
\mathcal{P}(\bs{\nu_{e,\Ll}}\rightarrow\bs{\nu_{\mu,\Ll}};t)=
\mathcal{P}(\bs{\nue}\rightarrow\bs{\numu};t)-
\mn{\frac{1}{4}\!\sin^2\!2\theta}
\Int{3}{\bp}\tilde{\psi}^\dag_\nue\!(\bp)\tilde{\psi}_\nue\!(\bp)
\Big(
\mn{\frac{m_1}{E_1}}\sin E_1t-\mn{\frac{m_2}{E_2}}\sin E_2t
\Big)^2 .
}\normalsize
The spinorial function $\tilde{\psi}_\nue\!(\bp)$ is the Fourier transform of
$\psi_{\nue}(\bx)$.
The conservation of total probability \eqref{prob:cons} no
longer holds because there is a probability loss due to the undetected right
handed component
\small\eq{
\label{KD:em:R}
LK_{\mu e}^{D\dag}RK_{\mu e}^D(\bp,t)L
=
\frac{1}{4}\sin^22\theta\bigg(
\frac{m_1}{E_1}\sin E_1t-\frac{m_2}{E_2}\sin E_2t\!\bigg)^2\!L
~,
}\normalsize
where $R=(1+\gamma_5)/2$ is the projector to right chirality. We can see that
the probability loss \eqref{KD:em:R} is proportional to the ratio $m_n^2/E_n^2$
which is negligible for UR neutrinos.
The probability loss of Eq.~\eqref{KD:em:R} is ultimately induced by the time
evolution of massive Dirac fermions which flips chirality from left to right
according to
\small\eq{
Re^{-iH^D_nt}L=(-i)\frac{m_n}{E_n}\sin(E_nt)\gamma^0 L\,.
}\normalsize
The total probability loss for an initial left-handed electron neutrino
turning into right-handed neutrinos, irrespective of the final flavor, is given
by the kernel
\small\eq{
\label{KD:LR}
LK_{\mu e}^{D\dag}RK_{\mu e}^D(\bp,t)L
+
LK_{ee}^{D\dag}RK_{ee}^D(\bp,t)L
=
\Big[\cos^2\!\theta \Big(\frac{m_1}{E_1}\Big)^2\sin^2E_1t
+\sin^2\!\theta \Big(\frac{m_2}{E_2}\Big)^2\sin^2E_2t\Big]L
~.
}\normalsize

To obtain the unphysical complementary kernels responsible for the conversion
of right-handed component to right-handed and left-handed components, it is
enough to make the substitution $L\leftrightarrow R$ in all formulas.

We can interpret Eq.~\eqref{quiral:prob:emuLL} as a mixture of chiral and flavor
oscillations.
To that end we recall the formalism of section \ref{subsec:Dirac+1},
restricted to one spatial dimension, where the normalized mass-eigenstate wave
functions $\psi_{\1,\2}\bb{z,t}$ are created  at time $t=0$ as a chiral eigenstate
with eigenvalue $-1$. We can calculate
\footnotesize\eqarr{
\mathrm{Re}\left\{\int_{_{-\infty}}^{^{+\infty}}\hspace{-0.4 cm} dz
\psi^{\dagger}_{\ii}\bb{z,t}   \gamma^{\5}  \psi_{\jj}\bb{z,t}\right\}
&=&
\int_{_{-\infty}}^{^{+\infty}}\hspace{-0.2 cm} \frac{d p_{\z}}{2 \pi}
\varphi^{\2}\bb{p_{\z} - p_{\0}}
\times\nonumber\\&&
\left\{\left[ \, 1 - f( p_{\z},m_{\ii},m_{\jj})  - \, \frac{m_{\ii}
m_{\ii}}{E\bb{p_{\z}, m_{\ii}}E\bb{p_{\z}, m_{\jj}}} \, \right]
\cos[\epsilon_{\mi}( p_{\z},m_{\ii},m_{\jj}) \, t ]\right. +
\nonumber\\  & & ~~~~~~
\left[~ \, f( p_{\z},m_{\ii},m_{\jj}) +
\frac{m_{\ii} m_{\jj}}{E\bb{p_{\z}, m_{\ii}}E\bb{p_{\z}, m_{\jj}}} \, \right]
\left.\cos [ \epsilon_{\pl}( p_{\z},m_{\ii},m_{\jj}) \, t] \right\}
}\normalsize
with $i,j \,= \,1,2$.
From the above integral, it is readily seen that an initial $-1$ chiral mass-eigenstate will evolve with time changing its chirality.
Once we know the time evolution of the chiral operator, we are able to construct an
{\em effective} oscillation probability which takes into account both flavor and
chiral conversion effects, i.\,e.,
\small\eqarr{
\mathcal{P}(\bs{\nu_{e,\Ll}}\rightarrow\bs{\nu_{\mu,\Ll}};t)=
 &=&
\int_{_{-\infty}}^{^{+\infty}}\hspace{-0.2 cm} dz
|\psi_{\numu,\Ll} (z, t;\theta)|^{\2}
= \int_{_{-\infty}}^{^{+\infty}}\hspace{-0.2 cm} dz
\psi^{\prime}_{\numu} (z, t;\theta) \, \frac{1-\gamma^{\5}}{2}
 \, \psi_{\numu}(z, t;\theta)
 \nonumber \\
 & = &
 \frac{\sin^{\2}{(2\theta)}}{2}
 \left\{ \, \frac{1}{2}\,\sum_{i=1}^{\2} \,
\left[\int_{_{-\infty}}^{^{+\infty}}\hspace{-0.2 cm} dz
|\psi_{\ii,\Ll} (z, t)|^{\2} \right]
-\mathrm{Re}\left[\int_{_{-\infty}}^{^{+\infty}}\hspace{-0.2 cm} dz
\psi_{\1,\Ll}^{\prime} (z, t) \,
\psi_{\2,\Ll}(z, t) \right]\right\}
\nonumber \\
& = &
\frac{\sin^{\2}{(2\theta)}}{2}
\left[\mbox{\sc Dco}\bb{t} - \mbox{\sc Dfco}\bb{t}\right]
.
\label{V41}
}\normalsize

As done before, the terms $\mbox{\sc Dco}\bb{t}$ and $\mbox{\sc Dfco}\bb{t}$ can be rewritten by using a $ p_{\z}$-integration,
\small\eqarr{
\mbox{\sc Dco}\bb{t}  &=&
\frac{1}{2} \sum_{i=1}^{\2}\,
\int_{_{-\infty}}^{^{+\infty}}\hspace{-0.2 cm} \frac{dp_z}{2\pi}
\varphi^{\2}\bb{p_{\z} - p_{\0}}
\left\{1 - c\bb{p_{\z},m_{\ii},m_{\ii}}\left[1 - \cos[2 E\bb{p_{\z}, m_{\ii}} \,
t]\right] \right\} ~~~~
\label{V41A}}\normalsize
and
\small\eqarr{
\mbox{\sc Dfco}\bb{t} &=&
\int_{_{-\infty}}^{^{+\infty}}
\frac{dp_z}{2 \, \pi}\,
\varphi^{\2}(p_{\z} - p_{\0})
\left\{ \, \left[ \, 1 -
c\bb{p_{\z},m_{\1},m_{\2}} \, \right] \, \cos [
\epsilon_{\mi}\bb{p_{\z},m_{\1},m_{\2}} \, t ] +
\right.\nonumber\\&&~~~~~~~~~~~~~~~~\left.
c\bb{p_{\z},m_{\1},m_{\2}} \, \cos [
\epsilon_{\pl}\bb{p_{\z},m_{\1},m_{\2}} \, t] \right\},~~
\label{V41B}}\normalsize
where
\[
c\bb{p_{\z},m_{\ii},m_{\jj}} =  f\bb{p_{\z},m_{\ii},m_{\jj}} + \frac{m_{\ii}
m_{\jj}}{2 \, E\bb{p_{\z}, m_{\ii}}E\bb{p_{\z},m_{\jj}}}~.
\]
The functions $c\bb{p_{\z},m_{\ii},m_{\jj}}$ have a common maximum at $ p_{\z}=0$ which, in opposition to what happens for $f\bb{p_{\z},m_{\1},m_{\2}}$, do not depend on the mass values, $c_{\mbox{\tiny max}}\bb{0,m_{\ii},m_{\jj}}=\frac{1}{2}$ and, following the same asymptotic behavior of $f\bb{p_{\z},m_{\1},m_{\2}}$, goes rapidly to zero for $ p_{\z}\gg m_{\1,\2}$.
As a consequence of the first order approximation (\ref{III34AA}), we get
\begin{eqnarray*}
c\bb{p_{\z},m_{\ii},m_{\jj}} &\approx&
 \mbox{$\frac{\left[
1 - \mbox{v}_{\ii}\mbox{v}_{\jj} + \mbox{v}_{\ii}\mbox{v}_{\jj} \,
\left(\mbox{v}_{\ii}^{\2} + \mbox{v}_{\jj}^{\2}
-2 \right) \, \frac{ p_{\z}- p_{\0} }{ p_{\0}} \, \right]}{2}$}
\nonumber\\
& \approx & \mbox{$\frac{m_{\1}^{\2}+
m_{\2}^{\2}}{4\, p_{\0}^{\, \2}} \, \left( \, 1 -  2 \, \,
\frac{ p_{\z}- p_{\0} }{ p_{\0}} \, \right)$}
\end{eqnarray*}
where we have considered the UR approximation in the second term.
By substituting $c( p_{\z},m_{\ii},m_{\jj})$ in the above integrations
(\ref{V41A}-\ref{V41B}), and after some algebraic manipulations, we explicitly obtain
\footnotesize\eqarr{
\mbox{\sc Dco}\bb{t} & \approx & \mbox{$1 -
 \frac{m_{\1}^{\2}}{4\, p_{\0}^{\, \2}}  +
 \exp \left[ - \left( \, \frac{ 2 \, p_{\0}^{\, \2} - m_{\1}^{\2}}{\sqrt{2} \,a \,
p_{\0}^{\, \2}} \, t \right)^{\2} \, \right] \,
 \frac{m_{\1}^{\2}}{4\, p_{\0}^{\, \2}}
\left.  \left\{ \cos \left[ \frac{2 \, p_{\0}^{\, \2} +
m_{\1}^{\2}}{ p_{\0}} \, t \right] + \frac{4 \, p_{\0}^{\, \2} - 2
\, m_{\1}^{\2}}{a^{\2}  p_{\0}^{\, \3}} \, t \, \sin \left[ \frac{2
\, p_{\0}^{\, \2} + m_{\1}^{\2}}{ p_{\0}} \, t \right]\, \right\} \,
\right\} \, $}\nonumber \\
&  &\mbox{$ -
 \frac{m_{\2}^{\2}}{4\, p_{\0}^{\, \2}}  +
 \exp \left[ - \left( \, \frac{ 2 \, p_{\0}^{\, \2} - m_{\2}^{\2}}{\sqrt{2} \,a \,
  p_{\0}^{\, \2}} \, t \right)^{\2} \, \right] \,
 \frac{m_{\2}^{\2}}{4\, p_{\0}^{\, \2}} \,  \left.  \left\{ \,  \cos \left[ \frac{2
\, p_{\0}^{\, \2} +
m_{\2}^{\2}}{ p_{\0}} \, t \right] + \frac{4 \, p_{\0}^{\2} - 2
\, m_{\2}^{\2}}{a^{\2}  p_{\0}^{\, \3}} \, t \, \sin \left[ \frac{2
\, p_{\0}^{\, \2} + m_{\2}^{\2}}{ p_{\0}} \, t \right]\, \right\} \,
\right\} \, \,$} ,\\
 \mbox{\sc Dfco}\bb{t} & \approx & \mbox{$\exp \left[ -
\left(  \frac{ \Delta m^{\2} }{2\sqrt{2} a   p_{\0}^{\2}}
 t \right)^{\2}
\right] \left\{
 \left[ 1 -  \frac{m_{\1}^{\2} + m_{\2}^{\2}}{4 p_{\0}^{
\2}}  \right]
 \cos \left[ \frac{\Delta m^{\2}}{2 p_{\0}} t \right]
 + \frac{m_{\1}^{\2} + m_{\2}^{\2}}{4 p_{\0}^{
\2}}
  \frac{\Delta m^{\2} }{a^{\2}  p_{\0}^{ \3}} t  \sin
\left[ \frac{\Delta m^{\2}}{2 p_{\0}}  t \right]
\right\}$} \nonumber \\
 & & \mbox{$ + \exp \left[ - \left(  \frac{ 4  p_{\0}^{ \2} - m_{\1}^{\2} -
m_{\2}^{\2}}{2\sqrt{2} a   p_{\0}^{ \2}}  t \right)^{\2}
\right]  \frac{m_{\1}^{\2} + m_{\2}^{\2}}{4 p_{\0}^{ \2}}
 \left\{
\cos \left[ \frac{4  p_{\0}^{ \2} + m_{\1}^{\2} +
m_{\2}^{\2}}{2   p_{\0}}  t \right] +  \frac{4  p_{\0}^{ \2}
- m_{\1}^{\2} - m_{\2}^{\2}}{a^{\2}  p_{\0}^{ \3}} t  \sin
\left[ \frac{4  p_{\0}^{ \2} + m_{\1}^{\2} + m_{\2}^{\2}}{2
 p_{\0}}  t \right] \right\} $} .
}\normalsize

Under the hypothesis of minimal decoherence between the mass-eigenstate wave
packets ($\Delta v L\ll a$), and for long distances between source and detector
($L \gg a$), i.\,e.,
\[
1 \, \ll \frac{L}{a} \ll \frac{ p_{\0}^{\, \2}}{\Delta \, m^{\2}}
\, , \,
\]
the standard flavor oscillation probability can be reobtained.
In fact,
\small\eqarr{
\mathcal{P}(\bs{\nu_{e,\Ll}}\rightarrow\bs{\nu_{\mu,\Ll}};L)=
& \approx &
\frac{\sin^{\2}{(2\theta)}}{2}
\left[ 1 - \, \frac{m_{\1}^{\2} +
m_{\2}^{\2}}{4\, p_{\0}^{\, \2}} \, \right]
\left\{ \, 1 - \left[ 1 - \left( \, \frac{ \Delta m^{\2}
}{2\sqrt{2} \,a \,  p_{\0}^{\, \2}} \, L \right)^{\2} \right] \,
\cos \left[ \frac{\Delta m^{\2}}{2\, p_{\0}} \, L \right] \,
\right\}
\nonumber \\
& \approx &
\frac{\sin^{\2}{(2\theta)}}{2}
\left\{ \, 1 - \cos \left[ \frac{\Delta
m^{\2}}{2\, p_{\0}} \, L \right] \,
\right\}
=
\sin^{\2}(2 \theta)
\sin^{\2} \left[ \frac{\Delta m^{\2}}{4  p_{\0}} \, L \right]
\,.}\normalsize

\subsection{Flavor conversion formulas for fermionic particles in an external
magnetic field}

After obtaining the time-evolution of the spinorial
mass-eigenstate wave packets in the presence of an external magnetic field in section
\ref{subsec:externalB} and the flavor conversion formula considering left-handed
neutrinos in section \ref{subsec:flavor+chiral}, we intend to investigate here how
flavor conversion is modified when chiral production/detection and external magnetic
fields are taken into account.
Consequently, the chiral nature of charged weak currents and the time-evolution of
the chiral operator must be aggregated to the flavor oscillation formula.

To have a complete description of the flavor conversion mechanism, taking
chiral conversions into account, we must write the complete oscillation
probability formula as
\small\eq{
\mathcal{P}(\bs{\nu_{e,\Ll}}\rightarrow\bs{\nu_{\mu,\Ll}};t) =
\ml{\frac{\sin^{\2}{\!(2\theta)}}{2}}
\left\{\, \mbox{\sc Dco}\bb{t} - \mbox{\sc Dfco}\bb{t} \,\right\}\,,
\label{V1AAA}
}\normalsize
where $\mbox{\sc Dco}\bb{t}$ corresponds exclusively to chiral oscillations.
Such term can be calculated by applying Eq.~(\ref{IVa15A}) to each mass-eigenstate
component yielding immediately
\small\eqarr{
\mbox{\sc Dco}\bb{t}
&=& \frac{1}{2} \int d^{\3}\bx
\left[\psi^{\dagger}_{\1}\bb{t, \bx} \frac{1 -
\gamma_{\5}}{2}\psi_{\1}\bb{t, \bx}
 + \psi^{\dagger}_{\2}\bb{t, \bx}\frac{1 -
\gamma_{\5}}{2}\psi_{\2}\bb{t, \bx}\right] \nonumber\\
&=& \frac{1}{2}\left(1 -
\frac{\langle\gamma_{\5}\rangle_{\1}\bb{t}+\langle\gamma_{\5}\rangle_{\2}\bb{t}}{2}
\right)
\label{V1AABC},
}\normalsize
where $\aver{~}_k$ refers to the average with respect to the mass-eigenstate WPs
$\psi_k$, $k=1,2$. From this point on we also turn back to the 3-dimensional
analysis in order to not constrain the spatial configuration of magnetic fields.
The average values of $\gamma_{\5}$ can be, for instance, explicitly calculated in
terms of the results of the Eqs.~(\ref{IVa15B}) and (\ref{IVa15C}).

Analogously, the mixed flavor and chiral oscillation term can be given by
\small\eq{
\mbox{\sc Dfco}\bb{t} = \frac{1}{2} \int d^{\3}\bx
\left[\psi^{\dagger}_{\1}\bb{t, \bx} \frac{1 -
\gamma_{\5}}{2}\psi_{\2}\bb{t, \bx}
 + \psi^{\dagger}_{\2}\bb{t, \bx}\frac{1 -
\gamma_{\5}}{2}\psi_{\1}\bb{t, \bx}\right],
\label{V1AABB}
}\normalsize
which deserves a more careful calculation.
We shall see how we can explicitly construct the complete oscillation formula
containing both ``flavor-appearance'' (neutrinos of a flavor not present in the
original source) and ``chiral-disappearance'' (neutrinos of wrong chirality)
probabilities for both of the particular cases of external magnetic fields
discriminated in section \ref{subsec:externalB}.
The following results are obtained after some straightforward but extensive
mathematical manipulations where, again, we have imposed an initial constraint which
establishes that the normalizable mass-eigenstate wave functions $\psi_{\1,\2}\bb{t,
\bx}$ are created  at time $t=0$ as a negative chiral eigenstate
($w_{\1,\2}^{\dagger}\gamma_{\5}w_{\1,\2} = \ms{-}1$).
All the subsequent calculations do not depend on the gamma matrix representation.

Considering the first case of section \ref{subsec:externalB}, where the
propagating momentum {\boldmath$p$} is parallel to the magnetic field {\boldmath$B$},
we can calculate
\footnotesize\eqarr{
\mbox{\sc Dfco}\bb{t} &=& \frac{1}{2}\int\frac{d^{\3} \bs{p}}{(2 \pi)^{\3}} \,
\varphi^{\2}\bb{\bs{p}}\sum_{\s\,\ig\,\1,\2}\left\{\,w^{\dagger}
\left(\frac{1-(\mfn{-}1)^{\s}\bs{\Sigma}\ponto\hat{\bs{a}}}{2}
\right)w\right.\times \nonumber\\
&&\left.\left[
\left(1+\frac{\bs{p}^{\2}}{E^{(\1)}_{\s}E^{(\2)}_{\s}}\right)
\cos[\left(E^{(\1)}_{\s}-E^{(\2)}_{\s}\right) \, t] +
\left(1-\frac{\bs{p}^{\2}}{E^{(\1)}_{\s}E^{(\2)}_{\s}}\right)
\cos[\left(E^{(\1)}_{\s}+E^{(\2)}_{\s}\right) \, t]\right]
\right\}
\label{V2AABB},
}\normalsize
where we have used the correspondence $\varphi^{\2}\bb{\bs{p}} \equiv
\varphi\bb{\bs{p}-\bs{p}_{\1}}
\varphi\bb{\bs{p}-\bs{p}_{\2}}$ and $E^{(\ii)}_{\s} =
\sqrt{\bs{p}^{\2}+ m^{(\ii)}_{\s}}$ with $i = 1,2$ corresponding to the mass indices.
In fact, in the UR limit, and for relatively weak magnetic fields
($|\mbox{\boldmath$a$}|\ll|\bs{p}|$), the contribution due to the very rapid
oscillations
does not introduce relevant modifications to the flavor conversion formula,
in analogy to the case of {\em purely} chiral oscillations.
On the other hand, by taking the NR limit, with a momentum distribution
sharply peaked around $|\bs{p}_{\ii}|\ll m_{\ii}$, the complete oscillation
probability formula can be written as
\small\eq{
\mathcal{P}(\bs{\nu_{e,\Ll}}\rightarrow\bs{\nu_{\mu,\Ll}};t) =
\ml{\frac{\sin^{\2}{(2\theta)}}{4}}
\sum_{\s\,\ig\,\1,\2}\left\{\,w^{\dagger}\left(\frac{1-\mfn{(-1)}^{\s}\bs{\Sigma}
\ponto\hat { \bs{a}}}{2}\right)w\,
\left(\cos{[m^{(\1)}_{\s} t]} - \cos{[m^{(\2)}_{\s} t]}\right)^{\2}\right\}
\,.
\label{V1AAAA}
}\normalsize
Equation \eqref{V1AAAA} introduces a completely different
pattern for flavor/chiral oscillations, despite of not being phenomenologically
verifiable.

On the other hand, a more interesting interpretation is provided when we
analyze the second case, where the propagating momentum {\boldmath$p$} is orthogonal
to the magnetic field {\boldmath$B$}.
The effects of the external magnetic field can be distinguished from the mass
interference term in the flavor/chiral oscillation formula in the sense that we
can write the complete expression for $\mbox{\sc Dfco}\bb{t}$ as
\small
\small\eqarr{
\lefteqn{\mbox{\sc Dfco}\bb{t} =\frac{1}{2} \int\frac{d^{\3} \bs{p}}{(2 \pi)^{\3}} \,
\varphi^{\2}\bb{\bs{p}}
\left\{\left(\cos[|\mbox{\boldmath$a$}^{(\1)}|\, t]
\cos[|\mbox{\boldmath$a$}^{(\2)}|\, t] \right)\right.}\nonumber\\
&&\times
\left[\left(1+\frac{\bs{p}^{\2} +
m^{(\1)}m^{(\2)}}{\varepsilon^{(\1)}_{\0}\varepsilon^{(\2)}_{\0}}\right)
\cos[\left(\varepsilon^{(\1)}_{\0}-\varepsilon^{(\2)}_{\0}\right) \, t] +
\left(1-\frac{\bs{p}^{\2} +
m^{(\1)}m^{(\2)}}{\varepsilon^{(\1)}_{\0}\varepsilon^{(\2)}_{\0}}\right)
\cos[\left(\varepsilon^{(\1)}_{\0}+\varepsilon^{(\2)}_{\0}\right) \, t]\right]\nonumber\\
&& ~~~~~~~~~~~~~~~~~~~~+
\frac{m^{(\1)}m^{(\2)}}{\varepsilon^{(\1)}_{\0}\varepsilon^{(\2)}_{\0}}\left[
\cos[\left(\varepsilon^{(\1)}_{\0}+\varepsilon^{(\2)}_{\0}\right) \, t]
\cos[\left(|\mbox{\boldmath$a$}^{(\1)}|-|\mbox{\boldmath$a$}^{(\2)}|\right)\, t]\right.
\nonumber\\
&&\left.\left.~~~~~~~~~~~~~~~~~~~~~~~~~~~~~~~~~~~~~~~~~~~~~~~~~~~~~ -
\cos[\left(\varepsilon^{(\1)}_{\0}-\varepsilon^{(\2)}_{\0}\right) \, t]
\cos[\left(|\mbox{\boldmath$a$}^{(\1)}|+|\mbox{\boldmath$a$}^{(\2)}|\right)\, t]
\right]
\right\},~~
\label{V3AABB}
}\normalsize
where we have used the correspondence of $\mbox{\boldmath$a$}^{(\ii)}$ and $\varepsilon^{(\ii)}$ with $m^{(\ii)}$.
Again, the UR limit reduces the impact of the modifications to residual effects which are difficultly detectable by experiments.
By taking the NR limit and following the same procedure for obtaining Eq.~(\ref{V1AAAA}), the complete oscillation probability formula can be written as
\small\eqarr{
\mathcal{P}(\bs{\nu_{e,\Ll}}\rightarrow\bs{\nu_{\mu,\Ll}};t)
&=&
\ml{\frac{\sin^{\2}{(2\theta)}}{4}}
\left(\cos^{\2}{[m^{(\1)} t]} + \cos^{\2}{[m^{(\2)} t]}\right.\nonumber\\
&&\left.~~~~~~~~~~ - 2\cos{[(|\mbox{\boldmath$a$}^{(\1)}|-|\mbox{\boldmath$a$}^{(\2)}|)t]} \cos{[m^{(\1)} t]}\cos{[m^{(\2)} t]} \right).
\label{V1AAAAA}
}\normalsize

If we had chosen a wave packet exclusively composed by positive frequency plane-wave
solutions, the high-frequency oscillation term would automatically vanish.
Such a peculiar oscillating behavior is similar to the quoted ZBW and
reinforces the argument that, when constructing Dirac wave packets, we cannot
simply forget the contributions due to negative frequency components.

Turning back to the foundations of the neutrino oscillation problem, we know that a
more sophisticated field-theoretical treatment is required.
Derivations of the oscillation formula resorting to field-theoretical methods are
not very popular and the existing quantum field computations of the oscillation
formula do not agree in all respects \cite{Beu03}.
The  Blasone and Vitiello model \cite{Bla95,Bla03} to neutrino/particle mixing
and oscillations seems to be an interesting trying to this aim.
They have attempt to define a Fock space of weak eigenstates to derive a
nonperturbative oscillation formula.
With Dirac wave packets, the flavor conversion formula can be reproduced
\cite{Ber04B} with the same mathematical structure.

In the context where we have intended to explore the Dirac formalism, we have
pointed out that a more satisfactory description for understanding chiral
oscillations of fermionic (spin one-half) particles like neutrinos requires the use
of the Dirac equation as evolution equation for the mass-eigenstates.
Within such a framework, we have introduced the non-minimal coupling of the
neutrino magnetic moment to an electromagnetic field in an external interacting
process \cite{Vol81}.
Although clear experimental evidences are still missing, we have reinforced the
idea that the spinorial form and the interference between positive and negative
frequency components of the mass-eigenstate wave packets can introduce small
modifications to the {\em standard} conversion formulas when chiral oscillations are
taken into account. More interestingly, when neutrinos interact with very
strong magnetic fields, completely different oscillation patterns may emerge.

\section{Inclusion of quantized fields}
\label{sec:qft}

This section considers the inclusion of some ingredients of
quantum field theory (QFT) to the description of the oscillation phenomenon.
There are divergent, and sometimes inconsistent, approaches in the literature.
We try to identify some common aspects to our previous relativistic first-quantized
treatments.

Four aspects will be studied in detail: (i) Firstly, we want to analyze if
considering quantum fields naturally eliminates the rapid oscillation effects coming
from the interference between positive and negative frequency states. (ii) Second, we
would like to know how EWP approaches are related to the IWP approaches studied so
far. (iii) Third, we want to relate the distinctive approach of constructing a flavor
Fock space\,\cite{Bla95} with the IWP approaches considered. (iv) Fourth, we intend
to
investigate the details of the neutrino flavor state that is produced through decays
when weak interactions, parent particle localization and finite decay width are
considered.

The first aspect (i) is considered in section \ref{subsec:simpleQFT} when
a simple second quantized description of flavor oscillations for Dirac
fermions is devised, based on free quantum field theory.
It is commonplace that going from first quantized Dirac theory to
the free quantum field theory of spin $1/2$ fermions eliminates the interference
between positive and negative frequency states by introducing a different notion of
quantum states through the construction of the $n-$particle state
space (Fock space)\,\cite{Zub80}. Such construction allows a meaningful
interpretation of energy which is positive for both positive or negative
frequency states and the notion of particle and antiparticle states
becomes meaningful as well (for Dirac fermions).
When interactions are considered through perturbation theory, the sectors that would
correspond to positive and negative frequencies in first quantized theories remain
separated by construction.
It is then natural to ask what happens if we quantize the free fields associated to
the mass-eigenstates that compose the flavor states.
We will show that such procedure indeed results in naturally eliminating the rapid
oscillation effects.
Moreover, such a description guarantees only particle or antiparticle propagation,
keeping the nice property of giving normalized oscillation probabilities, like the
first quantized examples treated in previous sections.
We will also recover the initial flavor violation obtained in section
\ref{subsec:IFV:D} without choice.

To treat the aspect (ii), we compare in section \ref{subsec:EWP} the different
descriptions of neutrino propagation subjected to flavor mixing from the point of
view of the propagators of the theories.
In first quantized theories, the role of the  propagator is played by the evolution
kernels introduced in section \ref{subsec:timeev}.
The latter are compared to the mixed Feynman propagators used in the EWP
approaches\,\cite{Beu03} to describe neutrino propagation.

Aspect (iii) is treated in section \ref{subsec:BV} where we briefly report about the
Blasone-Vitiello approach to flavor mixing, giving emphasis to the mathematical
similarities of the oscillation formula with the first quantized Dirac description of
section \ref{subsec:dirac}.

Finally, aspect (iv) is considered in section \ref{subsec:intrinsic} where we
calculate, within quantum field theory, the probability of detecting neutrino flavor
states produced from pion decay including explicitly the weak interactions
responsible for the decay, the pion localization and the pion finite decay width.
Within this setting, it is possible to quantify the amount of intrinsic flavor
violation that arises when neutrinos are created and its relation with the decay
width and pion localization. Such amount is still negligible but much larger than
other indirect lepton flavor violation effects, such as $\mu\rightarrow e\gamma$,
where neutrino mixing only contributes at loop order.

\subsection{Simple second quantized theory}
\label{subsec:simpleQFT}

Considering that only real neutrinos or antineutrinos (one of them exclusively)
should travel from production to detection, the possibility to use the free second
quantized theory for spin 1/2 fermions to describe flavor oscillations is now
investigated.

To accomplish the task of calculating oscillation probabilities in QFT we have to define the neutrino states that are produced and detected through weak interactions.
Firstly, we define the shorthand for the combination of fields appearing in the weak
effective charged-current Lagrangian \eqref{LW}
\small\eq{
\label{nu:f:field}
\nu_\alpha(x)\equiv
U_{\alpha i}\nu_i(x)~,~\alpha=e,\mu\,.
}\normalsize
We will restrict the problem to two flavor families and use the matrix $U$ as the same as in Eq.~\eqref{mixing}.
The mass eigenfields $\nu_i(x)$, $i=1,2$, are the physical fields for which the
mass-eigenstates $\ket{\nu_i(\bp)}$ are well defined asymptotic states.
The free fields $\nu_i(x)$ can be expanded in terms of creation and annihilation
operators (see appendix \ref{app:def}) and the projection to the one-particle space
defines the mass wave function \small\eq{ \label{wf:12}
\psi_{\nu_i}(\bx;g_i)=\bra{0}\nu_i(\bx)\ket{\nu_i\!:\!g_i}
\equiv
\underset{s}{\textstyle \sum}
\Int{3}{\bp}\frac{g_i^s(\bp)}{\sqrt{2E_i}}u_i^s(\bx;\bp)
~,~~i=1,2,
}\normalsize
where
\small\eq{
\label{s2q:nui}
\ket{\nu_i\!:\!g_i}
\equiv
\underset{s}{\textstyle \sum}
\Int{3}{\bp}g_i^s(\bp)\ket{\nu_i(\bp,s)}
~.
}\normalsize

Since the creation operators for neutrinos (antineutrinos) can be written in terms of
the free fields $\bar{\nu}_i(x)$ ($\nu_i(x)$), we can define the flavor states as the
superpositions of mass-eigenstates
\eqarr{
\label{nu:f:qft}
\ket{\nu_\alpha\!:\!\{g\}}
&\equiv&
U^*_{\alpha i}
\ket{\nu_i\!:\!g_i}
\cr
\ket{\bar{\nu}_\alpha\!:\!\{g\}}
&\equiv&
U_{\alpha i}
\ket{\bar{\nu}_i\!:\!g_i}
~.
}\normalsize
The details of creation are encoded in the functions $g_i$.

We can also define
\eqarr{
\label{wf:em}
\psi_{\nu_\alpha\nu_e}(x;\{g\})
&\equiv &
\bra{0}\nu_e(x)\ket{\nu_\alpha\!:\!\{g\}}
\cr
&=&
U_{ei}U^*_{\alpha i}\psi_{\nu_i}(x;g_i)
~,
}\normalsize
where $\psi_{\nu_i}(x)$ are then mass wave functions defined in Eq.~\eqref{wf:12}. We can see from Eq.~\eqref{wf:em} that if $\psi_{\nu_1}(\bx,t)=\psi_{\nu_2}(\bx,t)=\psi(\bx)$, for a given time $t$, $\psi_{\nu_e\nu_e}(\bx,t)=\psi(\bx)$ and $\psi_{\nu_\mu\nu_e}(\bx,t)=0$ due to the unitarity of the mixing matrix.

Although this approach does not rely on flavor Fock spaces and Bogoliubov transformations, one considers the following  observables to quantify flavor oscillation \cite{Bla03B}: the flavor charges, which are defined as
\small\eq{
\label{Q:f}
Q_\alpha(t)=\Int{3}{\bx}:\nu_\alpha^\dag(\bx,t)\nu_\alpha(\bx,t):
~,~~\alpha=e,\mu\,,
}\normalsize
where $:\;:$ denotes normal ordering.
Note that the $Q_e(t)+Q_\mu(t)=Q$ is conserved \cite{Bla95}, the two flavor charges are compatible for equal times, i.\,e., $[Q_e(t),Q_\mu(t)]=0$, and $\bra{\nu\!:\!\{g\}}Q\ket{\nu\!:\nolinebreak\!\{g\}}=\pm\braket{\nu\!:\!\{g\}} {\nu\!:\!\{g\}}$ for any particle state (+) or antiparticle state (-).
Notice that in the second quantized version the charges can acquire negative values,
despite the fermion probability density in first quantization is a positive definite
quantity. The conservation of total charge guarantees the conservation of total
probability \eqref{prob:cons}.

With the flavor charges defined, we can calculate, for example, the conversion probability
\eqarr{
\label{prob:em:qft}
\mathcal{P}(\ms{\bs{\nu_e}\!\rightarrow\!\bs{\nu_\mu}};t)
&\equiv &
\bra{\nu_e\!:\!\{g\}}Q_\mu(t)\ket{\nu_e\!:\!\{g\}}
\\
&=&
U_{\mu i}U^*_{\mu j}U_{e j}U^*_{e i}
\Int{3}{\bp}
e^{-i(E_i-E_j)t}
\tilde{\psi}_{\nu_j}^\dag(\bp;g_j)
\tilde{\psi}_{\nu_i}(\bp;g_i)
\label{prob:em:qft1}
~,
}\normalsize
where the neutrino wave functions $\psi_{\nu_i}$ are defined in terms of the function $g_i(\bp)$ in Eq.~\eqref{wf:12}.
Eq.~ \eqref{prob:em:qft1} is exactly equal to Eq.~\eqref{Fdef:prob}, if Eq.~\eqref{Fdef:Psi:0} is used with $\alpha=\theta$.
If we could equate the two mass wavefunctions in momentum space $\tilde{\psi}_{\nu_1}(\bp;g_1)= \tilde{\psi}_{\nu_2}(\bp;g_2)= \tilde{\psi}_{\nu_e}(\bp)$ we would obtain, from Eq.\;\eqref{prob:em:qft1}, the standard two family conversion probability \eqref{prob:em:S}, with $|\tilde{\varphi}_{\nue}(\bp)|^2$ replaced by $\tilde{\psi}_{\nu_e}^\dag(\bp)\tilde{\psi}_{\nu_e}(\bp)$.
However, the equality can not hold as proved in section\,\ref{subsec:IFV:D}: two wavefunctions with only positive energy components with respect to two bases characterized by different masses can not be equal.
In fact, Eq.~\eqref{prob:em:qft1} is equivalent to Eq.~\eqref{Fdef:Psi:0} with
Eq.~\eqref{Fdef:cosXi:+}, and the states \eqref{s2q:nui} adopted in this second
quantized formulation automatically accomplishes \textbf{cond.\,B}: there is no rapid
oscillations.
The flavor conversion probability is given by Eq.~\eqref{Fdef:prob:0} with Eq.~\eqref{Fdef:cosXi:g+}
\small\eq{
\label{s2q:prob:+}
\mathcal{P}(\ms{\bs{\nu_e}\!\rightarrow\!\bs{\nu_\mu}};t)=
\mn{\frac{1}{2}}\sin^2\!2\theta
\Big\{1-
\mathrm{Re}\underset{s}{\textstyle\sum}\!\Int{3}{\bp}
g_1^*(\bp,s)g_2(\bp,s)e^{i\Delta E(\bp)t}[1-2f_2(\bp)]
\Big\}\,,
}\normalsize
where $g_i(\bp,s)$ are the probability amplitudes in momentum $\bp$ and spin $s$.
One can see that the probability \eqref{s2q:prob:+} is never null and one can not
avoid initial flavor violation but one can still adopt equal momentum
distributions as in Eq.~\eqref{g1=g2}: $g_1(\bp,s)=g_2(\bp,s)=g(\bp,s)$.
In this case we have initial flavor violation given by Eq.~\eqref{Fdef:fvio:+2}:
\small\eq{
\label{s2q:prob:+0}
\mathcal{P}(\ms{\bs{\nu_e}\!\rightarrow\!\bs{\nu_\mu}};0)=
\sin^22\theta
\Int{3}{\bp}f_2(\bp)|g(\bp)|^2
~,
}\normalsize
where $|g(\bp)|^2=\sum_s|g(\bp,s)|$ is the spin-independent momentum distribution.
The conversion probability \eqref{s2q:prob:+} yields
\small\eq{
\label{s2q:prob:+2}
\mathcal{P}(\ms{\bs{\nue}\!\rightarrow\!\bs{\numu}};t)
=
\mn{\frac{1}{2}}\sin^2\!2\theta
\Big\{1-
\Int{3}{\bp}|g(\bp)|^2\cos[\Delta E(\bp)t][1-2f_2(\bp)]
\Big\}\,.
}\normalsize
An approximate version of the formulas \eqref{s2q:prob:+0} and
\eqref{s2q:prob:+2} was previously presented in Ref.\,\cite{ccn:no12}.

The conversion probability for antineutrinos $\bar{\nu}_e\rightarrow\bar{\nu}_\mu$
is exactly the same as Eq.~\eqref{s2q:prob:+} by using Eq.\,\eqref{ket:nub:t}.

The formulas obtained in this second quantized version do exhibit the
interference terms between positive and negative energies like in
Eq.\;\eqref{prob:em:D2}.  Such interference terms are absent because the
possible mixed terms like $b_2(\bp)a^\dag_1(\bp)\ket{0}$ are null for an initial
flavor state superposition that contains only particle states (or only
antiparticles states).

\subsection{Connection with the EWP approach}
\label{subsec:EWP}

Now we try to establish a correspondence between our results and the quantum
field theory (QFT) treatments, notably the EWP approaches.
These approaches follow the idea that the oscillating particle cannot be treated in
isolation\,\cite{Ric93,Giu93,Giu02}, i.\,e., the oscillation process must be considered
globally: the oscillating states become
intermediate states, not directly observed, which propagate between a {\em
source} ($A$) and a {\em detector} ($B$).
This idea can be implemented in QFT when the intermediate oscillating states
are represented by internal lines of Feynman diagrams and the interacting particles
at source/detector are described by external wave packets \cite{Giu93,Beu03}.

Let us consider the weak flavor-changing processes occurring through
the intermediate propagation of a neutrino,
\eqarr{
\label{external:2}
 A+l_\alpha \rightarrow A' + \nu_{\alpha}
 ~~&\text{(oscillation)}&~~
 \nu_{\beta} + B \rightarrow l_\beta + B'
 \,,\\
\label{IIIA15}
 A \rightarrow A' + \bar{l}_\alpha + \nu_{\alpha}
 ~~&\text{(oscillation)}&~~
 \nu_{\beta} + B \rightarrow l_\beta + B'
 \,,
}\normalsize
where $A$ and $A'$ ($B$ and $B'$) are respectively the initial and final production
(detection) particles.
The process \eqref{external:2} may describe the process\,\cite{Dolgov,Giu02}
where a charged lepton $l_\alpha$ hit a nucleus A turning it into another nucleus
A$'$ with emission of a neutrino (this process happens around $x_A=(t_A,\bx_A)$).
The process
\eqref{IIIA15} describes the similar process where particle $A$ decays into $A'$
with emission of $\bar{l}_\alpha,\nu_\alpha$; for processes such as pion decay,
$A'$ is absent.
Subsequently the neutrino travels a long distance and hit a nucleus B which
transforms into B$'$ emitting a lepton $l_\beta $ (this process happens around
$x_B=(t_B,\bx_B)$). The whole processes for Eqs.\,\eqref{external:2} and
\eqref{IIIA15} look like
\eqarr{
\label{process:4}
(l_\alpha +A)+(B)&\rightarrow&\hspace{4.6ex} (A')+(B'+l_\beta)
\\
\label{process:3}
(A)+(B)&\rightarrow& (\bar{l}_\alpha+A')+(B'+l_\beta)
}\normalsize
with transition amplitudes given generically by
\eqarr{
\label{A:S->D:2}
\mathcal{A}_{\alpha\beta}&=&\bra{A',B',l_\beta}S
\ket{A(x_A),l_\alpha(x_A),B(x_B)}
\,,\\
\label{A:S->D:1}
\mathcal{A}_{\alpha\beta}&=&
\bra{A',B',l_\beta,\bar{l}_\alpha}S
\ket{A(x_A),B(x_B)}\,,
}\normalsize
where $S$ is the scattering matrix.
The initial states are localized\,\cite{Beu03} while the final states might be
localized states\,\cite{Beu03} or momentum eigenstates\,\cite{Dolgov}.

After some mathematical manipulations \cite{Beu03}, both amplitudes in
Eqs.\,\eqref{A:S->D:2} and \eqref{A:S->D:1} can be
represented by the integral
\small\eq{
\mathcal{A}_{\alpha\beta} = \int{\frac{d^4k}{(2\pi)^{\4}}}\,
\bar{\Psi}_{B}(k)G_{\beta\alpha}(k)\Psi_{A}(k)\,
\exp{[-ik\ponto(x_B - x_A)]}\,,
\label{IIIA17}
}\normalsize
where the spinorial function $\Psi_{A}(k)$ represents the {\em overlap} of
the incoming wave packets at the source while $\bar{\Psi}_{B}(k)$ represents the {\em
overlap} of outgoing wave packets at the detector. Notice that Eq.~\eqref{IIIA17} is
an adaptation for propagating Dirac fermions of the formula obtained in
Ref.\,\cite{Beu03} for scalar neutrinos and scalar interactions.
The {\em Green} function in momentum space, $G_{\beta\alpha}(k)$, represents the
mixed neutrino propagator which carries the information of the oscillation process:
\small\eq{
\label{Gab}
G_{\beta\alpha}(k)=\sum_jU_{\beta j}G_{j}(k)U^\dag_{j\alpha}\,,
}\normalsize
where $G_{j}(k)$ represents the propagators for each neutrino $\nu_j$.
Thus Eq.~\eqref{IIIA17} can be also rewritten as the superposition of amplitudes,
\small\eq{
\label{Aab->Ai}
\mathcal{A}_{\beta\alpha}=\sum_jU_{\beta j}\mathcal{A}_jU^\dag_{j\alpha}\,.
}\normalsize
The overlap functions are independent of production/detection times
$\{t_A,t_B\}$, and production/detection positions $\{\bx_A,\bx_B\}$, but depends on
the directions of incoming and outgoing momenta.
In a certain way, the physical conditions of source and detector, in terms of time
and space intervals, are better defined in this framework than in the {\em
intermediate} wave packet framework. Anyway, to understand the oscillation process we
must turn back to the definition of mixing in quantum mechanics. It is similar in
field theory, except that it applies to fields, not to physical states. This
difference allows us to bypass the problems arising in the definition of flavor and
mass bases \cite{Beu03}.

Let us analyze some Green functions $G_{\alpha\beta}(k)$.
The main improvement of the covariant approaches developed in
section\;\ref{sec:relativistic} is that the propagation kernels governed by Dirac and
Sakata-Taketani Hamiltonians are causal, i.\,e., are null for spacelike separations
(see Eqs.~\eqref{S:D} and \eqref{S:ST} and Refs. \cite{thaller,Zub80,Roman}). On the
contrary, the kernel of spinless particles restricted only to positive energies is
not null for spacelike intervals \cite{thaller}. From the point of view of
relativistic classical field theories, a causal kernel guarantees, by the Cauchy
theorem, the causal connection between the wave-function in two spacelike surfaces at
different times \cite{Roman}.

To compare the IWP and EWP approaches it
is useful to rewrite the Dirac evolution kernel for a fermion of mass $m_j$,
present in Eq.~\eqref{f:evol:D} of section\,\ref{sec:relativistic}, in the
form \cite{Zub80}
\eqarr{ \label{S}	
K_j^D(x-y)&=&
\underset{s}{\textstyle \sum}\int\!
\frac{d^3\!\bp\,}{2E_n}
[u^s_j(x;\!\bp)\bar{u}^s_j(y;\bp)+v^s_j(x;\!\bp)\bar{v}^s_j(y;\bp)]\gamma_0
\cr
&\equiv&iS(x-y;m_j)\gamma_0
~,~~j=1,2,
}\normalsize
where $(x-y)^0=t, (x-y)^i=(\bx-\bx')^i$ when compared to the notation
of Eq.~\eqref{f:evol:D}.
The spinorial functions $u,v$, are the free solutions of the Dirac equation and
they are explicitly defined in appendix \ref{app:def}. (More familiar forms
for the function $S$ are also shown in appendix \ref{app:def}.)
Clearly the function $iS(x-y;m_j)=\bra{0}\{\psi_j(x),\bar{\psi}_j(y)\}\ket{0}$
satisfies the homogeneous Dirac equation with mass $m_j$ \eqref{nu:n:eq} and it
is known to be null for spacelike intervals $(x-y)^2<0$\linebreak ensuring
microcausality\,\cite{thaller,Roman}.

In contrast, the Feynman propagator $iS_F(x-y)$ appears in QFT. It is a
Green function for the inhomogeneous Dirac equation obeying particular
boundary conditions. The EWP approaches use this Green function for the
propagation of virtual neutrinos.
In fact we have for the Green function in Eq.~\eqref{Gab},
\small\eq{
G_{j}(k)=iS_F(k;m_j)=\frac{i}{\sla{k}-m_j+i\epsilon}\,.
}\normalsize
To directly compare the Feynman propagator to
the kernel in Eq.~\eqref{S} we can write $iS_F$ in the form
\eqarr{
\label{S:F}
iS_F(x-y;m_j)&\equiv &
\bra{0}T(\psi_j(x),\bar{\psi}_j(y))\ket{0}
\cr
&=&
\underset{s}{\textstyle \sum}\int\!
\frac{d^3\!\bp}{2E_n}\,
[u^s_j(x;\!\bp)\bar{u}^s_j(y;\bp)\theta(x_0-y_0)
\cr
&&\phantom{\sum_s\int\!\frac{d^3\!\bp}{2E_n}\,}\hs{-1.9ex}
- v^s_j(x;\!\bp)\bar{v}^s_j(y;\bp)\theta(y_0-x_0)]
~.
}\normalsize
Although the function $S_F$ is called causal propagator, it is not null for
spacelike intervals, and it naturally arises in QFT when interactions are present
and treated in a covariant fashion.
Equation~\eqref{S:F} shows that the propagator $S_F$
describes positive energy states propagating forward in time and negative energy
states propagating backward in time \cite{Zub80}. At a first glance,
both neutrino and antineutrino parts of Eq.~\eqref{S:F} seem to contribute to
the space-time integrations present in covariant perturbation theory, as
neutrino-antineutrino contributions in Eq.~\eqref{S} have led to
Eq.~\eqref{prob:em:D2}.

However, it was shown that in an EWP approach, only on-shell contributions
contribute for large separations between production and detection (see the review
in Ref.\,\cite{Beu03}). Moreover, although both neutrino and antineutrino
parts may contribute as intermediate neutrinos for certain
situations\,\cite{ccn:no12}, the wrong state contributions are very much
suppressed. Therefore, the intermediate neutrino states propagating in the EWP
approach are dominated by real and on-shell neutrino (antineutrino) states.

Let us consider the process \eqref{external:2} and analyze more carefully some
calculations using, instead of the scalar interaction\,\cite{Dolgov,Beu03}, the
effective charged-current weak Lagrangian
\eqarr{
\label{LW}
\mathscr{L}_W&=&
G\sum_{i,\alpha=1}^{N=3}
\,[
\bar{l}_\alpha (x)\gamma^\mu L\,U_{\alpha i}\nu_i(x)J_\mu(x)
\cr&&\hs{3em}
+\
\bar{\nu}_i(x)U^*_{\alpha i}\gamma^\mu L\, l_\alpha (x)J_\mu^\dag(x)
\,]
\\
&=&\mathscr{L}_1+\mathscr{L}_1^\dag
~,
}\normalsize
where $G=\sqrt{2}G_F$ and $J_\mu$ is the sum of any effective leptonic or hadronic
current. The Lagrangian \eqref{LW} is written only in terms of physical
mass-eigenstate fields, which coincides with flavor state fields only for the
charged leptons: $l_1(x)\equiv e(x), l_2(x)\equiv \mu(x), \ldots$\,.

The lowest order nonzero contribution of the scattering matrix $S$ to
Eq.\;\eqref{A:S->D:2} is second order in the Lagrangian \eqref{LW}. More explicitly,
the term that contributes to the amplitude
\eqref{A:S->D:2} comes from
\eqarr{
S^{(2)}&=&\frac{i^2}{2}T\aver{\lag_W}^2
=-\frac{1}{2}T\aver{\lag_1+\lag_1^\dag}^2
\\
\label{S2:1}
&\sim&-T\aver{\lag_1}\aver{\lag_1^\dag}
\\
\label{S2:2}
&\sim&
-G^2\int d^4x d^4y \sum_{\beta\alpha}\lag_{\beta\alpha}(x,y)
~,
}\normalsize
where $\aver{~}$ stands for space-time integration and
\small\eq{
\lag_{\beta\alpha}(x,y)
\equiv
\sum_i
\!:\!J_\mu(x)\bar{l}_\beta (x)\gamma^\mu L U_{\beta i}\,iS_{F}(x-y;m_i)U^*_{\alpha
i} \gamma^\nu L l_\alpha (y)J_\nu^\dag(y)\!:
~.
}\normalsize
In Eq.~\eqref{S2:1} we kept only the mixed product and in Eq.~\eqref{S2:2} we
kept from all possible terms in Wick expansion\,\cite{Zub80} only
the term responsible for the transition of interest.

We can write in explicit form the production and detection spinorial wave functions
in Eq.~\eqref{IIIA17} as\,\cite{Beu03}
\eqarr{
\bar{\Psi}_{B}(k)&=&
iG\Int{4}{y}e^{-ik\ponto(y-x_B)}\,
\bra{B',l_\beta}J_\mu(y)\bar{l}_{\beta}(y)\gamma^\mu L\ket{B}\,,
\\
\Psi_{A}(k)&=&
iG\Int{4}{x}e^{ik\ponto(x-x_A)}\,
\bra{A'}\gamma^\nu
Ll_\alpha(x)J^\dag_\nu(x)\ket{A,l_\alpha}\,.
}\normalsize

Since the intermediate neutrinos in the process \eqref{external:2} can be
considered real on-shell neutrinos in the EWP approach for large separations between
production and detection, both processes in $x_A$ and $x_B$ should be considered as
real scattering processes with real neutrinos involved.
These informations permit us to rewrite Eq.~\eqref{Aab->Ai} in a slightly
different form
\eqarr{
\sum_iU_{\beta i}U^*_{\alpha i}\mcal{A}_i&=&
\sum_i\int \frac{d^3\!\bp}{2E_i\bb{\bp}}
\int d^4y\, \bra{B',l_\beta}\lag_1(y)e^{i(P-p_i)\ponto x_B}\ket{B,\nu_i\bb{\bp}}
\cr&&\hs{6em}
\int d^4x\,\theta(y-x)
\bra{A',\nu_i\bb{\bp}}\lag_1^\dag(x)e^{iP\ponto x_A}\ket{A,l_\alpha}
\label{A:7}
~,~~
}\normalsize
where $P=(H,\mathbf{P})$ is the energy-momentum operator.
A change of notation were made here: in Eq.~\eqref{A:7} the states
$\ket{B}$ and $\ket{A,l_\alpha }$ are centered around the origin while in
Eq.~\eqref{A:S->D:2} they are respectively centered around $x_B$ and
$x_A$; the translation is explicitly performed by the translation operator
$e^{iP\ponto x}$. Additionally, the step function $\theta(y-x)$ is necessary to
ensure that the contributions of points $y$ around $x_B$ should always be after
the contributions of points $x$ around $x_A$.

Equation \eqref{A:7} shows us the amplitude of the entire process from
production to detection in ``decomposed'' form (apart from the step
function in time): the amplitude of production process multiplied by the
amplitude of detection process summed over all possible intermediate real
neutrinos of different masses $m_i$ and momentum $\bp$. (The sum over spins are
implicit.) Thus the EWP approach is not conceptually different from IWP approaches
when the three amplitudes of production, propagation and detection of neutrinos can
be factored out independently from each
other\,\cite{Beu03,Cardall.00,akhmedov:paradoxes}.
For those cases, for most of the
situations, IWP approaches provide the same results as EWP approaches if the
localization aspects can be transposed as inputs to neutrino wave packets.
A situation where the decomposition can not be performed simply and the
localization aspects have to be explicitly taken into account is exemplified by the
unusual case of Mossbauer neutrinos recoilessly emitted and
detected from bound state electrons\,\cite{lindner:mossbauer}.

\subsection{The bridge to the quantum mixing}
\label{subsec:BV}

In one-dimensional spatial coordinates, the mixing is illustrated by the unitary
transformation
\small\eq{
\psi_{\sigma}(z,t;\theta) = \mathcal{G}^{\mi \1}(\theta; t)\,
\psi_i\bb{z,t}\,\mathcal{G}(\theta; t)
\label{IIIA00}
}\normalsize
as the result of the noncoincidence of the flavor basis ($\sigma =\,\alpha, \,
\beta$) and the mass basis ($i =\, 1,\, 2$).
The Eq.~(\ref{IIIA00}) gives Eq.~(\ref{III0B}) when the generator of mixing
transformations $\mathcal{G}(\theta; t)$ is given by
\small\eqarr{
\mathcal{G}(\theta; t) &=& \exp\{\theta \int\,dz \,
[\psi_{\1}\bb{z,t}\psi_{\2}\bb{z,t}
-\psi_{\2}\bb{z,t}\psi_{\1}\bb{z,t}]\}
\label{IIIA00B}.
}\normalsize
By taking the one-dimensional representation of Eq.~(\ref{IIIA17}), the propagator
$G(E, p_{\z},t_B,t_A)$ can also be written in the flavor basis as
\small\eqarr{
G^{\alpha\beta}(\theta; E, p_{\z},T) &=&
 \mathcal{G}^{\mi \1}(\theta; t)\,G(E, p_{\z},T)\,\mathcal{G}(\theta; t)
\nonumber\\
&=&
 \mathcal{G}^{\mi \1}(\theta; t)\,G(E, p_{\z},t_D,t_P)\,\mathcal{G}(\theta; t)~~
\label{IIIA00C}
}\normalsize
with $T = t_B-t_A$.

In particular, the definition of a Fock space of weak eigenstates becomes
possible and a nonperturbative flavor oscillation amplitude can be derived \cite{Bla95,Bla03B}.
In this case, the complete Lagrangian (density) is split in a propagation Lagrangian,
\small\eqarr{
\mathcal{L}_{p}\bb{z,t} &=&
\bar{\psi}_{\1}\bb{z,t}\,\left(i \,\partial\hspace{-0.2cm}\slash\hspace{0.1cm} -
m_{\1}\right)\,\psi_{\1}\bb{z,t}
+\bar{\psi}_{\2}\bb{z,t}\,\left(i \,\partial\hspace{-0.2cm}\slash\hspace{0.1cm} -
m_{\2}\right)\,\psi_{\2}\bb{z,t},
\label{IIIA21}
}\normalsize
and an interaction Lagrangian
\small\eqarr{
\mathcal{L}_{\ii}\bb{z,t}&=&
\bar{\psi}_{\alpha}(z,t;\theta)\,\left(i
\,\partial\hspace{-0.2cm}\slash\hspace{0.1cm} -
m_{\alpha}\right)\,\psi_{\alpha}(z,t;\theta)
+\bar{\psi}_{\beta}(z,t;\theta)\,\left(i
\,\partial\hspace{-0.2cm}\slash\hspace{0.1cm} -
m_{\beta}\right)\,\psi_{\beta}(z,t;\theta)
\nonumber\\&&~~~~~~~~~~~~~~~~~~~~~~~~~~~~
- m_{\alpha\beta} \,\left(\bar{\psi}_{\alpha}(z,t;\theta)\psi_{\beta}(z,t;\theta)
+ \bar{\psi}_{\beta}(z,t;\theta)\psi_{\alpha}(z,t;\theta)\right),
\label{IIIA20}
}\normalsize
where
\small\eqarr{
m_{\alpha (\beta)} = m_{\1(\2)}\, \cos^{\2}{\theta} + m_{\2(\1)}\,\sin^{\2}{\theta}
\nonumber
}\normalsize
and
\small\eqarr{
m_{\alpha\beta} = (m_{\1} - m_{\2})\, \cos{\theta}\sin{\theta}.\nonumber
}\normalsize
In general, the two subsets of the Lagrangian can be distinguished if there is a
flavor transformation which is a symmetry of $\mathcal{L}_{\ii}\bb{z,t}$ but not of
$\mathcal{L}_{p}\bb{z,t}$.
Particle mixing occurs if the propagator built from $\mathcal{L}_{p}\bb{z,t}$, and
representing the creation of a particle of flavor $\alpha$ at point $z$ and the
annihilation of a particle of flavor $\beta$ at point $z^{\prime}$, is not diagonal,
i.\,e. not zero for $\beta = \alpha$.
The free fields $\psi_{\ii}\bb{z,t}$ can be quantized in the usual way by rewriting
the momentum distributions $b^s\bb{p_{\z}, m_{\ii}}$ and $d^{s*}\bb{\mi p_{\z},
m_{\ii}}$ in Eq.~(\ref{III23}) as creation and annihilation operators ${\mathit
B}^s\bb{p_{\z}, m_{\ii}}$ and ${\mathit D}^{s\dagger}\bb{\mi p_{\z}, m_{\ii}}$.
The interacting fields are then given by
\small\eqarr{
\psi_{\sigma}\bb{z,t}&=& \mbox{$\int_{_{\infm}}^{^{\infp}}\frac{d p_{\z}}{2\pi}
\exp{[i  p_{\z} z]}$}
\sum_{s=1,2}\{{\mathit B}^s_{\sigma}( p_{\z}; t)\,u^s_{\sigma}( p_{\z}; t)
		+ {\mathit D}^{s*}_{\sigma}(\mi  p_{\z}; t)\,v^s_{\sigma}(\mi
p_{\z}; t)\}
\label{IIIA22}
}\normalsize
where the new flavor creation and annihilation operators which satisfy canonical
anticommutation relations are defined by means of Bogoliubov transformations
\cite{Bla03B} as
\small\eqarr{
{\mathit B}^s_{\sigma}( p_{\z}; t) &=& \mathcal{G}^{\mi \1}(\theta; t)\,{\mathit
B}^s\bb{p_{\z}, m_{\ii}}\,\mathcal{G}(\theta; t)
\nonumber\\
{\mathit D}^s_{\sigma}(\mi  p_{\z}; t) &=& \mathcal{G}^{\mi \1}(\theta; t)\,{\mathit
D}^s\bb{\mi p_{\z}, m_{\ii}}\,\mathcal{G}(\theta; t).
\nonumber
}\normalsize

By following \cite{Bla95,Bla98}, the flavor conversion formula can be written as
\small\eqarr{
\mathcal{P}(\bs{\nu_\alpha}\!\rightarrow\!\bs{\nu_\beta};t)
            &=& \left|\left\{{\mathit B}^s_{\beta}\bb{p_{\0}; t},\,{\mathit
B}^s_{\alpha}\bb{p_{\0}; t}\right\}\right|^{\2}
			+ \left|\left\{{\mathit D}^s_{\beta}\bb{\mi p_{\0}; t},
\,{\mathit B}^s_{\alpha}\bb{p_{\0}; t}    \right\}\right|^{\2}
\label{IIIA25}
}\normalsize
which is calculated without considering the localization conditions imposed by wave
packets, i.\,e., by assuming that $ p_{\z} \approx  p_{\0}$.
When the explicit form of the flavor annihilation and creation operators are
substituted in Eq.~(\ref{IIIA25}), it was also demonstrated \cite{Bla03B}
that the flavor oscillation formula becomes
\small\eqarr{
\mathcal{P}(\mbox{\boldmath$\nu_\alpha$}\!\rightarrow\!\mbox{\boldmath$\nu_\beta$};t)
			&=& \ml{\frac{\sin^{\2}{(2\theta)}}{2}}
			\big\{\!\left[1 - f\bb{p_{\0},m_{\1,\2}} \, \right] \,
			\cos [ \epsilon_{\mi}\bb{p_{\0},m_{\1,\2}} \, t ]
			+ f\bb{p_{\0},m_{\1,\2}} \, \cos [
\epsilon_{\pl}\bb{p_{\0},m_{\1,\2}} \, t] \big\}\nonumber\\
		    &\approx& \mbox{$\sin^{\2}{(2\theta)}
			\left\{\left[1 - \left(\frac{\Delta m}{2
p_{\0}}\right)^{\2}\right]\sin^{\2}{\left[\frac{\Delta m^{\2}}{4  p_{\0}}t\right]}
			\right.$}
			\mbox{$\left.
			+ \left(\frac{\Delta m}{2 p_{\0}}\right)^{\2}
\sin^{\2}{\left[ p_{\0}t\left(1 + \frac{m_{\1}^{\2} + m_{\2}^{\2}}{4
p_{\0}^{\2}}\right)\right]}\right\}~~$}\,,
			\label{IIIA26}
}\normalsize
where the last approximation takes place in the relativistic limit $ p_{\0} \gg
\sqrt{m_{\1} m_{\2}}$.
After some simple mathematical manipulations, Eq.~(\ref{IIIA26}) gives exactly the
oscillation probability $
\mathcal{P}_{\Dirac}(\ms{\bs{\nu_\alpha}\!\rightarrow\!\bs{\nu_\beta}};L)$
calculated from Eq.~(\ref{IIIA14}) when it is assumed that the wave packet width $a$
tends to infinity and $t \approx L$.

This new oscillation formula tends to the standard one (\ref{II27A}) in the
UR limit.
If the mass-eigenstates were nearly degenerate, we could have focused on the case of
a NR oscillating particle having {\em very} distinct mass-eigenstates.
Under these conditions, the quantum theory of measurement says that interference
vanishes.
Therefore, as we have already pointed out, the effects are, under realistic
conditions, far from observable.
Besides, in spite of working on a QFT framework, the lack of observability conditions
must be overcome by implementing external wave packets, i.\,e., by calculating
the explicit form of Eq.~(\ref{IIIA17}) for fermions.
Such a procedure was applied by Beuthe for scalar particles \cite{Beu03} and, in a
very particular analysis, based on our {\em intermediate} wave packet
results, it could be extended to the fermionic case.

\subsection{Intrinsic flavor violation}
\label{subsec:intrinsic}
\providecommand{\sww}{_{_{\rm WW}}}
\providecommand{\pe}[1]{#1{\cdot}}

In this subsection we will try to quantify the initial flavor
violation\,\eqref{Fdef:fvio:+}, i.\,e., the amount by which neutrino flavor is not
well defined at creation or detection. To accomplish this task, it is crucial to
analyze two aspects: (i) the definition of neutrino flavor and (ii) the interactions
responsible for neutrino creation and detection. We will focus on the second aspect
and justify the use of the usual neutrino flavor states\,\cite{endnote2}
\small\eq{
\label{intrinsic:fstates}
\ket{\nu_\alpha(\bp)}=U^*_{\alpha i}\ket{\nu_i(\bp)}~, ~~
\ket{\bar{\nu}_\alpha(\bp)}=U_{\alpha i}\ket{\bar{\nu}_i(\bp)}~, ~~
}\normalsize
where $\braket{\nu_i(\bp)}{\nu_j(\bp')}=\delta_{ij}\delta^3(\bp-\bp')$, as
approximately defining a neutrino (antineutrino) state of momentum $\bp$ and flavor
$\alpha$, associated to charged lepton $l_\alpha$.

We will treat explicitly here the antineutrinos created through pion decay:
$\pi^-\rightarrow l_\alpha\bar{\nu}_\alpha$, $\alpha=e,\mu$. Although, many of the
conclusions drawn within this process may be extended to more general contexts. The
two processes $\alpha=e,\mu$ should be considered as different superpositions of the
six channels $\pi^-\rightarrow l_i\bar{\nu}_j$, $i=1,2$, $j=1,2,3$, where
$l_1=l_e=e,l_2=l_\mu=\mu$. These six channels contribute slightly differently for
different neutrinos and for each channel with fixed $l_\alpha$, the creation of a
pure neutrino flavor state $\bar{\nu}_\alpha$, as defined in
Eq.~\eqref{intrinsic:fstates}, is not guaranteed. Indeed, we will show that
processes such as $\pi\rightarrow\mu\bar{\nu}_e$, are possible with a branching
ratio much greater than loop induced processes such as $\mu\rightarrow e\gamma$ in
the SM, considering the known massive neutrinos and leptonic mixing.

Let us define the free two-particle states with definite flavor
\small\eq{
\label{alphabeta}
\ket{l_\alpha(\bq)\bar{\nu}_\beta(\bk)}\equiv
\delta_{\alpha i}U_{\beta j} \ket{l_i(\bq)\bar{\nu}_j(\bk)}
\,.
}\normalsize
The charged lepton states remain as mass-eigenstates while the neutrino states are
mixed through $U_{\beta j}$. We will see, in accordance to usual expectations, that
pions decay mainly into the states $\ket{l_\alpha(\bq)\bar{\nu}_\beta(\bk)}$ with
$(\alpha,\beta)=(\mu,\mu)$. However, we will also see that there is a
non-null probability of the pion to decay into the neutrino flavor violating states
with $(\alpha,\beta)=(\mu,e)$ or $(\alpha,\beta)=(e,\mu)$.
For that purpose, we want to ultimately calculate the probability
\small\eq{
\label{Plnu:0}
\mathcal{P}(\pi\rightarrow l_\alpha\bar{\nu}_\beta;t)=
\int\! d^3\!\bq\Int{3}{\bk} \sum_{\rm spins}
|\braket{l_\alpha(\bq)\bar{\nu}_\beta(\bk)}{\pi(t)}|^2
\,,
}\normalsize
where $\ket{\pi(t)}$ describes the pion state that decays as time $t$ evolves. The
initial condition is
\small\eq{
\ket{\pi(0)}=\ket{\pi_\psi}\equiv\Int{3}{\bp}\psi(\bp)\ket{\pi(\bp)}\,,
}\normalsize
where $\ket{\pi(\bp)}$ is the free pion state normalized to
$\braket{\pi(\bp)}{\pi(\bp')}=\delta^3(\bp-\bp')$. This normalization will be used
throughout this subsection for free one-particle states. The function $\psi(\bp)$
characterizes the initial momentum distribution for the pion.

The calculation of $\ket{\pi(t)}$ can be performed at lowest order by using time
dependent perturbation theory and the approximation method of
Wigner-Weisskopf\,\cite{WW,Coh77}.
The details can be found in appendix \ref{ap:WW} and
Ref.\,\cite{ccn:intrinsic}. The important result is summarized by the time
dependent transition amplitude $\chi_{ij}(\bq,\bk;t)$ for
$\ket{\pi_\psi}\rightarrow \ket{l_i(\bq)\bar{\nu}_j(\bk)}$,
\eqarr{
\label{intrinsic:chiij}
\chi_{ij}(\bq,\bk;t)&=&
\tilde{\chi}_{ij}(\bp,\bq,\bk;t)
\,\psi(\bp)\big|_{\bp=\bq+\bk}
\,,
\\
\label{intrinsic:tchiij}
\tilde{\chi}_{ij}(\bp,\bq,\bk;t)
&\equiv&
\big[1-e^{-i(\Delta E_{ij}-i\Gamma/2\gamma)t}\big]
N_{ij}^{-1/2}
\frac{f\mathscr{M}_{ij}(\bp,\bq,\bk)}
{\Delta E_{ij}-i\frac{\Gamma}{2\gamma}}
\,,
}\normalsize
where
$N_{ij}=(2\pi)^3 2E_{l_i}(\bq)2E_{\nu_j}(\bk)2E_\pi(\bp)$,
$\Delta E_{ij}\equiv E_\pi-E_{l_i}-E_{\nu_j}$,
$\Gamma$ is the pion decay width
and
$\mathscr{M}_{ij}\equiv\mathscr{M}_{ij}(\bp,\bq,\bk)=
\mathscr{M}(\ms{\pi^-(\bp)\rightarrow l_i(\bq)\bar{\nu}_j(\bk)})$ is the
usual invariant matrix element, normalized to obey
$\bra{\pi(\bp)}S\ket{l_i(\bq)\bar{\nu}_j(\bk)}=(2\pi)^4\delta^4(p-q-k)
\ms{(-i)}\mathscr{M}_{ij}$.

By using Eqs.\,\eqref{intrinsic:chiij} and \eqref{intrinsic:tchiij}, for $t\gg
1/\Gamma$, we can rewrite Eq.~\eqref{Plnu:0} as
\small\eq{
\label{Plnu}
\mathcal{P}(\pi\rightarrow l_\alpha\bar{\nu}_\beta;t)=
\Int{3}{\bp}|\psi(\bp)|^2\!
\Int{3}{\bk} \sum_{\rm spins}
\Big|\sum_j
U_{\alpha j}e^{-iE_{\nu_j}t}U^{\dag}_{j\beta}F_{\alpha j}
\Big|^2_{\bq=\bp-\bk}
\,,
}\normalsize
where
\small\eq{
\label{F:def}
U_{\alpha j}F_{\alpha j}(\bp,\bq,\bk)\equiv
N_{\alpha j}^{-1/2}\frac{\mathscr{M}_{\alpha j}(\bp,\bq,\bk)}
{\displaystyle\Delta E_{\alpha j}-i\frac{\Gamma}{2\gamma}}
\,.
}\normalsize
We see the exponential $e^{-iE_{\nu_j}t}$ is responsible for the neutrino
oscillation phenomenon. In fact, if we neglect the neutrino mass $m_j$ in every term
of Eq.~\eqref{Plnu}, except in the exponential, we get
\small\eq{
\label{Plnu:factor}
\mathcal{P}(\pi\rightarrow l_\alpha\bar{\nu}_\beta;t)=
\Int{3}{\bp}|\psi(\bp)|^2\!
\Int{3}{\bk}
\mathcal{P}\ms{(\bar{\nu}_\alpha\rightarrow\bar{\nu}_\beta;t)}
|F_{\alpha}|^2
\,,
}\normalsize
where $F_{\alpha}=(F_{\alpha j})_{m_j\rightarrow 0}$.

Notice the usual oscillation probability,
\small\eq{
\mathcal{P}(\ms{\bar{\nu}_\alpha\!\rightarrow\!\bar{\nu}_\beta};t)=
\Big|\sum_j
U_{\alpha j}e^{-iE_{\nu_j}(\bk)t}U^{\dag}_{j\beta}\Big|^2
,
}\normalsize
factors out from the creation probability of $l_\alpha\bar{\nu}$, $|F_{\alpha}|^2$, for massless neutrinos. Such factorization is what allows the definition of the state \eqref{alphabeta} as a flavor state, since
\small\eq{
\label{Plnu:m=0}
\mathcal{P}(\pi\rightarrow l_\alpha\bar{\nu}_\beta;t)
\approx \delta_{\alpha\beta}\frac{\Gamma_\alpha}{\Gamma}
,
}\normalsize
for $1/\Gamma\ll t\ll L_{\rm osc}$, where $L_{\rm osc}$ is the typical flavor oscillation length (period). Therefore, the antineutrino flavor state $U_{\alpha j}\ket{\bar{\nu}_j}$ is only created jointly with the charged lepton $l_\alpha$\,\cite{flavorLee,giunti:torino04}.
Notice Eq.~\eqref{Plnu:m=0} correctly coincides with the branching ratio of the decay $\pi\rightarrow l_\alpha\bar{\nu}$.

For the sake of completeness let us rewrite the relevant part of Eq.~\eqref{Plnu:m=0} as
\small\eq{
\Int{3}{\bk}|F_{\alpha}|^2\big{|}_{\bq=\bp-\bk}
=
\Int{3}{\bk}
\frac{|\sum_jU^*_{\alpha j}\mathscr{M}_{\alpha j}|^2}
{N_{\alpha j}|\Delta E_{\alpha j}-i\frac{\Gamma}{2\gamma}|^2}
\bigg{|}_{\bq=\bp-\bk,m_j=0}
\approx
\frac{\Gamma_\alpha}{\Gamma}
\,,
}\normalsize
where the last approximation considers $|\Delta E_{\alpha
j}-i\frac{\Gamma}{2\gamma}|^2\approx \frac{2\pi\gamma}{\Gamma}\delta(\Delta
E_{\alpha j})$ for small enough $\Gamma$. A regularization function $f$ might be
necessary to guarantee the convergence of the integral for large
$\bk$\,\cite{ccn:intrinsic}.

Neutrinos, however, are not strictly massless and we may have initial flavor
violation because different neutrino masses contribute differently to each channel
$\pi\rightarrow l_i\bar{\nu}_j$\,\cite{flavorLee}. We will focus on initial flavor
violation and denote the interval of time satisfying $1/\Gamma\ll t\ll L_{\rm osc}$
by $t=0$.

We can make the flavor violating contributions explicit by rewriting the term inside
the squared modulus in Eq.~\eqref{Plnu} as
\small\eq{
\sum_{j=1}^{3}U_{\alpha j}U^*_{\beta j}F_{\alpha j}=
\delta_{\alpha\beta}F_{\alpha 1}+\sum_{j=2}^{3}U_{\alpha j}U^*_{\beta j}\Delta
F_{\alpha j}
\,,
}\normalsize
where $\Delta F_{\alpha j}\equiv F_{\alpha j}-F_{\alpha 1}$.
Thus the squared modulus becomes
\small\eq{
\label{sum23}
\big|\sum_{j=1}^{3}U_{\alpha j}U^*_{\beta j}F_{\alpha j}\big|^2=
\delta_{\alpha\beta}|F_{\alpha 1}|^2+
\delta_{\alpha\beta}2\mathrm{Re}
\Big[F_{\alpha 1}^*\sum_{j=2}^{3}U_{\alpha j}U^*_{\beta j}\Delta F_{\alpha j}
\Big]
+
\Big|\sum_{j=2}^{3}U_{\alpha j}U^*_{\beta j}\Delta F_{\alpha j}\Big|^2
\,.
}\normalsize
Notice there is no summation over repeated indices $\alpha$ or $\beta$.
We recognize that only the last term of Eq.~\eqref{sum23} is flavor
non-diagonal. The second term, which is flavor diagonal, can be shown to be of
the same order of the flavor violating contribution, but negative in sign.

Specializing to $\alpha\neq\beta$, under the approximation of $U_{\alpha3}U_{\beta
3}^*\approx 0$ (which is valid if $\alpha=e$ or $\beta=e$), the initial creation
probability yields
\small\eq{
\label{Pmunue}
\mathcal{P}_{l_\alpha\nu_\beta}=
\Int{3}{\bp}|\psi(\bp)|^2\!
\Int{3}{\bk}
|U_{\alpha 2}U_{\beta 2}^*|^2
|\Delta F_{\alpha 2}|^2
\,,
}\normalsize
where $\mathcal{P}_{l_\alpha\nu_\beta}=\mathcal{P}(\pi\rightarrow
l_\alpha\bar{\nu}_\beta;t)$, for $1/\Gamma\ll t\ll L_{\rm osc}$.
For the two family parametrization, we have $|U_{\alpha 2}U_{\beta
2}^*|^2=\frac{1}{4}\sin^2\!2\theta$, thus indicating that this phenomenon is
indeed mixing dependent.

To determine the most dominant contribution to Eq.~\eqref{Pmunue}, we have to
analyze the dominant contribution to $\Delta F_{\alpha 2}$ due to $\Delta
m=m_1-m_2\neq 0$. We anticipate that the dominant contribution is
due to $(\Delta E_{\alpha j}-i\Gamma/2\gamma)^{-1}$ in $F_{\alpha j}$.
Therefore
\eqarr{
\label{DeltaF:1}
|\Delta F_{\alpha 2}|^2&\approx&
\frac{|\mathscr{M}_{\alpha 0}|^2}{N_{\alpha 0}}
\bigg|
\frac{1}{\Delta E_{\alpha 2}-i\frac{\Gamma}{2\gamma}}
- \frac{1}{\Delta E_{\alpha 1}-i\frac{\Gamma}{2\gamma}}
\bigg|^2
\\
&\approx&
\label{DeltaF:final}
\frac{|\mathscr{M}_{\alpha 0}|^2}{N_{\alpha 0}}
\frac{(\Delta E_\nu)^2}
{\Big[(\Delta E_{\alpha 0}+\meio\Delta E_\nu)^2 + \frac{\Gamma^2}{4\gamma^2}\Big]
\Big[(\Delta E_{\alpha 0}-\meio\Delta E_\nu)^2 + \frac{\Gamma^2}{4\gamma^2}\Big]}
\,,
}\normalsize
where
\small\eq{
\Delta E_\nu\equiv E_{\nu_1}(\bk)-E_{\nu_2}(\bk)
=\frac{\Delta m^2}{E_{\nu_1}(\bk)+E_{\nu_2}(\bk)}\,,
}\normalsize
the subscript $0$ means we assume the massless limit, $m_1,m_2\rightarrow0$,
for neutrinos in those terms and $|\mathscr{M}_{\alpha 0}|^2$ refers to
$|\mathscr{M}_{\alpha j}|^2_{m_j\rightarrow 0}$ without the mixing matrix element
$|U_{\alpha j}|^2$  [see Eq.~\eqref{F:def}]. We have also the equivalence
$|\mathscr{M}_{\alpha 0}|^2=\sum_j|\mathscr{M}_{\alpha j}|^2_{m_j\rightarrow0}$.

To justify Eq.~\eqref{DeltaF:1} we note that $\Gamma\gg
 |E_{\nu_1}(\bk)-E_{\nu_2}(\bk)|=|\Delta E_{\alpha 2}-\Delta
E_{\alpha 1}|$ for neutrino momentum close to the energy conserving value
$|\bk|_\nu=E_\nu=\frac{M^2_\pi-M^2_\alpha}{2M_\pi}$, for massless neutrinos and pion
at rest $\bp\approx 0$. Retaining small neutrino masses does not alter the
conclusion.
Neutrino energy (momentum) for $\nu_1$ and $\nu_2$ can be distinct from the energy
conserving values only by an amount $\Gamma\ll E_\nu$ and such variation are not
relevant in
other terms besides the last one in Eq.~\eqref{DeltaF:1}.
The contribution of neutrino masses to the invariant matrix elements
$|\mathscr{M}_{\alpha j}|$ and other kinematical terms are negligible since
neutrinos are produced UR. Neutrino masses can be neglected in most
of the terms except for
\small\eq{
\Delta E_\nu \approx \frac{\Delta m^2}{2E_{\nu}}
\,.
}\normalsize
We recall the numeric values for the relevant quantities $\Gamma=2.53\times
10^{-8}\mathrm{eV}$\,\cite{pdg} and $\Delta
m^2/2E_\nu\sim\frac{1}{6}\times10^{-7}\mathrm{eV}\frac{\Delta m^2}{1\rm
eV^2}$, where $\Delta m^2$ is either $|\Delta m^2_{12}|\sim 0.8\times 10^{-4}\rm
eV^2$ or $|\Delta m^2_{23}|\sim 2.5\times 10^{-3}\rm eV^2$\,\cite{vissani:review}.
A more detailed justification can be found in Ref.\,\cite{ccn:intrinsic}.

The flavor violating creation probability in Eq.~\eqref{Pmunue}, $\alpha\neq\beta$,
can be calculated by changing the neutrino momentum variable $|\bk|$ to energy
$E_\nu$ and extending the lower integration limit to $-\infty$, with negligible
contribution, giving
\small\eq{
\label{Pmunue:final}
\mathcal{P}_{l_\alpha\nu_\beta}\approx
\frac{1}{2}\sin^2\!2\theta
\frac{\Gamma_\alpha}{\Gamma}\Big(\frac{\Delta m^2}{2E_\nu\Gamma}\Big)^2_{\rm EC}
\,,
}\normalsize
where the two family parametrization, $|U_{\alpha 2}U_{\beta
2}^*|^2=\frac{1}{4}\sin^2\!2\theta$, was employed and $\bp\approx 0$ (pion at rest)
was considered by adjusting $\psi(\bp)$.
The subscript EC denotes that energy conservation was assumed.
The following integral was also necessary,
\small\eq{
\label{int:[]^2}
\int_{-\infty}^{\infty}d\lambda
\frac{1}{\big[\lambda^2+\frac{\Gamma^2}{4\gamma^2}\big]^2}
=
\frac{2\pi}{\Gamma}\Big(\frac{2\gamma^2}{\Gamma^2}\Big)
\,.
}\normalsize
One can recognize the term inside parenthesis in
Eqs.\,\eqref{DeltaF:final} and \eqref{int:[]^2} as the additional contribution that
appears in Eq.~\eqref{Pmunue:final}.

Let us estimate some specific flavor violation probabilities (branching ratios):
\small\eq{
\label{Plnu:num}
\frac{\mathcal{P}_{\mu\nu_e}}{\sin^2\!2\theta_{12}} \sim
10^{-9}
\,,~~
\frac{\mathcal{P}_{e\nu_\mu}}{\sin^2\!2\theta_{12}} \sim
3\times 10^{-15}\frac{\Gamma_e}{\Gamma}
\,,~~
\frac{\mathcal{P}_{\mu\nu_\tau}}{\sin^2\!2\theta_{23}} \sim
10^{-6}
\,.
}\normalsize
To compute the last value in Eq.~\eqref{Plnu:num}, we considered $|\Delta
m^2_{13}|\approx |\Delta m^2_{23}|\gg |\Delta m^2_{12}|$ and replaced $\Delta
F_{\alpha 3}$ by $\Delta F_{\alpha 2}$ in Eq.~\eqref{Pmunue}.

For completeness, let us consider non-realistic cases that do not satisfy
$\frac{\Delta m^2}{2E_\nu\Gamma}\ll 1$. In that case Eq.~\eqref{Pmunue:final}
should be corrected to
\small\eq{
\label{Pmunue:final:B}
\mathcal{P}_{l_\alpha\nu_\beta}\approx
\ml{\frac{1}{2}}\sin^2\!2\theta
\frac{B^2}{1+B^2}
\,,
}\normalsize
where
\small\eq{
B\equiv \frac{\Delta m^2}{2E_\nu\Gamma}\bigg|_{\rm EC}\,.
}\normalsize
The expression in Eq.~\eqref{Pmunue:final:B} is obtained if we retain the terms
$\Delta E_\nu$ in the denominator of \eqref{DeltaF:final} and consider its energy
conserving value $B$. In particular, Eq.~\eqref{Pmunue:final:B} have the correct
limit for
\small\eq{
\mathcal{P}_{l_\alpha\nu_\beta}\stackrel{|B|\rightarrow \infty}{\longrightarrow}
\ml{\frac{1}{2}}\sin^2\!2\theta
\,,
}\normalsize
which corresponds to the incoherent creation limit. This limit can be also seen in
Eq.~\eqref{Plnu}, considering that Eq.~\eqref{F:def} for such limit implies there
is no overlap in the sum. Equation \eqref{Plnu} could be rewritten
\small\eq{
\mathcal{P}_{l_\alpha\bar{\nu}_\beta}\rightarrow
\Int{3}{\bp}|\psi(\bp)|^2\!
\Int{3}{\bk} \underset{j}{\mn{\sum}}
\big|U_{\alpha j}U^{\dag}_{j\beta}\big|^2
\underset{\rm spins}{\mn{\sum}}\big|F_{\alpha j}\big|^2_{\bq=\bp-\bk,m_j=0}
\,,
}\normalsize
where
$\sum_{j=1}^{2}|U_{\alpha j}U^{\dag}_{j\beta}|^2=\frac{1}{4}\sin^2\!2\theta$ in the
two family approximation.
In other words, the condition $|B|\gg 1$ corresponds to the limit where the energy
uncertainty is much smaller than the energy separation of mass-eigenstates, which
allows the distinction of mass-eigenstates and consequent loss of coherence for
oscillation.
We can see a close relationship between intrinsic flavor violation and loss of
coherence by wave packet separation: if intrinsic flavor violation is large, flavor
oscillation is also suppressed.

We should remark we do not analyze the effects to oscillation due to
finite size in detection\,\cite{Kie96} but if the microscopic time of
detection for any interaction is much shorter than $\tau=1/\Gamma$, then
intrinsic flavor violation should remain.
The intrinsic uncertainty in momentum $\sigma_p$ of the parent pion, encoded in
$\psi(\bp)$, was also not considered in the discussion but the conclusions
remain unchanged as long as $\sigma_p\lesssim \sigma_E=\Gamma$. In fact the limit
$|\psi(\bp)|^2\rightarrow \delta^3(\bp)$ is easily calculable.
Other subtleties of the approximation used can be seen in
Ref.\,\cite{ccn:intrinsic}. The same conclusion of requirement of coherent
superpositions of neutrinos can be drawn in an exactly solvable QFT that does not
rely on perturbation theory\,\cite{flavorLee} and then all the aspects of flavor
oscillation and flavor violation can be calculated exactly.
It is important to emphasize that the effect of intrinsic flavor violation is more
robust than flavor oscillation effects because no propagation is needed and
averaging over the source or detector does not affect the results.

It should be also emphasized that the intrinsic flavor violation effect calculated in
Eq.\,\eqref{Pmunue:final} assumes two facts: (1) neutrinos are directly detected and
(2) they are detected as flavor states, which are coherent superpositions of
mass-eigenstates, as defined (approximately) by Eq.\,\eqref{alphabeta}.
On the one hand, the coherent creation of neutrino flavor states is indeed guaranteed
from the observations of neutrino oscillations, implying that intrinsic flavor
violation effects should be
small\,\cite{ccn:intrinsic,flavorLee,Kay81,giunti:torino04}.
If mass-eigenstates were created and detected incoherently, flavor violating effects
would be analogous to flavor changing processes for quarks, at tree level, without
the explicit appearance of the $\Delta m^2$ dependence. On the other hand, when
neutrinos are not explicitly detected, their effects can be computed from an
incoherent sum of the contributions of each neutrino mass-eigenstate\,\cite{shrock},
as in the intended direct measurements of absolute neutrino mass.
Another aspect of assumption (2) is that the definition of the usual flavor states in
Eq.\,\eqref{alphabeta} is only approximate \cite{giunti:torino04} and it may depend
on the details of the creation and detection. Usually, however, the error made is of
the order of $\Delta m^2/E_\nu^2$ and it can be neglected\,\cite{giunti:torino04}.

\section{Conclusions and Outlook}
\label{sec:conclusion}

In order to understand some subtle changes that appear in the standard flavor oscillation probability\,\cite{Kay04} when one goes deep in the theoretical fundamentals of the quantum oscillation phenomenon, we have quantified the consequences of assuming the quantum mixing and oscillation in the Dirac theory.

We have started our study by analyzing the consequences of including the spatial localization in the standard {\em effective} treatment of quantum flavor oscillations.
The classification of first and second order effects were relevant in establishing a satisfactory criterion for quantifying the effects of including spin and relativistic completeness for fermionic particles in the Dirac theory.
By taking into account the spinorial form of neutrino wave functions and imposing an initial constraint where a {\em pure} flavor state is created at $t = 0$, it was possible, for a constant spinor $w$, to obtain the oscillation probability containing the contribution of positive and negative frequency solutions of the Dirac equation.
We have noticed a term of very high oscillation frequency depending on the sum of energies in the new oscillation probability formula (rapid oscillations).

In addition to rapid oscillations, the spinorial form of the wave functions and their chiral oscillating character subtly modify the coefficients of the oscillating terms in the flavor conversion formula.
To describe the time evolution of the mass-eigenstates, we have assumed an initial gaussian localization and performed integrations by considering a strictly peaked momentum distribution.
Under the particular assumption of UR particles, we have been able to obtain an analytic expression for the coupled chiral and flavor conversion formula.
For the case of Dirac wave packets, these modifications introduce corrective factors which are negligible in the UR limit.
Nevertheless, strictly from the theoretical point of view, we have confirmed that the {\em fermionic} character of the particles modify the standard oscillation probability which was previously obtained by implicitly assuming a {\em scalar} nature of the mass-eigenstates, restricted to positive frequencies.
These results concerning chiral oscillations coupled with the flavor conversion mechanism of free propagating wave packets \cite{Ber04,DeL04,ccn:no12}, are therefore relevant in quantifying the role played by the intrinsic spinorial structure in flavor oscillations.

The quantum-mechanical treatment which associates Dirac wave packets with the propagating mass-eigenstates is indeed rich in physical insights which were extensively discussed.
Besides the review of analytical calculations done with gaussian wave packets for {\em scalar} and {\em fermionic} particles, the main conceptual aspects related to the very rapid oscillations lead to the study of chiral oscillations.
In the standard model flavor-changing interactions, neutrinos with positive chirality are decoupled from the neutrino absorbing charged weak currents.
A state with {\em left-handed} helicity can be approximated by a state with negative chirality in the UR limit.
Once we have assumed the interactions at the source and detector are chiral, only the component with negative chirality contributes to the propagation.
Since the chiral state affects the results at the detection process, the mechanism of chiral oscillation results in modifications to the standard flavor conversion formula, with correction factors proportional to $m_{\1,\2}^{\2}/p_{\0}^{\2}$.
The effect of such factors are, however, practically undetectable by the current experimental analysis.
It leads to the conclusion that, in spite of often being criticized, the standard flavor oscillation formula resorting to the plane wave derivation can be reconsidered when {\em properly} interpreted, but a satisfactory description of {\em fermionic} (spin one-half) particles requires the use of the Dirac equation as evolution equation for the mass-eigenstates.
Correlated constructions involving more general spatial configurations for the magnetic fields might be considered as well as
extensions of the corrections discussed here to more general chiral conversion rates, polarization effects and neutrino
propagation in magnetized media.
Some of these investigations has already been under consideration, primarily focused on observable effects or phenomenological implications \cite{Ber08A,ccn:helicity}.
Another difficult task in the same framework is to describe the nonstationary evolution of neutrinos in supernovae or in the early universe, where interaction rates are in competition with the flavor oscillation frequencies, and where the wave packet effects seems to be maximized.

In parallel to the description in various contexts of the phenomenon of neutrino flavor oscillation based on the Dirac theory of fermion propagation (sections \ref{sec:relativistic} and \ref{sec:consequences}), a dichotomy between rapid oscillations and initial flavor violation was introduced.
Initially discussed in the context of the free Dirac theory, we could conclude that at least one of the phenomena of rapid oscillations or initial flavor violation should invariably occur in neutrino flavor oscillations.
Rapid oscillations could be avoided but only at the expense of not having an initially well defined flavor.
In fact, as discussed in subsection \ref{subsec:flavorvio:I}, initial flavor violation could arise even in the context of the usual positive-energy scalar wave packet description of neutrinos, reviewed in section \ref{sec:localization}.
However, in the latter context, it is usually eliminated completely, by setting $\phi_1(\bx,0)=\phi_2(\bx,0)$, or approximately, by taking $|\phi_1(\bx,0)-\phi_2(\bx,0)|$ small enough.
Using Dirac theory, however, equal wave packets for the mass-eigenstates, $\psi_1(\bx,0)=\psi_2(\bx,0)$, imply rapid oscillations.
Quantitatively, both phenomena were unimportant because their effects could be as small as of the order of $(\Delta m/E_\nu)^2$.
Moreover, in the context of first quantized Dirac theory, there are no sufficient ingredients to decide which effect is taking place in neutrino oscillations and with which size.

Seeking a more sophisticated approach towards more realistic descriptions, we have considered the inclusion of field-theoretical elements in section \ref{sec:qft}.
Derivations of the oscillation formula resorting to field-theoretical methods, however, are not very popular.
They are thought to be very complicated and the existing quantum field computations of the oscillation formula do not agree in all respects \cite{Beu03}.
For instance, in the  Blasone and Vitiello model \cite{Bla95,Bla03} they have attempted to define a Fock space of weak eigenstates and to derive a nonperturbative oscillation formula.
Flavor creation and annihilation operators, satisfying canonical (anti)commutation relations, are defined by means of Bogoliubov transformations.
As a result, new oscillation formulas were obtained for fermions and bosons, with the oscillation frequency depending not only on the difference but also on the sum of the energies of the different mass-eigenstates.
The BV model, however, is not unanimously accepted throughout the community.
Among the arguments against it, it has been argued that Fock states of flavor neutrinos are unphysical \cite{GiuBla} because their use to describe the neutrinos produced or detected in charged-current weak interaction processes imply that measurable quantities depend on the arbitrary unphysical ``flavor neutrino masses.''
In general terms, if one imagines a stationary plane wave state, the position-dependence of the complex phase is different for each mass-eigenstate component, so when the state is probed at different locations in space, different flavor mixtures should be found. Thus, if the state is a pure flavor at the origin, it will be a mixture of all three flavors far from the origin.
One concludes that flavor creation and annihilation operators are problematic and that it does not make sense to talk about states that are simultaneously momentum and flavor eigenstates, as prescribed by Blasone and Vitiello.

Despite the mathematical similarities between the flavor conversion formula obtained with Dirac wave packets \cite{Ber04B} and the one obtained in the BV model \cite{Bla95,Bla03}, the former one does not have the problematic interpretation of delocalization.
Moreover each new effect present in the oscillation formula can be depicted and separately quantified.
Likewise, the dichotomy between rapid oscillations and initial flavor violation can
be solved through a more standard application of field quantization.
Excluding the BV model, second quantization is easily capable of eliminating rapid oscillations, as positive and negative frequency components are disconnected by associating creation and annihilation operators.
Rapid oscillations are indeed absent in EWP approaches.
More simply, a free second quantized theory can be devised and an oscillation formula without rapid oscillations could be found in section \ref{subsec:simpleQFT}.
But in that case, the conclusion from first quantized Dirac theory remains: there is initial flavor violation.
The amount of violation, however, could not be calculated a priori because of the lack of information about the neutrino creation aspects, mainly, localization.
Such lack of information is then filled by taking the creation process into account.
Under the assumption that the usual superposition of mass-eigenstates \eqref{intrinsic:fstates} are detected as neutrino flavor states, we specifically calculated in section \ref{subsec:intrinsic} the probability (branching ratio) of pion decay processes with flavor violation, such as $\pi\rightarrow \mu\bar{\nu}_e$, showing nonzero results.
In such context, initial flavor violation was renamed \textit{intrinsic flavor violation}.
The effect is very small but much greater than the naive estimate $\Delta m^2/E_\nu^2$ (a similar result is indeed obtained in Refs.\,\onlinecite{blasone:short} and \onlinecite{liliu}), typical for equal momentum distributions for the mass-eigenstates, or the branching ratio of indirect flavor violating processes such as $\mu\rightarrow e\gamma$ within the SM, considering the known massive neutrinos and leptonic mixing.
The effect is indeed relatively large because of the presence of the finite but small decay width for pions, which was not considered in works previous to Ref.\,\cite{ccn:intrinsic}, and there is no loop suppression as in indirect flavor changing processes.
Compared to the quark sector, the large mixing angles also contribute to the effect.

The important point of the detailed calculation of section \ref{subsec:intrinsic} is the confirmation of the occurrence of intrinsic neutrino flavor violation when neutrinos are created and the quantification of its magnitude as a function of quantities, such as the parent particle decay width, that have been known qualitatively to play a role in neutrino oscillations\,\cite{Kay81}.
The effect is the consequence of the slightly different creation amplitudes, functions of different neutrino masses, that have to be summed coherently.
The smallness of the effect explains why neutrino flavor is an approximately well defined concept in the SM and it is directly related to the smallness of the neutrino mass differences.
At the same time, small mass splitting allows the coherent creation of neutrino flavor states that is required for the phenomenon of neutrino flavor oscillations.
Although tiny, the effect of intrinsic flavor violation also deserves further attention in scenarios with sterile neutrinos\,\cite{minos:10,miniboone:sterile} or genuine non-standard interactions\,\cite{grossman:NI}.

Taking the reverse route from second quantized to first quantized treatments, one can conclude that, in the strict sense, we should not use \emph{generic} Dirac wave packets to describe the mass-eigenstates that compose the flavor states within IWP approaches.
In the light of field theoretical treatments that naturally do not exhibit rapid oscillations, one can solve the dichotomy between rapid oscillations and initial flavor violation in IWP approaches by giving up strict initial flavor definition and avoiding rapid oscillations by imposing the auxiliary condition \eqref{Fdef:cond:+} to the initial mass-eigenstate wave packets.
Such a solution is only relevant in qualitative terms.
Quantitatively, both initial flavor violation and rapid oscillation effects can be taken to be very small if no information on the creation and detection of neutrinos is taken into account.

Even though the existence of intrinsic flavor violation can be supported by many
arguments, its magnitude can still be subjected to large uncertainties
concerning the localization aspects of the created and detected neutrino states.
The quantification of such effect in more general contexts clearly deserves a
careful examination.
As a related issue, one should notice some debate in the recent literature\,\cite{glashow:no,Akh10} about
the importance of intrinsically quantum phenomena, such as quantum entanglement, in
the phenomenon of neutrino flavor oscillations: an indication that further
progress on the subject can be expected in the future.

To conclude, it is evident that field-theoretical aspects should be fully considered in describing the fascinating phenomenon of neutrino oscillations and their consequences to other physical contexts.
Nevertheless, first quantized descriptions, specially when confronted to second quantized treatments containing similar ingredients, can help us to understand certain aspects of the phenomenon without including some diverting mathematical machinery.
We hope that the formal developments presented here, from which one can confront the
differences and depict some fundamental overlaps between distinct frameworks, could
stimulate further investigation on the theoretical foundations of flavor
oscillations.

\section*{Acknowledgments}

The authors would like to thank the financial support from the Brazilian Agencies FAPESP, through the grants 08/50671-0 (AEB), 09/11309-7 (CCN) and 04/00220-1 (MMG), and CNPq, through the grants 300233/2010-8 (AEB), 309455/2009-0 (CCN) and 303937/2009-2 (MMG).

\appendix
%
\providecommand{\rp}{\mathrm{p}}
\providecommand{\q}{\mathrm{q}}
\providecommand{\rk}{\mathrm{k}}
\providecommand{\rx}{\mathrm{x}}

\section{Notation and definitions}
\label{app:def}

The (scalar, spinorial or ST) wave functions
related by Fourier transforms are denoted as
\eqarr{
\label{fourier:xp}
\varphi(\bx)
&=&\frac{1}{(2\pi)^{\mt{3/2}}}
\Int{3}{\bp}\tilde\varphi(\bp)\,e^{i\bp\ponto\bx}
~,
\\
\label{fourier:px}
\tilde\varphi(\bp)
&=&\frac{1}{(2\pi)^{\mt{3/2}}}
\Int{3}{\bx}\varphi(\bx)\,e^{-i\bp\ponto\bx}
~.
}
The tilde denotes the inverse Fourier transformed function.

Using the property of the Dirac or ST Hamiltonian, $H_n^2=(\bp^2+m_n)^2\id$, we
can write the evolution operator in the form
\eq{
e^{-iH_nt}=\cos(E_nt)-i\frac{H_n}{E_n}\sin(E_nt)
~,
}
where the momentum dependence have to be replaced by $-i\nabla$ in coordinate
space.

The free neutrino field expansion used is ($i=1,2$)
\eq{
\label{psi:qft}
\nu_i(x)=
\underset{s}{\textstyle \sum}\int\! \frac{d^3\bp}{2E_{\bp}}\,
[u_i^s(x;\!\bp)a_i(\bp,s)+v_i^s(x;\!\bp)b_i^\dag(\bp,s)]
~,
}
where the creation and annihilation operators satisfy the
canonical anticommutation relations
\eqarr{
\{a_i(\bp,r),a_j^\dag(\bp',s)\}&=&\delta_{ij}\delta_{rs}2E_i(\bp)
\delta^3(\bp-\bp ')
~,
\\
\{b_i(\bp,r),b_j^\dag(\bp',s)\}&=&\delta_{ij}\delta_{rs}2E_i(\bp)
\delta^3(\bp-\bp ')~;
}
all other anticommutation relations are null.
The functions $u,v$ are defined as
\eqarr{
\label{u:x}
u_i^s(x;\!\bp)&=&u_i^s(\bp)\frac{e^{-ip_i\ponto x}}{(2\pi)^{\mt{3/2}}}
~,
\\
\label{u:p}
u_i^s(\bp)&=&
\ml{\frac{m_i+E_i\gamma^0-\bp\ponto\bs{\gamma}}{\sqrt{E_i+m_i}}}u_0^s
=\ms{\sqrt{2E_i}}\,u^s\bb{\bp,m_i}
~,
\\
\label{v:x}
v_i^s(x;\!\bp)&=&v_i^s(\bp)\frac{e^{ip_i\ponto x}}{(2\pi)^{\mt{3/2}}}
~,
\\
\label{v:p}
v_i^s(\bp)&=&
\ml{\frac{m_i-E_i\gamma^0+\bp\ponto\bs{\gamma}}{\sqrt{E_i+m_i}}}v_0^s
=\ms{\sqrt{2E_i}}\,v^s\bb{\bp,m_i}
~,
}
where $p_i\ponto x=E_i\ms{(\bp)}t-\bp\ponto\bx$ and they satisfy the properties
\eqarr{
\bar{u}_0^ru_0^s=u_0^{r\dag}u_0^s&=&
-\bar{v}_0^rv_0^s=v_0^{r\dag}v_0^s=\delta_{rs}
~,
\\
v_0^{r\dag}u_0^s&=&u_0^{r\dag}v_0^s=0 ~~\forall~ r,s
~,
\\
\label{uv:complete}
\underset{s}{\textstyle \sum}u_i^{s}(\bp)\bar{u_i}^s(\bp)
&=&{\sla p + m_i}
=2E_i(\bp)\Lambda^D_{i+}(\bp)\gamma^0
~,
\\
\underset{s}{\textstyle \sum}v_i^{s}(\bp)\bar{v_i}^s(\bp)
&=&{\sla p - m_i}
=2E_i(\bp)\Lambda^D_{i-}(-\bp)\gamma^0
~.
}
For comparison, Eqs.\,\eqref{u:p} and \eqref{v:p} also show the relation with the
spinors $u$ and $v$ defined in Eq.\,\eqref{III02}

The Feynman propagator for fermions is
\eqarr{
iS_F(x-y)&\equiv& \bra{0}T(\psi(x)\bar{\psi}(y))\ket{0}
\\
&=&\int\! \frac{d^4\!p}{\ms{(2\pi)}^4}\,\frac{i}
{\sla{p}-m+i\epsilon}\,e^{-ip\ponto(x-y)}
\\
&=&(i\sla{\partial}+m)i\Delta_F(x-y;m)
~.
}

The function $S$ in Eq.~\eqref{S} and its equivalent for the Sakata-Taketani
Hamiltonian can be written as
\eqarr{
\label{S:D}
iS(x;m)&=&(i\sla\partial+m)i\Delta(x;m)
~,
\\
\label{S:ST}
K^{\ST}(x;m)&=&[i\partial_t-\frac{\nabla^2}{2m}(\tau_3\!+\!i\tau_2)+m^2]
i\Delta(x; m )
~,
\\
\label{Delta}
i\Delta(x;m)&=&\frac{1}{(2\pi)^3}\Int{4}{p}
\delta(p^2-m^2)\epsilon(p_0)e^{-ip\ponto x}
\cr
&=&
\frac{1}{(2\pi)^3}\int
\frac{d^3\bp}{2E_p}\,[e^{-ip\ponto x}-e^{+ip\ponto x}]
~.
}

The free neutrino eigenstates are defined as
\eqarr{
\ket{\nu_i(\bp,s)}&\equiv &\frac{a_i^\dag(\bp,s)}{\sqrt{2E_i}}\ket{0}
\\&=&
\Int{3}{\bx}\nu_i^\dag(x)\ket{0}\frac{u_i(x;\bp)}{\sqrt{2E_i}}
~,
\\
\ket{\bar{\nu}_i(\bp,s)}&\equiv&\frac{b_i^\dag(\bp,s)}{\sqrt{2E_i}}\ket{0}
\\&=&
\Int{3}{\bx}\frac{v_i^\dag(x;\bp)}{\sqrt{2E_i}}\nu_i(x)\ket{0}
~,
}
whose normalization is
$\braket{\nu_j(\bp,r)}{\nu_i(\bp',s)}=\delta_{ij}\delta_{rs}
\delta^3(\bp -\bp')$. The same normalization is valid for the antiparticle
states.
The states with finite momentum distributions are defined as
\eqarr{
\ket{\nu_i\!:\!g}&=&
\underset{s}{\textstyle \sum}
\Int{3}{\bp}g^s(\bp)\ket{\nu_i(\bp,s)}
\\&=&
\Int{3}{\bx}\nu_i^\dag(x)\ket{0}\psi_{\nu_i}(x)~,
\\
\label{psi:nu}
\psi_{\nu_i}(x)&\equiv&
\underset{s}{\textstyle \sum}
\Int{3}{\bp}g^s(\bp)\frac{u_i^s(x;\bp)}{\sqrt{2E_i}}~,
\\
\label{ket:nu:t}
e^{-iHt}\ket{\nu_i\!:\!g}&=&
\Int{3}{\bx}\nu_i^\dag(\bx,0)\ket{0}\psi_{\nu_i}(\bx,t)
~,\\
\label{ket:nub:t}
\ket{\bar{\nu}_i\!:\!g}&=&
\underset{s}{\textstyle \sum}
\Int{3}{\bp}g^{s*}(\bp)\ket{\bar{\nu}_i(\bp,s)}
\\&=&
\Int{3}{\bx}\psi_{\bar{\nu}_i}^\dag(x)\nu_i(x)\ket{0}~,
}\eqarr{
\label{psi:nub}
\psi_{\bar{\nu}_i}(x)&\equiv&
\underset{s}{\textstyle \sum}
\Int{3}{\bp}g^{s}(\bp)\frac{v_i^s(x;\bp)}{\sqrt{2E_i}}
~,
\\
e^{-iHt}\ket{\bar{\nu}_i\!:\!g}&=&
\Int{3}{\bx}\psi_{\bar{\nu}_i}^\dag(\bx,t)\nu_i(\bx,0)\ket{0}
~.
}

\section{Wigner-Weisskopf approximation in pion decay}
\label{ap:WW}

Consider the pion decay $\pi^-\rightarrow l_i^-+\bar{\nu}_j$, $i=1,2$ ($l_1\equiv
e,l_2\equiv \mu$) and $j=1,2,3$.
The detailed description of this decay will be made by applying the Wigner-Weisskopf
(WW) approximation method\,\cite{WW}. The WW method is essentially an improved method
of second order time dependent perturbation theory which can describe the dynamics of
decaying and decayed states at intermediate times (exponential behavior).

To calculate the decaying pion state at any time $t$, within the applicable
approximation that only $l_i\bar{\nu}_j$ states appear as decay states, it
suffices to discover the functions $\psi$ and $\chi$ in
\eq{
\label{pi(t)}
\ket{\pi(t)}\sww=
\int\! d^3\rp\,\psi(\bp,t)e^{-iE_\pi t}\ket{\pi(\bp)}
+\sum_{ij}\int\! d^3\q d^3\rk\,\chi_{ij}(\bq,\bk;t)\,e^{-i(E_{l_i}+E_{\nu_j})t}
\ket{l_i(\bq)\nu_j(\bk)}~,
}
where the spin degrees of freedom are omitted and the states
$\{\ket{\pi(\bp)},\ket{l_i(\bq)\nu_j(\bk)}\}$, $i,j=1,2$, refer to the free
states, eigenstates of $H_0$, normalized as
\eqarr{
\label{norm}
\braket{\pi(\bp')}{\pi(\bp)}&=&\delta^3(\bp-\bp')~,\cr
\braket{l_i(\bq')\nu_j(\bk')}{l_i(\bq)\nu_j(\bk)}
&=&\delta^3(\bq-\bq')\delta^3(\bk-\bk')
~.
}
The expansion \eqref{pi(t)} means we are restricted to
the lowest order of perturbation theory.

The free Hamiltonian is characterized by the free energy of the states with
physical masses
\eqarr{
H_0\ket{\pi(\bp)}&=&E_\pi(\bp)\ket{\pi(\bp)}~,
\\
H_0\ket{l_i(\bq)\nu_j(\bk)}&=&\big(E_{l_i}(\bq)+E_{\nu_j}(\bk)\big)
\ket{l_i(\bq)\nu_j(\bk)}
~,
}
where $E_\alpha(\bp)=\sqrt{\bp^2+M_\alpha^2}$ ($\alpha=\pi,l_i,\nu_j$),
and we will denote $M_{l_i}\equiv M_i$ and $M_{\nu_j}\equiv m_j$.
The interaction Hamiltonian is given by
\eq{
V= -\int d^3\rx \lag_F(\bx) + \text{counter terms}~,
}
where $\lag_F$ is the Fermi interaction Lagrangian.

Considering the total Hamiltonian
\eq{
H=H_0+V\,,
}
we can write a Schrödinger-like equation
\eqarr{
(i\frac{d}{dt}-H_0)\ket{\pi(t)}\sww&=&
\int d^3\rp\,i\frac{\partial \psi(\bp,t)}{\partial t}e^{-iE_\pi t}\ket{\pi(\bp)}
\cr
&&
+\sum_{ij}\int\! d^3\q d^3\rk\,
i\frac{\partial \chi_{ij}(\bq,\bk;t)}{\partial t}\,e^{-i(E_{l_i}+E_{\nu_j})t}
\ket{l_i(\bq)\nu_j(\bk)}~,
\\&=&
V\ket{\Psi(t)}~.
}
Contraction with the appropriate states yields
\eqarr{
\label{dpsidt1}
i\frac{\partial}{\partial t}\psi(\bp,t)&=&\frac{\delta M^2}{2E_\pi}
\psi(\bp,t)+ \sum_{ij}\int\! d^3\q d^3\rk\,
\chi_{ij}(\bq,\bk;t)\,
\bra{\pi(\bp)}V(t)\ket{l_i(\bq)\nu_j(\bk)}~,
\\
\label{dchidt1}
i\frac{\partial}{\partial t}\chi_{ij}(\bq,\bk;t)&=&
\int d^3\rp \psi(\bp,t)
\bra{l_i(\bq)\nu_j(\bk)}V(t)\ket{\pi(\bp)}~,
}
where $V(t)=e^{iH_0t}Ve^{-iH_0t}$ and $\delta M^2$ is a counter term.


From the initial conditions
\eqarr{
\label{ic:psi}
\psi(\bp,0)&=&\psi(\bp)\,,
\\
\label{ic:chi}
\chi_{ij}(\bq,\bk;0)&=&0~,
}
we can formally solve
\eq{
\chi_{ij}(\bq,\bk;t)=
-i\int_{0}^{t}\!dt'
\int \!d^3\rp\, \psi(\bp,t')
\bra{l_i(\bq)\nu_j(\bk)}V(t')\ket{\pi(\bp)}~,
}
and obtain
\eqarr{
\label{dpsidt2}
\frac{\partial}{\partial t}\psi(\bp,t)&=& -i\frac{\delta M^2}{2E_\pi}
\psi(\bp,t)
+ \int\!d^3\rp' d^3\q d^3\rk\int_{0}^{t}\!dt'\,
\bra{\pi(\bp)}V(t)\ket{l_i(\bq)\nu_j(\bk)}
\cr &&
\phantom{-i\frac{\delta M^2}{2E_\pi}\psi(\bp,t) +\int\!d^3\rp'}
\times
\bra{l_i(\bq)\nu_j(\bk)}V(t')\ket{\pi(\bp')}\psi(\bp',t')
~.
}
This is the key equation for the WW approximation.

Notice that only momentum conservation holds for the matrix elements, in particular,
\eq{
\label{<V>}
\bra{l_i(\bq)\nu_j(\bk)}V\ket{\pi(\bp)}=
N_{ij}^{-1/2}
\mathscr{M}_{ij}
\,\delta^3(\bp-\bq-\bk)~,
}
where
$N_{ij}=(2\pi)^3 2E_{l_i}(\bq)2E_{\nu_j}(\bk)2E_\pi(\bp)$ and
$\mathscr{M}_{ij}\equiv\mathscr{M}_{ij}(\bp,\bq,\bk)=
\mathscr{M}(\ms{\pi^-(\bp)\rightarrow l_i^-(\bq)\bar{\nu}_j(\bk)})
$.
Replacing Eq.\,\eqref{<V>} into Eq.\,\eqref{dpsidt2} yields
\eq{
\label{dpsidt3}
\frac{\partial}{\partial t}\psi(\bp,t)=
-i\frac{\delta M^2}{2E_\pi} \psi(\bp,t)
-\frac{1}{2E_\pi(\bp)}\int_0^{t}\!dt'\,
\psi(\bp,t-t')K(\bp,t')
~,
}
where
\eq{
\label{K}
K(\bp,t')=\frac{1}{(2\pi)^3}\sum_{ij}
\int \frac{d^3\q}{2E_{l_i}}\frac{d^3\rk}{2E_{\nu_j}}
e^{i\Delta E_{ij}t'}
|\mathscr{M}_{ij}|^2
\delta^3(\bp-\bq-\bk)~,
}
where $\Delta E_{ij}\equiv E_\pi-E_{l_i}-E_{\nu_j}$ and the respective $\bp,\bq,\bk$
dependence of $E_\pi,E_{l_i},E_{\nu_j}$ is implicit.
The expression in Eq.\,\eqref{K}, however, does not provide a convergent
integral since $|\mathscr{M}_{ij}|^2$ behaves as $\bk^2$ for
$\bq=\bp-\bk$ and $|\bk|\rightarrow\infty$.
However, a cutoff function $f(\bp,\bq,\bk)$ multiplying $\mathscr{M}_{ij}$ is
understood to regularize the expression. Such function can arise effectively from the
pion form factor and vertex corrections in higher orders\,\cite{DGH}. Such cutoff
function is necessary to ensure the convergence of Eq.\,\eqref{K} and the production
rate of $\pi(\bp)\rightarrow l_i(\bq)\bar{\nu}_j(\bk)$ to be more probable for the
energy conserving states and do not grow indefinitely for high $|\bk|$.
We will assume that the cutoff function $f$ satisfies the properties
\begin{itemize}
\item[(P1)] the functional form of $f$ is broad for
$E_{l_i}$ or $E_{\nu_j}$ and it varies very slowly for values close to the energy
conserving values, in particular $f=1$ for $\Delta E_{ij}=0$.
\item[(P2)] the suppression of high momentum $|\bk|$ or $|\bq|$ (with $\bq+\bk$
fixed) occurs only significantly at an scale $\Lambda$ which satisfies
$\Gamma\ll\Lambda\ll M^2_\pi/\Gamma$, where $\Gamma$ is the pion decay width.
\end{itemize}
Only these properties will be necessary for most of the calculations in this article.
The inclusion of an explicit cutoff function to justify the property (P2) can be
found in Ref.\,\cite{ccn:intrinsic}.

With the introduction of $f$ we can argue that the dominant contribution of
$K(\bp,t)$ is for $t \sim 0$, since Eq.\,\eqref{K} corresponds to a Fourier transform
in $E_{\nu_j}$ and the integrand is a very broad function, which leads to a narrow
function in time. We can then approximate Eq.\,\eqref{dpsidt2} as
\eqarr{
\label{dpsidt4}
\frac{\partial}{\partial t}\psi(\bp,t)\approx
-i\frac{\delta M^2}{2E_\pi} \psi(\bp,t)
-\frac{1}{2E_\pi(\bp)}\left[\int_0^{\infty}\!dt'\,K(\bp,t')\right]
\psi(\bp,t)
~.
}
The Eq.\,\eqref{dpsidt4} corresponds to the WW approximation and it is valid
for intermediate times, i.e., $t$ should be greater than the time width of
$K(\bp,t)$, since for such short time the original expression \eqref{dpsidt3}
can be significantly different.
Within the WW approximation the expression inside the bracket in Eq.\,\eqref{dpsidt4}
gives
\eq{
\label{intK}
\int_0^{\infty}\!dt'\,K(\bp,t')
=
\frac{i}{(2\pi)^3}\sum_{ij}
\int \frac{d^3\q}{2E_{l_i}}\frac{d^3\rk}{2E_{\nu_j}}
\frac{|f\mathscr{M}_{ij}|^2}
{\Delta E_{ij}+i\epsilon}
\delta^3(\bp-\bq-\bk)~.
}
Using the relation
\eq{
\frac{1}{E\pm i\epsilon}=\mathcal{P}\frac{1}{E}\mp i\pi\delta(E)~,
}
we obtain
\eqarr{
\label{ReK}
\mathrm{Re}\,\text{Eq.\,\eqref{intK}}&=&
\frac{\pi}{(2\pi)^3}\sum_{ij}
\int \frac{d^3\q}{2E_{l_i}}\frac{d^3\rk}{2E_{\nu_j}}
|f\mathscr{M}_{ij}|^2
\delta^4(\bp-\bq-\bk)~,
\\
\label{ImK}
\mathrm{Im}\,\text{Eq.\,\eqref{intK}}&=&
\frac{1}{(2\pi)^3}\sum_{ij}
\mathcal{P}\int \frac{d^3\q}{2E_{l_i}}\frac{d^3\rk}{2E_{\nu_j}}
\frac{|f\mathscr{M}_{ij}|^2}{\Delta E_{ij}}
\delta^3(\bp-\bq-\bk)~.
}
Using the property (P1) of $f$ we can identify Eq.\,\eqref{ReK} as proportional
to the pion decay rate at rest\,\cite{DGH}
\eq{
\mathrm{Re}\,\text{Eq.\,\eqref{intK}}=M_\pi \Gamma
~,
}
while Eq.\,\eqref{ImK} can be absorbed by the counterterm
\eq{
\mathrm{Re}\,\text{Eq.\,\eqref{intK}}= -\delta M^2~.
}

We can finally find the functions $\psi$ and $\chi$.
Equation \eqref{dpsidt4} gives
\eq{
\frac{\partial}{\partial t}\psi(\bp,t)=
-\frac{\Gamma}{2\gamma}\psi(\bp,t)~,
}
which can be readily solved to give
\eq{
\psi(\bp,t)=\psi(\bp)e^{-\Gamma t/2\gamma}~,
}
in accordance to the expected exponential decay law.
The factor $\gamma=E_\pi(\bp)/M_\pi$ accounts for the Lorentz dilatation of
time.
At the same time, the production wave function can be obtained from
Eq.\,\eqref{dchidt1}
\eqarr{
\chi_{ij}(\bq,\bk;t)&=&
\tilde{\chi}_{ij}(\bp,\bq,\bk;t)
\,\psi(\bp)\big|_{\bp=\bq+\bk}
\,,
\\
\tilde{\chi}_{ij}(\bp,\bq,\bk;t)
&\equiv&
\big[1-e^{-i(\Delta E_{ij}-i\Gamma/2\gamma)t}\big]
N_{ij}^{-1/2}
\frac{f\mathscr{M}_{ij}(\bp,\bq,\bk)}
{\Delta E_{ij}-i\frac{\Gamma}{2\gamma}}
\,.
}
Thus $|\chi_{ij}(\bq,\bk;t)|^2$ is the production probability density.

From the conservation of probability at any time $t$, we must check if
\eq{
\label{psi+chi}
\int \!d^3\rp|\psi(\bp,t)|^2
+\sum_{ij}\int \!d^3\q d^3\rk|\chi_{ij}(\bq,\bk;t)|^2
=1\,.
}
The calculation can be found in Ref.\,\cite{ccn:intrinsic}. The important point is
that Eq.\,\eqref{psi+chi} is satisfied if we neglect the terms that does not conserve
energy in the squared amplitude $|\mathscr{M}_{ij}|^2$, i.e., the second term in
\eq{
\label{M+dM}
\sum_{\rm spins}|\mathscr{M}_{ij}|^2=|\mathscr{M}_{ij}^{\rm EC}|^2+
|\delta\mathscr{M}_{ij}|^2
\,,
}
where the upperscript EC stands for energy conservation.
Notice that the usual energy conserving term $|\mathscr{M}_{ij}^{\rm EC}|^2$ is
positive definite while $|\delta\mathscr{M}_{ij}|^2$ has no definite sign. The
cutoff function $f$ is responsible for controlling such contributions. Therefore
we retain only the energy conserving parts of $|\mathscr{M}_{ij}|^2$ further on.

For future use, we also define
\eq{
\label{Gammaij}
M_\pi\Gamma_{ij}=
\frac{\pi}{(2\pi)^3}|\mathscr{M}_{ij}^{\rm EC}|^2
\int\! d\Omega_k\Big[
\Big(\frac{k^2}{2E_{l_i}2E_{\nu_j}}
\Big(\frac{d(E_{l_i}+E_{\nu_j})}{dk}\Big)^{-1}
\Big]_{\rm EC}
\,,
}
and
\eq{
\label{Gammai}
\Gamma_{i}=\sum_j\Gamma_{ij}\,.
}
The ratio $\Gamma_i/\Gamma$ corresponds to the branching ratio of the reaction
$\pi\rightarrow l_i+\bar{\nu}$, independent of neutrino flavor, and it practically
coincides with the usual branching ratio calculated with massless neutrinos, since
$\sum_j|U_{ij}|^2=1$ and the kinematical contribution of neutrino masses are
negligible. Obviously, $\sum_i\Gamma_i=\Gamma$.

As a last remark, we should emphasize that nowhere in this section the precise form
of the interaction was used, except in the asymptotic behavior of
$|\mathscr{M}_{ij}|^2$. Therefore, this approximation can be used in any two-body
decay for which the interaction Hamiltonian is known, as long as a proper cutoff
function is understood.


\pagebreak
\newpage
\begin{figure}
\begin{center}
\epsfig{file= 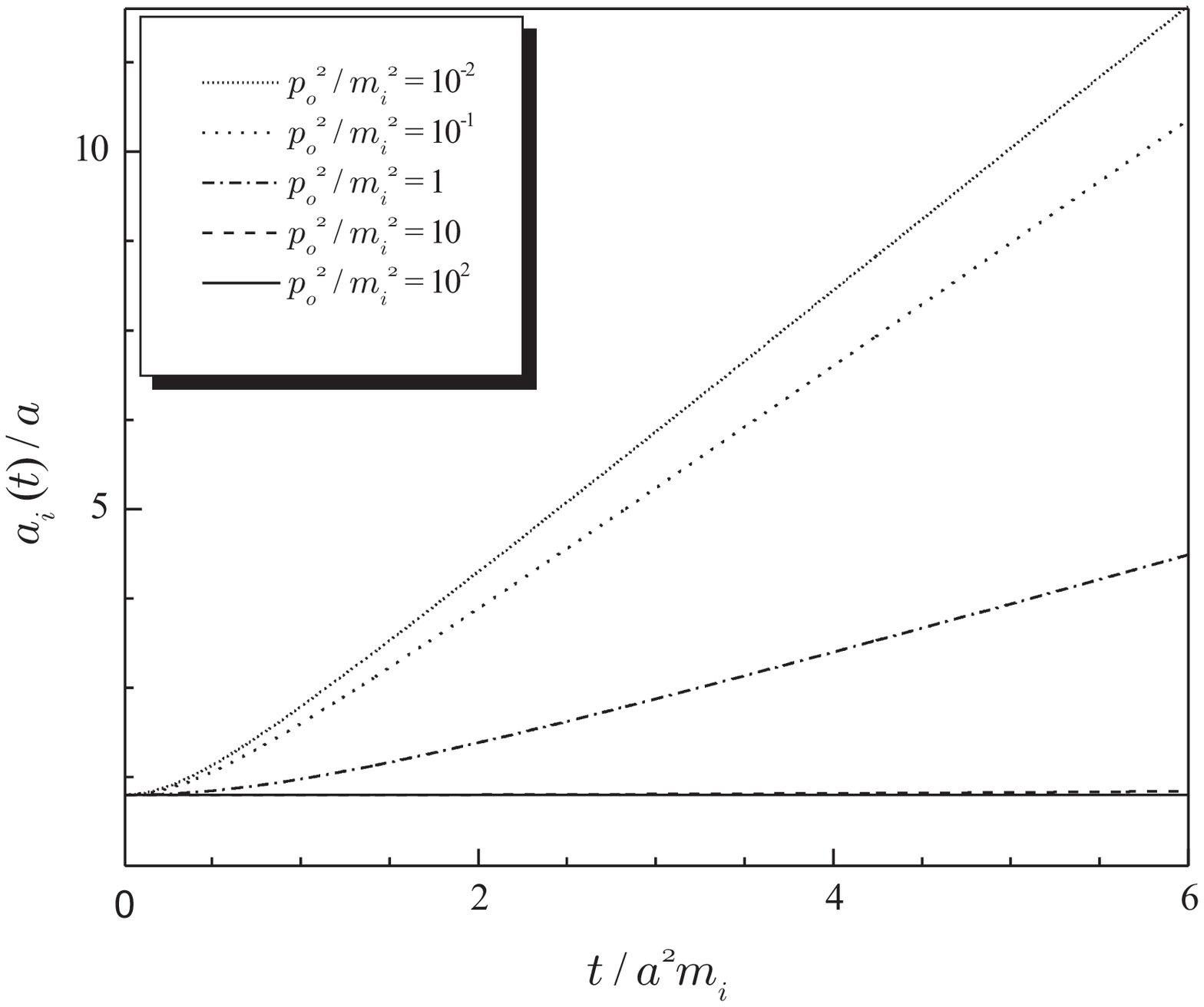, width= 11 cm}
\vspace{-0.8cm}
\end{center}
\caption{\label{an1} The time-dependence of the wave packet width $a_{\ii}\bb{t}$ is
given for different values of the ratio $p_{\0}\, / \, m_{\ii}$.
By considering a fixed mass value $m_{\ii}$, we compare the NR $(p_{\0} \ll
m_{\ii})$ and the UR $(p_{\0} \gg  m_{\ii})$ propagation regimes.
We observe that the spreading is much more relevant in the former case.
In the UR limit $(m_{\ii} = 0)$, the wave packet does not spread and $a_{\ii}\bb{t}$
assumes a constant value $a$.}
\end{figure}

\pagebreak
\newpage
\begin{figure}
\begin{center}
\epsfig{file= 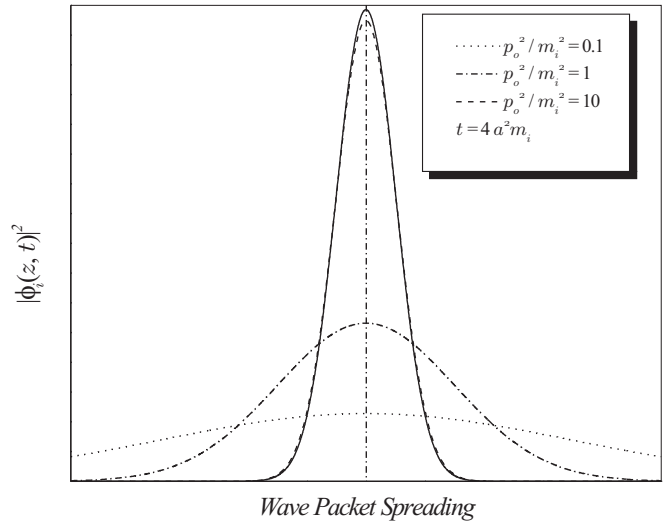, width= 11 cm}
\vspace{-0.8cm}
\end{center}
\caption{\label{an2} The wave packet spreading in both NR and UR propagation regimes
is described at time $t = 4 a^{\2} m_{\ii}$ in correspondence with Fig.\,\ref{an1}.
The solid line represent the shape of the wave packet at time $t = 0$.
In the case of an UR propagation expressed in terms of
$\frac{p_{\0}^{\2}}{m_{\ii}^{\2}} = 10$, the spreading is indeed irrelevant.}
\end{figure}

\pagebreak
\newpage
\begin{figure}
\begin{center}
\epsfig{file= 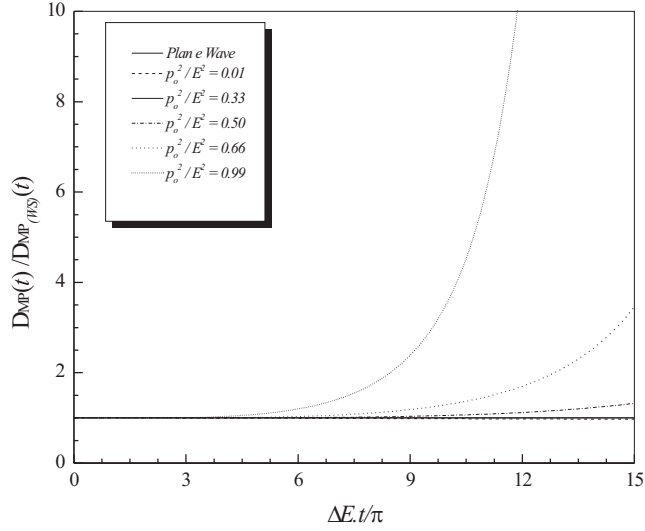, width= 11 cm}
\vspace{-0.8cm}
\end{center}
\caption{\label{an3} The comparison between the damping behavior {\em with}
($\mbox{\sc Dmp}(t)$) and {\em without} ($\mbox{\sc Dmp}_{\mbox{\tiny $WS$}}(t)$)
the second order corrections for different propagation regimes.
In order to have a realistic interpretation of the information carried by the second
order corrections we arbitrarily fix $a \, \bar{E} = 10$.
The second order corrections could be indeed effective for both NR and
(ultra)relativistic propagation regimes, however, the oscillations are destroyed
much more rapidly in the latter case.
If $\frac{p_o^{\2}}{\bar{E}^{\2}} \approx \frac{1}{3}$, the second order corrections
are minimal.
}
\end{figure}

\pagebreak
\newpage
\begin{figure}
\begin{center}
\epsfig{file= 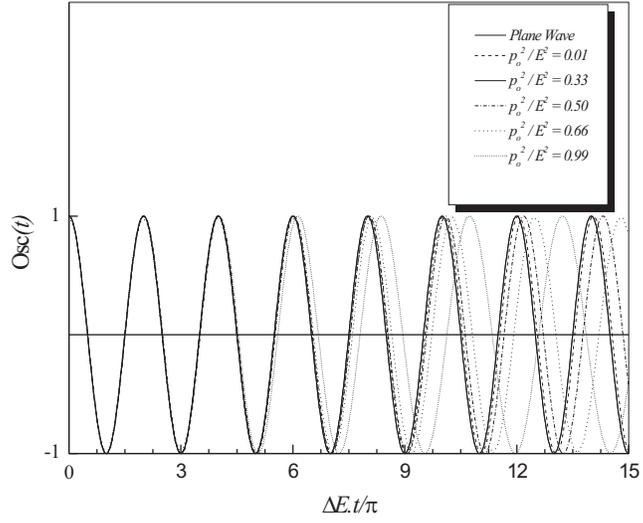, width= 11 cm}
\vspace{-0.8cm}\end{center}
\caption{\label{an4} The time-behavior of $\mbox{\sc Osc}\bb{t}$ compared with the
{\em standard} plane-wave oscillation given by $\cos{[\Delta E \, t]}$ for different
propagation regimes.
The additional phase $\Theta\bb{t}$ changes the oscillating character after some
time of propagation.
The minimal deviation occurs for $\frac{p_{\0}^{\2}}{\bar{E}^{\2}} \approx
\frac{1}{3}$ which is represented by a solid line superposing the plane-wave case.}
\end{figure}

\pagebreak
\newpage
\begin{figure}
\begin{center}
\epsfig{file= 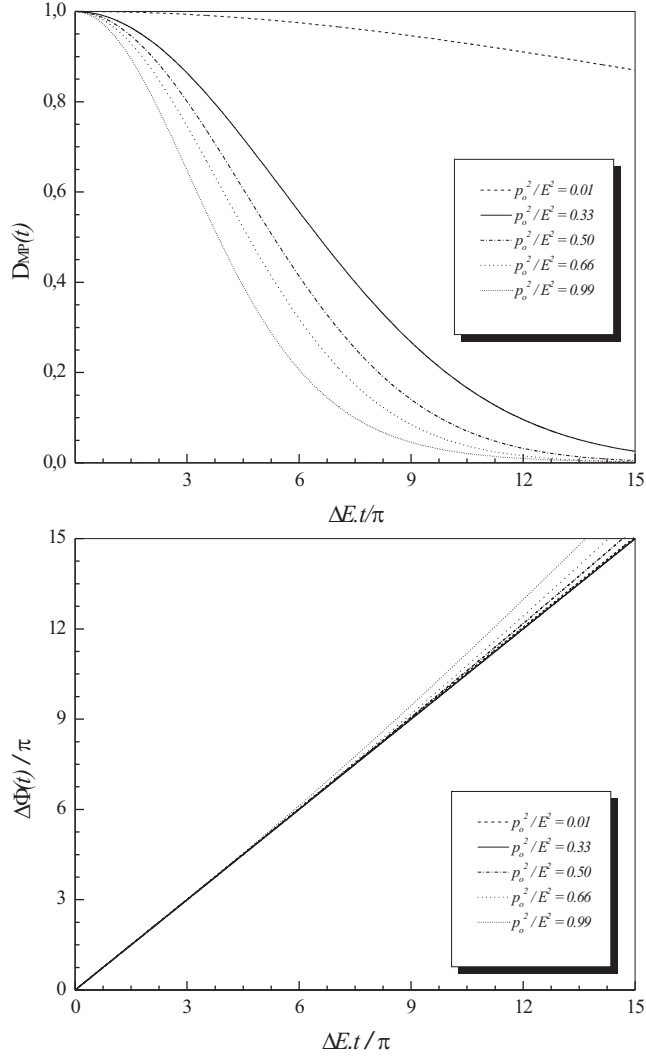, width= 11 cm}
\vspace{-0.8cm}\end{center}
\caption{\label{an5} The behavior of the corrected phase $\Delta \Phi\bb{t}
= \Delta E \, t + \Theta\bb{t}$ for different propagation regimes and we observe
that the values assumed by $\Theta\bb{t}$ are {\em  effective} only when the
interference boundary function $\mbox{\sc Dmp}\bb{t}$ does not vanish.
By diminishing the value of the wave packet parameter $a \, \bar{E}$ (we have used
$a \, \bar{E} = 10$ for this plot) the amortizing behavior is attenuated and the
range of modifications introduced by the additional phase $\Theta\bb{t}$ increases.
}
\end{figure}

\pagebreak
\newpage
\begin{figure}
\begin{center}
\epsfig{file= 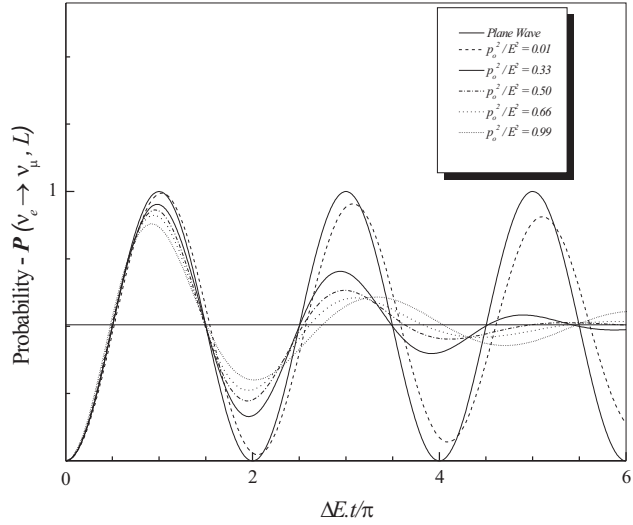, width= 11 cm}
\vspace{-0.8cm}
\end{center}
\caption{\label{an8} The time-dependence of the flavor conversion probability
obtained with the introduction of second-order corrections in the series expansion
of the energy for a strictly peaked momentum distribution
($\mathcal{O}(\sigma_{\ii}^{\3})$).
By comparing with the plane wave predictions, depending on the propagation regime, the
additional time-dependent phase $\Delta \Phi\bb{t} \equiv\, \Delta E \, t +
\Theta\bb{t}$ produces a delay/advance in the local maxima of flavor detection.
Phenomenologically, we shall demonstrate that such modifications allow us to
quantify small corrections to the averaged values of neutrino oscillation
parameters, i.\,e., the mixing-angle and the mass-squared difference.
Essentially, it depends on the product of the wave packet width $a$ by the averaged
energy $\bar{E}$.
Here again  we have used $a\bar{E} = 10$.}
\end{figure}

\pagebreak
\newpage
\begin{figure}
\begin{center}
\epsfig{file= 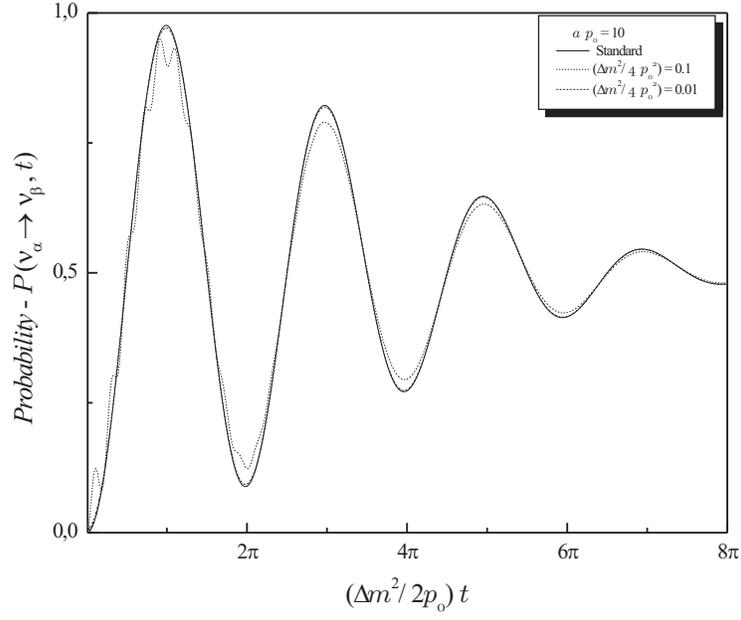, width= 11 cm}
\end{center}
\caption{The flavor oscillation probability modified by the ultra-rapid oscillations
which arise from the interference between positive and negative frequency solutions
of the Dirac equation.
For initial times, the propagating particle exhibits a violent quantum fluctuation
of its flavor quantum number around a flavor average value.
In this plot we have conveniently fixed the ratio $\frac{m_{\1}-m_{\2}}{m_{\1} +
m_{\2}} = 0.99$.}
\label{zwb}
\end{figure}

\pagebreak
\newpage
\begin{figure}
\begin{center}
\epsfig{file= 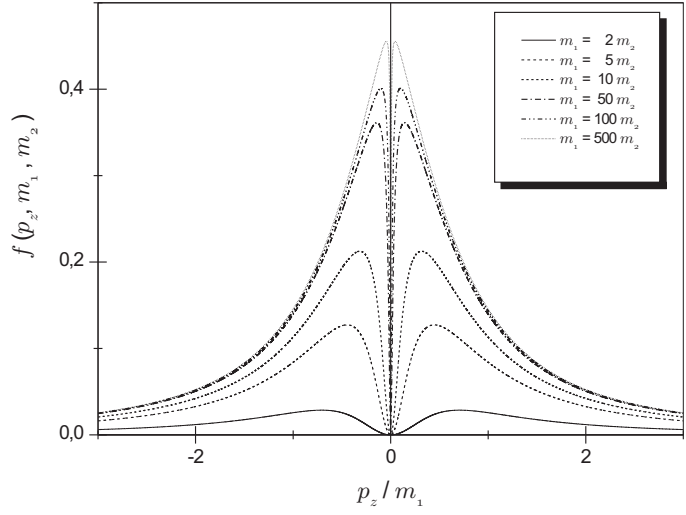, width=11 cm}
\end{center}
\caption{The function $f\bb{p_{\z},m{\1},m_{\2}}$ is plotted for different values of
the ratio between $m_{\1}$ and $m_{\2}$.
For a momentum distribution sharply peaked around $p_{\0} \gg m_{\1,\2}$ (UR limit),
$f\bb{p_{\z},m{\1},m_{\2}}$ does not play a significant role in the ``modified''
oscillation formula.
When the value of $m_{\1}$ tends to the value of $m_{\2}$, independently of the
value of $p_{\0}$ and of the width of the momentum distribution, the maximal value
assumed by $f\bb{p_{\z},m{\1},m_{\2}}$ tends to be negligible.}
\label{fig1}
\end{figure}

\pagebreak
\newpage
\begin{figure}
\begin{center}
\epsfig{file= 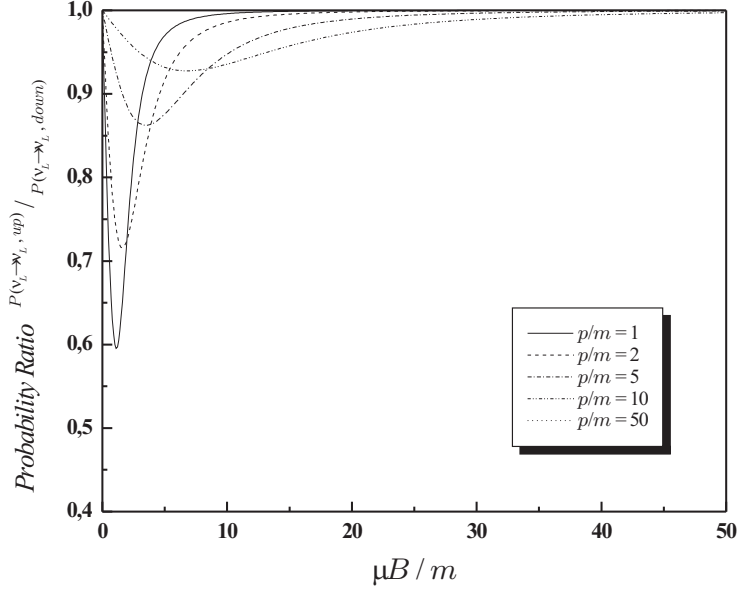, width= 11 cm}
\vspace{-0.8 cm}\end{center}
\small\caption{\label{Chi1} Ratio between the detection probabilities of
spin-up and spin-down negative chiral eigenstates as a function of the magnetic
interaction energy $\mu\mbox{\boldmath{$B$}}/m$ for different values of
$\mbox{\boldmath{$p$}}/m$ representing the propagation regime.
The results are for $w^{\dagger}\left(\frac{1-(\mi
1)^{\1}\bs{\Sigma}\ponto\hat{\bs{a}}}{2}\right)w =
w^{\dagger}\left(\frac{1-(\mi
1)^{\2}\bs{\Sigma}\ponto\hat{\bs{a}}}{2}\right)w=
\frac{1}{2}$ which corresponds to the assumption of neutrinos being created at $t=0$
with a completely random spin orientation.
It can be viewed as a collection of neutrinos where one-half are characterized by spin-up states and the remaining one
half by spin-down states.}
\end{figure}

\pagebreak
\newpage
\begin{figure}[t]
\begin{center}
\epsfig{file= 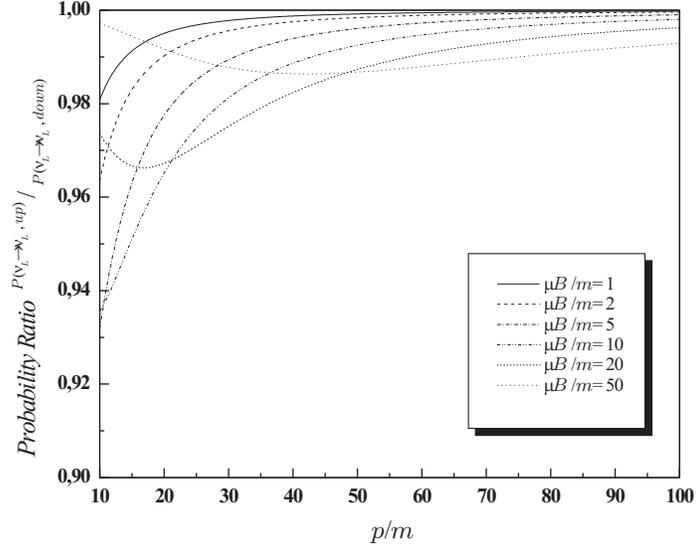, width= 11 cm}
\vspace{-0.8 cm}
\end{center}
\small\caption{\label{Chi2} Ratio between detection probabilities of
spin-up and spin-down negative chiral eigenstates as a function of $\mbox{\boldmath{$p$}}/m$ that parameterizes the propagation regime for different values of the magnetic interaction energy $\mu\mbox{\boldmath{$B$}}/m$.
We have also established that $w^{\dagger}\left(\frac{1-(\mi
1)^{\1}\bs{\Sigma}\ponto\hat{\bs{a}}}{2}\right)w =
w^{\dagger}\left(\frac{1-(\mi
1)^{\2}\bs{\Sigma}\ponto\hat{\bs{a}}}{2}\right)w=
\frac{1}{2}$.}
\end{figure}


\begin{thebibliography}{99}
\bibitem{endnote0}
Not to be confused with $\bar{E}$ in section \ref{sec:localization}, defined after
Eq.\,\eqref{II24A}.
\bibitem{endnote1}
The charge here is a global U(1) charge which is conserved for $n=1,2$ and does
not mean eletric charge. Obviously the ``flavor'' charges $\alpha=e,\mu$,
introduced by mixing, depend on time and quantifies the oscillation, as in
Eq.~\eqref{prob:em:ST}.
\bibitem{endnote2}
Recall that neutrino flavor states $\ket{\nu_\alpha}$ are superpositions of mass
eigenstates $\ket{\nu_j}$ with coefficients $U^*_{\alpha j}$ while antineutrino
flavor states $\ket{\bar{\nu}_\alpha}$ have coefficients$U_{\alpha j}$
\,\cite{giunti:torino04}.
\end{thebibliography}
\end{document}